%% file: main.tex
\documentclass[letterpaper,journal]{IEEEtran}
\usepackage{graphicx,url,textcomp,xcolor,microtype,placeins}
\usepackage{booktabs,tabularx,multirow,threeparttable,makecell}
\usepackage{mathtools,amsmath,amssymb,amsfonts,array,interval}
\usepackage{adjustbox}
\usepackage{cite}
\usepackage[inline]{enumitem}
\usepackage[ruled]{algorithm2e}
\usepackage[italic]{mathastext}
\usepackage[normalem]{ulem}
\usepackage[most]{tcolorbox}
\usepackage[simplified]{pgf-umlcd}
\usepackage[colorlinks=true, linkcolor=black, citecolor=black]{hyperref}
\usepackage[capitalise]{cleveref} 
\usepackage{marginnote}
\Crefname{figure}{Fig.}{Figs.}
\Crefname{equation}{Eq.}{Eq.}

\input{glossaries}

\newif\ifannotated%
\annotatedfalse
\newcommand{\annotate}[2][0cm]{\ifannotated\marginnote{\scriptsize\bfseries\color{blue}#2}[#1]\fi}
\newcommand{\addedtext}[1]{%
  \ifannotated{%
    \begingroup%
    \renewcommand*{\glstextformat}[1]{\textcolor{blue}{##1}}%
    \hypersetup{citecolor=blue,linkcolor=blue}%
    \color{blue}#1%
    \hypersetup{citecolor=black,linkcolor=black}%
    \endgroup%
  }%
  \else#1\fi%
}
\newcommand{\deletedtext}[1]{%
  \ifannotated%
    \begingroup%
    \renewcommand*{\glstextformat}[1]{\textcolor{red}{##1}}%
    \hypersetup{citecolor=red,linkcolor=red}%
    \color{red}\sout{#1}%
    \hypersetup{citecolor=black,linkcolor=black}%
    \endgroup%
  \fi%
}

\newcounter{definition}
\newcommand{\define}[1]{
    \stepcounter{definition}
    \textbf{Definition \thedefinition} (#1).
}

\newtcolorbox[auto counter]{finding}{enhanced,
  attach boxed title to top text left={yshift=-2mm},
  fonttitle=\bfseries, title=Answer to RQ\thetcbcounter}
\newcommand{\Finding}[1]{\begin{finding}#1\end{finding}}

\makeatletter
\def\pgfumlcd@enumeration{\pgfumlcd@enumeration@[]}
\def\pgfumlcd@enumeration@[#1]#2#3{%
  \pgfumlcd@classAndInterfaceCommon{#1}{#2}{#3}%
}
\def\endpgfumlcd@enumeration{%
    \pgfumlcd@calcuateNumberOfParts{}%
    \node[anchor=north,this umlcd style] (\pgfumlcd@ClassName) at (\pgfumlcd@ClassPos) {%
      \(<<\)enumeration\(>>\)\break{}
      {\pgfutil@font@bfseries\pgfumlcd@ClassName} 
      \pgfumlcd@insertAttributesAndOperations{}%
    };%
  \endpgfumlcd@classAndInterfaceCommon{}
}
\tikzaddtikzonlycommandshortcutlet{\enumeration}{\pgfumlcd@enumeration}
\tikzaddtikzonlycommandshortcutlet{\endenumeration}{\endpgfumlcd@enumeration}
\makeatother

\newcolumntype{Y}{>{\centering\arraybackslash}X}

\begin{document}

\title{Using Cooperative Co-evolutionary Search to Generate Metamorphic Test Cases for Autonomous Driving Systems}

\author{Hossein Yousefizadeh,
  Shenghui Gu,
  Lionel C. Briand,~\IEEEmembership{Fellow,~IEEE,}
  and Ali Nasr
  \thanks{Hossein Yousefizadeh and Shenghui Gu are with the School of EECS, University of Ottawa, Canada. E-mails: hyous028@uottawa.ca; sgu2@uottawa.ca}
  \thanks{Hossein Yousefizadeh and Shenghui Gu contributed equally to this work.}
  \thanks{Lionel C. Briand holds shared appointments with the School of EECS, University of Ottawa, Canada, and the Research Ireland Lero centre, University of Limerick, Ireland. E-mail: lbriand@uottawa.ca}
  \thanks{Ali Nasr is with Waterloo Research Center of Huawei Technologies Canada Co., Ltd., Canada. E-mail: ali.nasr2@huawei.com}
}

\markboth{IEEE Transactions on Software Engineering,~Vol.~14, No.~8, August~2021}%
{Shell \MakeLowercase{\textit{et al.}}: A Sample Article Using IEEEtran.cls for IEEE Journals}

\maketitle

\input{Abstract}
\input{Introduction}
\input{Problem}
\input{Related_work}
\input{Approach}
\input{Evaluation}
\input{Discussion}
\input{Conclusion}

\section*{Data Availability}
The replication package for our experiments, which includes the implementation of our approach, baseline methods, configuration files, and experimental results, is publicly available on Figshare at \url{https://doi.org/10.6084/m9.figshare.27910677}. 

\section*{Acknowledgments}
This work was supported by a research grant from Huawei Technologies Canada Co., Ltd., as well as the Canada
Research Chair and Discovery Grant programs of the Natural Sciences and Engineering Research Council of Canada (NSERC). Lionel C. Briand’s contribution was partially funded by the Research Ireland grant 13/RC/209.

\bibliographystyle{IEEEtran}
\bibliography{IEEEabrv,main}

\input{Appendices}







\end{document}

%% file: glossaries.tex
\usepackage[acronym,nogroupskip]{glossaries}

\renewcommand{\glossarysection}[2][]{}
\setacronymstyle{long-short}

\newacronym{ads}{ADS}{Autonomous Driving System}
\newacronym{dnn}{DNN}{Deep Neural Network}
\newacronym{dtw}{DTW}{Dynamic Time Warping}
\newacronym{ast}{AST}{Adaptive Stress Testing}
\newacronym{rl}{RL}{Reinforcement Learning}
\newacronym{rss}{RSS}{Responsibility-Sensitive Safety}
\newacronym{ours}{CoCoMEGA}{Cooperative Co-evolutionary MEtamorphic test Generator for Autonomous systems}
\newacronym{mt}{MT}{Metamorphic Testing}
\newacronym{mr}{MR}{Metamorphic Relation}
\newacronym{ccea}{CCEA}{Cooperative Co-Evolutionary Algorithm}
\newacronym{sga}{SGA}{Standard Genetic Algorithm}
\newacronym{rs}{RS}{Random Search}
\newacronym{rq}{RQ}{Research Question}
\newacronym{ds}{DS}{Distinct Solutions}
\newacronym{sd}{SD}{Solution Diversity}
\newacronym{apd}{APD}{Average Pairwise Distance}
\newacronym{mrc}{MRC}{\glsentrylong{mr} Coverage}
\newacronym{cmr}{CMR}{Combinations of \glsentrylongpl{mr}}
\newacronym{auc}{AUC}{Area Under the Curve}

%% file: Abstract.tex
\begin{abstract}
Autonomous Driving Systems (ADSs) rely on Deep Neural Networks, allowing vehicles to navigate complex, open environments. However, the unpredictability of these scenarios highlights the need for rigorous system-level testing to ensure safety, a task usually performed with a simulator in the loop.
Though one important goal of such testing is to detect safety violations, there are many undesirable system behaviors, that may not immediately lead to violations, that testing should also be focusing on, thus detecting more subtle problems and enabling a finer-grained analysis.
This paper introduces Cooperative Co-evolutionary MEtamorphic test Generator for Autonomous systems (CoCoMEGA), a novel automated testing framework aimed at advancing system-level safety assessments of ADSs. CoCoMEGA combines Metamorphic Testing (MT) with a search-based approach utilizing Cooperative Co-Evolutionary Algorithms (CCEA) to efficiently generate a diverse set of test cases. CoCoMEGA emphasizes the identification of test scenarios that present undesirable system behavior, that may eventually lead to safety violations, captured by Metamorphic Relations (MRs).
When evaluated within the CARLA simulation environment on the Interfuser ADS, CoCoMEGA consistently outperforms baseline methods, demonstrating enhanced effectiveness and efficiency in generating severe, diverse MR violations and achieving broader exploration of the test space.
\addedtext{Further expert assessments of these violations confirmed that most represent real safety risks, which validates their practical relevance.}
These results underscore CoCoMEGA as a promising, more scalable solution to the inherent challenges in ADS testing with a simulator in the loop. Future research directions may include extending the approach to additional simulation platforms, applying it to other complex systems, and exploring methods for further improving testing efficiency such as surrogate modeling.
\end{abstract}

\begin{IEEEkeywords}
Autonomous driving systems,
system-level testing,
metamorphic testing,
search-based testing,
cooperative co-evolutionary algorithms,
automated testing.
\end{IEEEkeywords}

%% file: Introduction.tex
\section{Introduction}
\IEEEPARstart{A}{utonomous} Driving Systems (\glsfirst{ads}s\glsunset{ads}) have made remarkable strides and have garnered substantial research interest in recent years~\cite{tahir2020coverage}.
Typically, autonomous driving involves using different perception sensors such as cameras, LiDAR, and radar to enable a vehicle to operate without human intervention~\cite{tian2018deeptest}.
\annotate{Comment 3.1}\addedtext{The vehicle equipped with these sensors and controlled by the \gls{ads} is referred to as the ego vehicle.
It continuously gathers information about its surroundings, such as the position and movement of other vehicles, and feeds this data into the \gls{ads} to facilitate decision-making.}
An \gls{ads} deployed in a\addedtext{n ego} vehicle usually integrates several complex modules, including perception, prediction, planning, and control.
Recent advances in \glspl{dnn} have further enhanced these modules, enabling \glspl{ads} to adapt their driving behavior to constantly changing and unpredictable environments~\cite{bojarski2016end} and numerous complex and hazardous situations in real-world scenarios~\cite{tahir2020coverage}.

In spite of their achievements, it remains crucial that \glspl{ads} undergo comprehensive system-level testing to ensure their safe functioning, particularly in uncommon situations where pedestrians or vehicles behave unpredictably~\cite{lou2022testing,mcphee2024safety}.
System-level testing refers to evaluating \glspl{ads} as a whole---making the system interact with its environment through realistic scenarios over a period of time---rather than focusing on individual modules such as perception or decision-making in isolation, based on single inputs.
Although developers and manufacturers strive to mitigate potential hazards throughout the development cycle, accidents involving \glspl{ads} regularly occur~\cite{banerjee2018hands,listoffatalities}.
Though fatal crashes are rare, non-fatal accidents are more common~\cite{carsurance2022selfdriving}, highlighting the ongoing challenges in ensuring the safety and reliability of \glspl{ads}~\cite{garcia2020comprehensive}.

\annotate{Comment 1.4}\addedtext{
In this paper, our focus is on validating the safety of the \glspl{ads}, aiming to assess whether the system behaves appropriately under realistic and challenging scenarios. In particular, we aim to identify subtle behavioral anomalies that may not result in immediate safety violations (e.g., collisions or traffic rule infractions) but reflect latent flaws in the system that could eventually compromise safety, especially in edge cases. Thus, this type of validation focuses on testing the external aspects of system behavior (e.g., safety), rather than evaluating the system's ability to function (e.g., failure rate measurement), which is a different objective and is typically addressed in \emph{reliability testing}~\cite{musa2004software}.
}

The system-level testing of \glspl{ads} is particularly challenging due to the huge and diverse set of hazards they face.
Recent testing techniques primarily focus on approaches based on safety metrics~\cite{tang2023survey,cheng2024evaluating}, which quantify how close an \gls{ads} is to dangerous situations (e.g., collisions), as well as the assessment of traffic rule violations.
These metrics can be calculated by analyzing system behavior during simulations and defined as formal specifications like Signal Temporal Logic~\cite{corso2020interpretable}.
Temporal metrics such as Time-to-Collision~\cite{tuncali2019rapidly} and Worst-Time-to-Collision~\cite{wakabayashi2003traffic}, assess the ego vehicle's proximity to potential collisions over time.
Other metrics, such as Time Headway and Time-to-React~\cite{tamke2011flexible}, estimate how much time the vehicle has to respond to potential hazards.
While safety metrics provide a useful quantitative framework for detecting violations of known safety properties, they may not fully capture subtle or indirect unsafe system behavior that may eventually lead to violations~\cite{cheng2024evaluating}, which we refer to as unsafe behavior hereafter.
For example, an \gls{ads} may unnecessarily apply excessive braking in low-risk scenarios, such as slowing down abruptly when another vehicle is far ahead.
Although this behavior does not immediately compromise safety, it can lead to negative consequences, including passenger discomfort, reduced fuel efficiency, and potentially even rear-end collisions.
Such issues emphasize the need for testing methods that explore such scenarios that are not entailing immediate safety risks but are still critical for evaluating the \gls{ads}.

However, existing software testing approaches lack the complexity and adaptability required to address the intricacies of \glspl{ads}~\cite{birchler2024roadmap,koopman2016challenges}.
A key challenge lies in specifying test oracles for \glspl{ads}, as they operate in the open contexts that constitute road traffic, where expected system behavior cannot always be precisely predicted or defined in advance.
Indeed, \glspl{ads} face an infinite number of potential traffic scenarios, making it nearly impossible to determine expected outcomes for every situation~\cite{haq2022efficient,zhong2023neural}.
The challenge is further compounded by the widespread adoption of \glspl{dnn} in \glspl{ads}. While \glspl{dnn} enable sophisticated decision-making and adaptability, they lack transparency and formal specifications, making their behavior difficult to evaluate~\cite{rajabli2021software}.
For example, determining whether an \gls{ads} has behaved correctly when encountering a pedestrian in foggy conditions is heavily influenced by environmental factors and sensor limitations. The lack of clear criteria for evaluating these decisions highlights the difficulty of defining reliable test oracles.
Another challenge stems from the open nature of the environment in which \glspl{ads} operate, where new and unexpected scenarios can continuously arise.
This openness means that test cases can never be exhaustive, as it is infeasible to cover all possible conditions in such a complex and dynamic domain.
Therefore, it is reasonable to conclude that getting absolute proof of \gls{ads} safety and reliability is unattainable~\cite{riedmaier2020survey}.
However, the effective and efficient identification of unsafe behavior emerges as an important and significant challenge.

To address these challenges, we propose \emph{\gls{ours}}, an effective and efficient automated test case generation method for system-level testing of \glspl{ads}.
Specifically, we combine \gls{mt} and a search-based testing approach using \gls{ccea}~\cite{potter1994cooperative} to automatically generate a comprehensive and diverse set of system test cases that can effectively identify potential unsafe behavior in the target \gls{ads}.
\gls{ccea} is used to efficiently explore the vast and complex input space of \glspl{ads}, increasing the likelihood of uncovering unsafe behavior.
\gls{mt}, in turn, provides a mechanism to automatically detect unsafe behavior, through the violation of \glspl{mr}, such as verifying that the ego vehicle's steering angle remains consistent when weather conditions change.

This approach allows us to test \gls{ads} more comprehensively, going beyond the sole reliance on safety properties \addedtext{and aligning with established industrial standards such as ISO 21448's Safety\annotate{Comment 1.2} of the Intended Functionality (SOTIF)~\cite{iso21448}, which emphasizes safety in the presence of functional or performance limitations. Indeed, \glspl{mr} allow us to capture conditions where a system may behave undesirably despite being fault-free, a core concern of SOTIF\@.
Another} key advantage of \gls{mt} is that it does not require a complete specification to derive useful \glspl{mr}, which is often unrealistic, particularly for \glspl{ads}.
Furthermore, given the complexity of \gls{ads} scenarios, with their large number of parameters and wide range of values, the search process and convergence to safety violations can be slow.
\gls{mt} expedites this convergence by selectively mutating only the parameters related to \glspl{mr}.
By leveraging \glspl{mr}, we can detect subtle, unsafe behaviors that may not immediately violate safety properties but are eventually expected to lead to problems, thus providing a more thorough assessment of system.
\addedtext{To ensure practical applicability, our method is compatible with simulation environments that support standard formats such as OpenScenario~\cite{openscenario}.}

We have evaluated \emph{\gls{ours}} on the \textsc{Carla} simulator using the \textsc{Interfuser} \gls{ads}.
Our evaluation results demonstrate that \emph{\gls{ours}} has significantly greater effectiveness and efficiency than baseline methods in identifying unsafe behavior and exploring diverse regions of the search space.
Its advantage is consistent across a wide range of search budgets and settings.
This superior performance highlights \emph{\gls{ours}} as a promising and scalable solution for testing \glspl{ads}.

The major contributions of this paper can be summarized as follows:

\begin{itemize}
    \item We proposed \emph{\gls{ours}}, the first automated testing method that combines \gls{mt} and \gls{ccea} to enable the effective and efficient system-level testing of \glspl{ads}.
    \item We applied \emph{\gls{ours}} to a complex case study utilizing an industry-grade simulator and a high-performing \gls{ads} incorporating \gls{dnn} modules.
    \item We evaluated the effectiveness and efficiency of \emph{\gls{ours}} through large scale experiments and comparisons with baseline methods, demonstrating its ability to identify a significant number of test cases that violate the \glspl{mr} and thus exhibit unsafe behavior.
\end{itemize}

\annotate{Comment 1.7}\addedtext{
Note that while we have dedicated considerable effort to comprehensively review the literature, identify relevant \glspl{mr} and validate them with experts, this is not intended to be a contribution of this paper.
Instead, this compilation of \glspl{mr} is simply used to support our experimental evaluation on a rich and diverse set of \glspl{mr} that were defined independently from our study, within feasible computational bounds.
Fully developing, validating, and integrating a more comprehensive set of \glspl{mr} would require collaboration among many domain experts, which is beyond the scope of this study.
}

The remainder of this paper is structured as follows.
\Cref{sec:problem} outlines the challenges and defines the problem in detail.
\Cref{sec:related_work} introduces related research works and existing gaps.
\Cref{sec:approach} describes \emph{\gls{ours}} in detail.
\Cref{sec:evaluation} presents an empirical evaluation of the proposed approach.
\Cref{sec:discussion} discusses the results and threats to validity.
\Cref{sec:conclusion} concludes the paper.

%% file: Problem.tex
\section{Problem and Challenges}\label{sec:problem}
In this section, we define the research problem and discuss its associated challenges.

\subsection{Problem Definition}
The primary problem to be addressed in this research is the development of an effective and efficient system-level testing methodology for \glspl{ads} to identify potential unsafe behavior, and thus determine when \glspl{ads} can be trusted under diverse and unpredictable scenarios and conditions.
This approach focuses on the \gls{ads} interactions with the environment within realistic scenarios, rather than testing individual modules in isolation.
To address this problem, we need to tackle the following specific issues.

\begin{itemize}
    \item \textbf{Scalability and coverage.}
          Real-world system testing is unable to exhaustively cover the vast space of possible execution scenarios.
          However, in the context of critical systems, such as \glspl{ads}, identifying unsafe scenarios is crucial for system assurance, even when they are unlikely.
    \item \textbf{Oracle problem.}
          Defining precise criteria to determine acceptable behavior in \glspl{ads}, particularly for complex tasks like perception and decision-making, is challenging.
          The ever-changing and unpredictable nature of real-world traffic scenarios makes it impossible to enumerate all potential outcomes of \glspl{ads} and define corresponding expectations.
          For instance, consider a situation where a pedestrian is standing at a crosswalk or a traffic light turns yellow as the ego vehicle approaches at high speed. Deciding whether to stop abruptly, proceed, or adjust speed depends on various factors such as weather conditions and nearby vehicles.
          Such complexities make it challenging to develop effective test oracles, potentially making system testing ineffective.
    \item \textbf{Computational costs and efficiency.}
          The computational expense associated with high-fidelity simulations presents a significant challenge, effectively limiting system testing in practice and the ability to rapidly iterate on system improvements.
\end{itemize}

Given the above, there is a critical need for innovative testing methodologies that can effectively and efficiently test \glspl{ads} across a wide range of scenarios and conditions to identify unsafe behavior.
The integration of \gls{mt} with advanced search techniques offers a promising solution, allowing for the systematic and efficient exploration of scenario spaces and the detection of undesirable system behaviors that do not necessarily and systematically lead to safety property violations but are indicative of potential safety problems.

The primary issue tackled in this work is the development of a system-level testing framework, assuming an environment simulator in the loop, that combines \gls{mt} with advanced search-based techniques to automatically generate test scenarios for \glspl{ads}.
\gls{mt} is especially suitable in this context because it facilitates the verification of system behavior without requiring an exhaustive system specification, focusing instead on identifying violations of \gls{mr} properties that must hold across different inputs.

Specifically, we re-express the problem as a search problem.
Let \(\mathcal{S}\) denote the space of all possible driving scenarios, and let \(\mathcal{Q}\) denote the space of all possible perturbations derived from a set of \glspl{mr}, all sharing the same output relation \(or\).
Let \(E_{or}(s, q)\) be a function that quantifies the extent of violation of \(or\) between the source scenario \(s \in \mathcal{S}\) and the follow-up scenario \(q(s)\) derived from applying \(q \in \mathcal{Q}\) to \(s\).
Imagine we are testing an \gls{ads} to check its behavior when encountering a pedestrian in various weather conditions.
Here, \(\mathcal{S}\) represents all possible driving scenarios, such as different road types, traffic conditions, and initial pedestrian positions.
The set \(\mathcal{Q}\) represents possible perturbations, like changes in weather conditions (e.g., fog or rain) based on an \gls{mr} with an output relation \(or\) specifying that the \gls{ads} should reduce speed when visibility decreases.
Now, suppose we have a source scenario \(s \in \mathcal{S}\) where the \gls{ads} encounters a pedestrian on a clear day.
By applying a perturbation \(q \in \mathcal{Q}\), we create a follow-up scenario \(q(s)\) in which fog or rain reduces visibility.
The function \(E_{or}(s, q)\) then quantifies the extent of violation in the speed reduction of the \gls{ads} in response to this visibility change, measuring how well the \gls{ads} meets the expected output relation \(or\).

\define{Problem}
The problem is to find a diverse set \(SP\) of scenario-perturbation pairs \((s, q) \in \mathcal{S} \times \mathcal{Q}\) such that the resulting pair \((s, q(s))\) violates \(or\):

\begin{equation}
    SP = \{(s, q) \in \mathcal{S} \times \mathcal{Q} \mid E_{or}(s, q) > 0\}.
\end{equation}

The challenge lies in efficiently searching the vast space \(\mathcal{S} \times \mathcal{Q}\) to discover pairs \((s, q)\) that maximize the detection of \gls{mr} violations, thereby exposing unsafe behavior in the \gls{ads}. This requires balancing thorough and efficient exploration of the search space, ensuring that a diverse set of test cases that violate the \glspl{mr} is generated.

\subsection{Challenges}
However, the development of this testing methodology entails addressing several key challenges.

\begin{itemize}
    \item \textbf{Efficiency.}
          Real-world driving scenarios are inherently complex, involving a multitude of dynamic and static objects, such as vehicles, pedestrians, and environmental conditions.
          Thus, the vast space of potential test scenarios is defined by various attributes such as vehicle speed, weather conditions, and object positions.
          Efficiently exploring this space to identify scenarios that violate \glspl{mr} is computationally intensive.
          This challenge is compounded by the high cost of simulations, necessitating a strategy that minimizes the number of scenarios to be tested while maximizing their violation-detection capability.
          For instance, instead of testing every possible combination of vehicle speeds and weather conditions, testing can prioritize key scenarios like a car making a sudden stop during heavy rain or encountering a pedestrian in foggy conditions.
          This approach allows testers to focus on critical, high-risk scenarios, rather than wasting resources on every minor variation.
          However, finding what such high-risk scenarios are in practice is not an easy endeavor.
    \item \textbf{Metamorphic relation definition.}
          \gls{mt} requires well-defined \glspl{mr} that describe expected behaviors or invariants across different test scenarios. Defining comprehensive and meaningful \glspl{mr} can be challenging, for example in \glspl{ads} where system behavior can be influenced by a wide range of factors.
          \glspl{mr} must be specific enough to detect subtle unsafe behavior while being general enough to widely apply across various scenarios. However, \glspl{mr} do not need to be perfect and complete to be useful and can be improved over time through the collective endeavor of domain experts.
    \item \textbf{Diversity of test scenarios.}
          Ensuring a diverse set of test scenarios is crucial for trustworthy testing.\glspl{ads} are likely to encounter many rare cases that were not considered during their training process~\cite{rajabli2021software}. A lack of diversity in test scenarios can lead to inadequate system evaluation and reduce the effectiveness of identifying violations. The challenge lies in developing mechanisms that promote diversity in the generated scenarios, thus covering a broader spectrum of potential real-world conditions, while still being guided towards problematic scenarios.
          For example, an \gls{ads} might need to be tested with a mix of pedestrian behaviors and varying weather conditions.
          One test could involve a pedestrian slowly crossing a busy road in clear daylight, while another might involve the same crossing at night in the rain.
          This variety ensures that the system is exposed to different lighting and environmental conditions, increasing the chances of identifying potential weaknesses in the system's response to real-world complexities.
\end{itemize}

Addressing these challenges requires a multifaceted approach that combines \gls{mt} with advanced search techniques.
Our proposed method employs \glspl{ccea} to generate and evaluate test scenarios, optimizing both the effectiveness and efficiency of the testing process, as justified in \cref{sec:approach}.

%% file: Related_work.tex
\section{Related Work}\label{sec:related_work}

In this section, we explore existing research pertaining to the issue of system-level testing for \glspl{ads}.
Given the components of our approach, we begin by reviewing search-based testing methods, followed by an examination of \gls{mt} approaches. Finally, we briefly highlight \gls{ast} techniques for system-level testing of \glspl{ads}.

\subsection{Search-Based Testing for ADSs}

One prominent approach to system-level testing of \glspl{ads} is search-based testing, which emphasizes the systematic exploration of the input parameter space of \glspl{ads} to identify sub-spaces that may lead to violations or unexpected behaviors.

Ben Abdessalem et al.~\cite{benabdessalem2016testing} introduced a multi-objective search-based testing approach that uses neural network-based surrogate models to test advanced driver assistance systems in a simulated environment. In a later investigation~\cite{benabdessalem2018testing}, the researchers utilized a similar search-based method to detect feature interaction failures, such as conflicts between automated functionalities, demonstrating that their approach could identify a significantly higher number of interaction failures compared to baseline methods.
Dreossi et al.~\cite{dreossi2019compositional} introduced a compositional framework for search-based testing, particularly designed for \glspl{ads} with machine learning components.
Their method combines constraints from both the perception input domain and the overall system input domain, which reduces computational effort and improves the efficiency of finding counterexamples.
In a practical setting, Li et al.~\cite{li2020av} developed \emph{AV-Fuzzer}, a testing framework that perturbs the maneuvers of traffic participants using a genetic algorithm to identify safety violations in the \textsc{Apollo} platform~\cite{baiduapollo}. Their results showed that \emph{AV-Fuzzer} outperformed random fuzzing and \gls{ast} by uncovering a broader range of unique safety-critical scenarios and detecting violations more efficiently.
Kolb et al.~\cite{kolb2021fitness} proposed a collection of fitness function templates for search-based testing of \glspl{ads} in intersection scenarios, extending previous work on highway scenarios~\cite{hauer2019fitness}. Through comparison with random testing baselines, their study demonstrated that the adapted fitness functions effectively guided the search toward generating diverse and safety-critical intersection scenarios.
Luo et al.~\cite{luo2021targeting} introduced \emph{EMOOD}, an evolutionary search-based testing approach designed to generate test scenarios that reveal diverse combinations of requirements violations in \glspl{ads}.
Birchler et al.~\cite{birchler2023cost, birchler2023single} explored test case prioritization methods to optimize the regression testing process in \glspl{ads}, showing that their approach improved early fault detection and testing efficiency over baseline methods.

The relationship between test inputs and system behavior has also been a research focus. Riccio and Tonella~\cite{riccio2020model} introduced the concept of a \emph{frontier of behaviors}, which marks the input boundary (threshold) at which the \gls{ads} begins to exhibit abnormal behavior.
Zohdinasab et al.~\cite{zohdinasab2021deephyperion,zohdinasab2023efficient} further refined this concept by using \emph{Illumination Search}~\cite{mouret2015illuminating} to explore the feature space, leading to more effective scenario identification.
Castellano et al.~\cite{castellano2021analysis} investigated the impact of different road representations on search-based testing for lane-keeping systems, concluding that road curvature and orientation are key factors influencing system behavior.

Further advancements in search-based testing have been achieved through the development of enhanced algorithms.
For instance, Gambi et al.~\cite{gambi2019automatically} proposed a novel approach that integrates procedural content generation (PCG) with search-based testing to systematically generate virtual road scenarios for evaluating lane-keeping functionalities.
Goss and Akba\c{s}~\cite{goss2022eagle} applied an \emph{Eagle Strategy}, which first broadly samples the state space to locate critical areas, then shifts to a focused local search around detected areas to define critical scenario boundaries more precisely. Zheng et al.~\cite{zheng2020rapid} proposed a quantum genetic algorithm to reduce the necessary population size for scenario generation.

Considering the high computational cost of \gls{ads} simulations, surrogate models have been developed to accelerate the testing process.
Ben Abdessalem et al.~\cite{benabdessalem2016testing} created a surrogate model that links scenario parameters to fitness functions, facilitating the reduction of non-critical parameters.
Likewise, Batsch et al.~\cite{batsch2021scenario}, Beglerovic et al.~\cite{beglerovic2017testing}, and Sun et al.~\cite{sun2020adaptive} explored different surrogate models to enhance the efficiency of search-based testing. More recently, Haq et al.~\cite{haq2022efficient} introduced \emph{SAMOTA}, a surrogate-assisted many-objective optimization approach that accelerates the detection of safety violations by using surrogate models to approximate the simulator, thereby reducing computational costs while effectively guiding the search toward critical scenarios.

In conclusion, search-based testing has proven effective in identifying critical failures in \glspl{ads} through systematic exploration of the input parameter space. Its strengths lie in the ability to generate diverse but targeted test cases, often uncovering edge-case scenarios that other testing methods may overlook.
However, current state-of-the-art search-based techniques for system-level testing of \glspl{ads} remain computationally intensive, particularly when applied to large-scale, complex scenarios with high-dimensional input spaces and high-fidelity \gls{ads} simulations.
While advancements such as surrogate models have helped reduce some of these computational costs, existing methods continue to face difficulties in efficiently exploring large parameter spaces to meet testing objectives. These challenges emphasize the need for further advancements to improve the scalability of search-based \gls{ads} testing frameworks.

\subsection{Metamorphic Testing for ADSs}

\acrfull{mt}~\annotate{Comment 1.1}\addedtext{\cite{chen2018metamorphic},}\cite{chen2020metamorphic} has become a vital technique for validating \glspl{ads}, particularly when one is struggling with the oracle problem.

At the module level, \gls{mt} has been widely used to evaluate the robustness and reliability of perception modules in \glspl{ads}.
Yang et al.~\cite{yang2024metalidar} introduced \emph{MetaLiDAR}, an automated \gls{mt} framework for LiDAR-based \glspl{ads}, focusing on object-level transformations to generate realistic follow-up point clouds. The framework defines three \glspl{mr}---object insertion, deletion, and affine transformations---targeting the evaluation of LiDAR-based object detection modules by ensuring consistency in detected object properties before and after transformations. Following up on this, they introduced \emph{MetaSem}~\cite{yang2024metasem}, an \gls{mt} approach that uses semantic-based transformations, such as adding, removing, or replacing scene elements (e.g., traffic signs, road markings, and traffic signals) to evaluate the consistency of system behaviors, resulting in the detection of more errors compared to prior methods.
Zhou and Sun~\cite{zhou2019metamorphic} demonstrated the effectiveness of \gls{mt} in detecting fatal errors in the LiDAR perception module of ADSs, revealing that even small, randomly added LiDAR data points outside the region of interest could cause the system to overlook nearby obstacles.

\gls{mt} has also been adapted for system-level testing. Tian et al.~\cite{tian2018deeptest} introduced \emph{DeepTest}, a tool that leverages \gls{mt} by defining transformation-specific \glspl{mr} (e.g., variations in lighting or weather conditions) to detect erroneous behaviors in \gls{dnn}-driven \glspl{ads}. Zhang et al.~\cite{zhang2018deeproad} extended this with \emph{DeepRoad}, using Generative Adversarial Networks (GAN) to generate realistic weather conditions for \gls{mt},which enabled the detection of more inconsistent steering behaviors across scenarios compared to \emph{DeepTest}.
Pan et al.~\cite{pan2021metamorphic} introduced a \gls{mt} approach for \glspl{ads} specifically in foggy conditions by defining \glspl{mr} based on fog density and direction.
In another study, Han and Zhou~\cite{han2020metamorphic} applied metamorphic fuzz testing, to automatically generate and evaluate unexpected scenarios. It uses \glspl{mr} as a filtering mechanism to differentiate true failures from false positives (alarms), when unexpected changes occur in the simulated environment.
Cheng et al.~\cite{cheng2024evaluating} introduced \emph{Decictor}, a scenario-based \gls{mt} framework designed to evaluate the optimal decision-making of \glspl{ads}.
\emph{Decictor} uses a novel \gls{mr} to detect non-optimal decision scenarios (NoDSs) by applying non-invasive mutations to the environment without altering the original optimal path of the \gls{ads}, thereby revealing instances where the \gls{ads} deviates from making optimal choices. While \emph{Decictor}'s methodology is categorized as system-level testing, its focus is primarily on evaluating the routing decisions of \glspl{ads} with limited types of scenarios and \glspl{mr}. This narrow focus restricts its ability to assess the system's behavior in a broader range of driving situations, such as those involving complex traffic interactions and varying weather conditions.

In summary, \gls{mt} has proven to be an effective technique for identifying errors in both the module-level and system-level testing of \glspl{ads}, especially when facing the oracle problem. Its primary advantage lies in its ability to define \glspl{mr} that capture expected system behaviors under transformed input conditions, allowing for the detection of subtle inconsistencies without the need for a full oracle.
However, \gls{mt} can be limited by the complexity of designing effective \glspl{mr} that fully \addedtext{and precisely} capture \deletedtext{realistic and }safety-critical behaviors, particularly in dynamic driving scenarios, which involve continuously changing traffic, environmental conditions, and agent interactions.
\annotate{Comment 2.2}\addedtext{
This challenge is further compounded by the difficulty of automatically identifying effective \glspl{mr}, which, as noted in a recent survey~\cite{li2024metamorphic}, remains an open research problem requiring deep domain expertise and a thorough understanding of system behavior.
Existing practices indicate that human expertise is still essential, as manual effort is often necessary to establish the foundation for generating \glspl{mr}~\cite{li2024metamorphic}.
However, unlike full system specifications, \glspl{mr} do not need to be complete or perfect to effectively drive testing.
Incomplete or imperfect \glspl{mr} can still successfully detect undesirable system behaviors, making them a valuable practical tool despite their limitations.}%
\deletedtext{This challenge is however no different from fully specifying systems, except that incomplete and imperfect \glspl{mr} are still effective in detecting undesirable system behavior.}

\subsection{Adaptive Stress Testing}

\acrfull{ast} is a test case generation approach that dynamically adjusts its focus by prioritizing certain test cases and allocating resources accordingly~\cite{tang2023survey}. The core of this approach lies in the policy used to assign priorities to different test scenarios.

Koren et al.~\cite{koren2018adaptive} implemented \gls{ast} for \glspl{ads} by designing a priority assignment policy that considers the disparity between the expected and the observed behavior of the system.
Their results demonstrated that using \gls{ast} with \gls{rl} effectively uncovered high-probability failure scenarios with fewer simulator calls, showing promising results in both efficiency and effectiveness.
In subsequent work, Corso et al.~\cite{corso2019adaptive} introduced an enhanced priority assignment policy based on the concept of \gls{rss}~\cite{shalev2018formal}, a formal framework that outlines idealized behaviors to prevent collisions in scenarios. The researchers used a reward augmentation technique to incorporate \gls{rss} into the \gls{ast}'s reward function, prioritizing scenarios where the \gls{ads} behavior significantly diverges from the idealized, collision-free behaviors defined by \gls{rss}. Their results showed that reward augmentation with \gls{rss} uncovered a broader range of diverse, high-relevance failure scenarios, improving its ability to identify critical safety violations.
Baumann et al.~\cite{baumann2021automatic} explored the use of \gls{rl}, to identify and generate more critical scenarios within the context of overtaking maneuvers. Moreover, \gls{rl} has been combined with \gls{rss} rules to further refine the generation of edge cases, as demonstrated by Karunakaran et al.~\cite{karunakaran2020efficient}.

In conclusion, \gls{ast} can be used to effectively generate critical scenarios by focusing on high-risk cases, especially through priority policies and \gls{rl}. However, \gls{ast} is limited by the complexity of defining priority policies, significant computational costs from repeated scenario similarity evaluations, and dependence on predefined safety models like \gls{rss}, which may restrict the exploration of unanticipated failures.

\subsection{Discussion}

The state-of-the-art suggests that \gls{mt} and search-based testing are two prominent approaches for system-level testing of \glspl{ads}, each with distinct strengths and limitations. Specifically, search-based testing, when well-designed, is effective at uncovering corner cases by systematically exploring the parameter space, while \gls{mt} is particularly useful for identifying subtle problems in system behavior through well-defined \glspl{mr}.

Therefore, a compelling research direction is the integration of these two approaches into a unified framework, designed to maximize their strengths while addressing their weaknesses. In such a combined approach, \gls{mt} could address the oracle problem by providing meaningful behavioral expectations to guide the search process. Conversely, the systematic exploration capabilities of search-based testing could enhance the effectiveness of \gls{mt}, turning mild violations identified through \glspl{mr} into severe ones by extending them along high-risk parameter dimensions. Therefore, this combination could potentially lead to a more effective and efficient testing framework.

The central question, which this paper is the first to address, is then how to optimally combine \gls{mt} and search-based testing into a cohesive framework that is both effective in detecting unsafe behavior and scalable for complex \glspl{ads}. The following sections describe our proposed novel approach in detail.

%% file: Approach.tex
\section{Our Approach}\label{sec:approach}

This section details our proposed solution to the problem described in \cref{sec:problem}.

Given the high-dimensional nature of the search space derived from real-world scenarios, a conventional search-based algorithm (e.g., evolutionary algorithm) might be insufficient to explore it effectively within practical time and computational limits~\cite{ma2019survey}.
To address this, we employ a \acrfull{ccea}, which decomposes the problem into lower-dimensional subproblems, each solved by independently evolving populations.
This decomposition not only enables parallelism but also increases the efficiency of the search~\cite{ma2019survey}.

We must therefore reframe our objective as a \gls{ccea}, where, based on the problem, source scenarios and perturbations are made to be two independent populations, but yet collaborate to discover complete solutions, i.e., test cases that violate predefined \glspl{mr}.

In the following subsections, we first describe the representation of \glspl{mr}, alongside the representation of the two separate populations within the context of the search problem (\cref{sec:representation}).
We then introduce a fitness evaluation framework, which incorporates both a joint fitness function~\cite{yang2008large,panait2006archive}, evaluating the extent of violation in a complete solution formed by collaboration between individuals from the two populations, and an individual fitness function~\cite{yang2008large,panait2006archive}, assessing the contribution of each individual within the two populations (\cref{sec:fitness}).
Ultimately, we unveil our innovative method founded on \glspl{ccea} that leverages the aforementioned representation and fitness functions (\cref{sec:algorithm}), and present the involved genetic operators (\cref{sec:operator}).

\subsection{Representations}\label{sec:representation}

Before detailing the representation of the two separate populations, we first introduce the representation of the \glspl{mr} used in the search, followed by a detailed description of the two populations.
We consider two populations: one representing source scenarios and the other representing perturbations derived from a predefined set of \glspl{mr}.
These perturbations are subsequently applied to the source scenarios to generate follow-up scenarios.

\subsubsection{Metamorphic relations}\label{sec:mr}

\annotate{Comment 1.1}\addedtext{Generally, an \gls{mr} can be expressed as a relation \(\mathcal{R}\) that defines a relationship between a sequence of inputs \(\langle x_1, x_2, \ldots, x_n\rangle \) and their respective outputs \(\langle f(x_1), f(x_2), \ldots, f(x_n)\rangle \), where \(f\) represents the target function or algorithm~\cite{chen2018metamorphic}.
A common special case occurs when \(n=2\), meaning that \(\mathcal{R}\) imposes a relationship between two inputs \(\langle x_1, x_2\rangle \) and their outputs \(\langle f(x_1), f(x_2)\rangle \) formulated as \(\langle x_1, x_2, f(x_1), f(x_2)\rangle \in \mathcal{R}\) or simply \(\mathcal{R}\left(x_1, x_2, f(x_1), f(x_2)\right)\).
This general formulation does not impose constraints on how the two inputs and outputs should relate to each other. A special case defines \(\mathcal{R}\) in terms of an input relation \(ir\) and an output relation \(or\), where \(ir\) specifies a relationship between inputs \(x_1\) and \(x_2\), and \(or\) dictates the corresponding relationship that must hold between \(f(x_1)\) and \(f(x_2)\) when \(x_1\) and \(x_2\) satisfy \(ir\):
\begin{align*}
    \mathcal{R}\left(x_1, x_2, f(x_1), f(x_2)\right) \coloneq ir\left(x_1, x_2\right) \land or\left(f(x_1),f(x_2)\right)
\end{align*}
In this formulation, an \gls{mr} can be perceived as a statement that determines how outputs should behave in response to specific changes in inputs.

In our work,} the input relation \(ir\) describes the relationship between the source scenario and the follow-up scenario, e.g., adding one pedestrian in the field of view of the ego vehicle within the source scenario.
Based on the input relation, we are able to generate corresponding metamorphic transformations to apply to the source scenario.
Note that the input relation is abstract, described in natural language, while the derived metamorphic transformations are precise, specifying exact perturbations to objects (e.g., position or speed).
Therefore, distinct metamorphic transformations can be derived from the same input relation.

The output relation \(or\) specifies how the outputs of the source inputs relate to those of the follow-up inputs~\cite{deng2023declarative}.
When the output relation is not fulfilled, the \gls{mr} is violated, indicating unsafe behavior from the \gls{ads}.
We employ three output relations in our search process, i.e., \emph{invariance}, \emph{increasing}, and \emph{decreasing} relations.
\emph{Invariance} output relations stipulate that the difference between the results of source and follow-up scenarios should remain within a predefined, typically small, threshold.
\emph{Increasing} output relations stipulate that the results of follow-up scenarios should exceed those of source scenarios, while \emph{decreasing} output relations imply the opposite.

Formally, the representation of an \gls{mr} can be defined as the tuple \(mr \coloneq (ir, or)\).
In our search, we use a set of \glspl{mr} that share the same output relation \(or\) to guide the search, which can be defined as \(MR \coloneq \{mr_i = (ir_i, or) \mid i \in \mathbb{N}^\ast \} \).
The set of metamorphic transformations derived from a set of \glspl{mr} constitutes a perturbation as elaborated later in this subsection.

\subsubsection{Scenarios}

\begin{figure*}
    \centering
    \begin{adjustbox}{width=.8\linewidth}
        \begin{tikzpicture}
            \begin{class}{Scenario}{0,0}
                \attribute{+ mission: Mission}
                \attribute{+ vehicles: List[Vehicle]}
                \attribute{+ pedestrians: List[Pedestrian]}
                \attribute{+ static\_objects: List[StaticObject]}
                \attribute{+ weather: Weather}
                \attribute{+ brightness: Float}
            \end{class}
            \begin{class}{Mission}{-7,3}
                \attribute{+ map: Map}
                \attribute{+ trajectory: List[Waypoint]}
                \attribute{+ ego\_vehicle: Vehicle}
            \end{class}
            \draw[umlcd style,<-diamond] (Mission.east) node[above right,sloped,black]{1} -| (Scenario);
            \begin{class}{Vehicle}{-7,0}
                \attribute{+ model: Model}
                \attribute{+ position: Position}
                \attribute{+ rotation: Rotation}
                \attribute{+ speed: Float}
            \end{class}
            \draw[umlcd style,<-diamond] (Vehicle.east) node[above right,sloped,black]{0..*} -- (Vehicle.east-|Scenario.west);
            \composition{Mission}{}{1}{Vehicle}
            \begin{class}{Pedestrian}{-7,-3.6}
                \attribute{+ model: Model}
                \attribute{+ position: Position}
                \attribute{+ rotation: Rotation}
                \attribute{+ speed: Float}
            \end{class}
            \draw[umlcd style,<-diamond] (Pedestrian.east) node[above right,sloped,black]{0..*} -- ++(1,0) |- (Scenario.west);
            \begin{class}{StaticObject}{0,-4}
                \attribute{+ model: Model}
                \attribute{+ position: Position}
                \attribute{+ rotation: Rotation}
            \end{class}
            \composition{Scenario}{}{0..*}{StaticObject}
            \begin{enumeration}{Weather}{7,3}
                \attribute{Clear}
                \attribute{Cloudy}
                \attribute{WetCloudy}
                \attribute{Wet}
                \attribute{HardRain}
                \attribute{MidRain}
                \attribute{SoftRain}
                \attribute{ExtraStrongFog}
                \attribute{StrongFog}
                \attribute{DenseFog}
                \attribute{HeavyFog}
                \attribute{LightFog}
            \end{enumeration}
            \draw[umlcd style,<-diamond] (Weather.west|-Scenario.east) node[above left,sloped,black]{1} -- (Scenario.east);
        \end{tikzpicture}
    \end{adjustbox}
    \caption{The scenario domain model for the case study.}\label{fig:domain-model}
\end{figure*}

A scenario can be modeled as a vector consisting of real and integer values.
In the context of \glspl{ads}, a scenario encompasses the ego vehicle, the trajectory, the environment (e.g., the weather), and other dynamic (e.g., vehicles and pedestrians) and static (e.g., obstacles) objects~\cite{haq2021offline,sharifi2023identifying}.
Each of them has multiple attributes of various types, including real numbers and categorical values that can be encoded as integers.
The scenario domain model for our case study is illustrated in \cref{fig:domain-model}. In practice, this is (partly) determined by what can be observed and controlled in the underlying simulator, i.e., \textsc{Carla} simulator~\cite{dosovitskiy17carla}.
From the model, we can see that a scenario comprises one ego vehicle, zero or more other objects (including vehicles, pedestrians, and static objects), as well as a defined mission and weather conditions.
In contrast to other studies~\cite{sharifi2023identifying,deng2022scenario}, we use a finer-grained level of simulation control, implying a larger scenario size due to the increased number of parameters that must be adjusted by the search algorithm. In practice, this is necessary if one wants to have precise \glspl{mr}.
For instance, we exert precise control over the specific positions of vehicles within the scenario by defining the precise three-dimensional coordinates, while other studies often limit their control to the number or approximate direction of vehicles, in the field of view of the ego vehicle or in the opposite lane.

Formally, let \(e\) be the ego vehicle and its attributes, \(W\) be a set of waypoints that characterizes the trajectory, \(A\) be a set of global attributes, e.g., weather and brightness, \(B\) be a set of static objects, e.g., traffic signs, and \(D\) be a set of dynamic objects, e.g., pedestrians and vehicles.
The representation of a scenario can be defined as the tuple \(s \coloneq (e, W, A, B, D)\), where:

\begin{align*}
    e & \coloneq (mod_e, pos_v, rot_e, v_e)                                               \\ 
    W & \coloneq \{w_i \mid i \in \mathbb{N}^\ast \}                                      \\
    A & \coloneq \{a_i \mid i \in \mathbb{N}^\ast \}                                      \\
    B & \coloneq \{b_i \coloneq (mod_i, pos_i, rot_i) \mid i \in \mathbb{N}^\ast \}       \\ 
    D & \coloneq \{d_i \coloneq (mod_i, pos_i, rot_i, v_i) \mid i \in \mathbb{N}^\ast \}. \\ 
\end{align*}
Here, \(mod\) represents the type of the object, e.g., the model of the vehicle such as a motorcycle or a car, \(pos\) represents the position of the object in three-dimensional space, \(rot\) represents the direction the object is facing in three-dimensional space, and \(v\) represents the speed of the respective objects. 

\annotate{Comment 3.21}\addedtext{
Each attribute (weather, brightness, object speed, etc.) is limited to a predefined range that reflects realistic conditions and is aligned with \textsc{Carla}'s valid parameters. These ranges are also configurable, allowing testers to constrain or expand the space of possible values to suit different testing requirements or fidelity levels.

To facilitate scenario evolution, we explicitly separate global scenario attributes \(A\) from object attributes \(B\) and \(D\). This independence helps ensure that modifying global attributes does not affect individual objects' parameters and vice versa, thus simplifying crossover and mutation.
Furthermore, the attributes within \(B\) and \(D\) are designed to be independent from each other, so modifying one object does not unintentionally influence others.
Additionally,}
\deletedtext{Note that }we use relative object positions w.r.t.\ the ego vehicle to ensure scenario consistency when the ego vehicle's starting position changes.

\subsubsection{Perturbations}\label{sec:perturbation}

A perturbation represents the changes applied to a source scenario, which can be used to generate a follow-up scenario by applying these changes sequentially.
In particular, given a set of \glspl{mr}, we encode a perturbation as a sequence of metamorphic transformations to a source scenario: \(q \coloneq \langle c_1, c_2, \ldots, c_k \rangle \).
Each entry \(c_i\) in \(q\) is derived from a unique input relation \(ir_i\) from the set of \glspl{mr}, indicating a metamorphic transformation applied to that source scenario.
Each entry \(c_i\) represents no change or a change to either a global attribute in \(A\), a static object in \(B\), or a dynamic object in \(D\).
Such changes can involve the \emph{addition} of an object, or the \emph{deletion} or \emph{replacement} of an existing object in the scenario.
\annotate{Comment 3.2}\addedtext{Note that an entry \(c_i\) can be a no-op, meaning that no change is applied to the source scenario for that specific metamorphic transformation.
This allows perturbations to be selectively constructed, as not every available transformation needs to be included in a given perturbation.
However, we ensure that at least one metamorphic transformation is applied to the source scenario to generate a follow-up scenario, ensuring meaningful variation.}

\subsection{Fitness Function}\label{sec:fitness}

Our objective is to define a fitness function that can both effectively direct the search towards solutions violating the predefined \glspl{mr} and maximize the diversity of generated solutions.
To achieve this, we first define how to measure the extent of \gls{mr} violations between source and follow-up scenarios, followed by an optimization strategy to maximize diversity.

In the context of scenario-based simulation, a scenario's outcome comprises time series data capturing ego vehicle properties at regular intervals, e.g., speed or position at each timestamp.
Consequently, the extent of violation is determined by comparing these time series across scenarios.
Nevertheless, the time series data of the ego vehicle in source and follow-up scenarios may not always be synchronized.
For instance, differences in pedestrian distances from the ego vehicle can lead to variations in the timing of stops, resulting in desynchronized speed data between the scenarios.
To overcome such an effect and enable meaningful comparisons, we use \emph{\gls{dtw}}, a method that finds the optimal alignment between two time series, even when they are not perfectly synchronized~\cite{berndt1994using}.
\annotate{Comment 1.5}\addedtext{
Unlike direct pointwise comparisons, \emph{\gls{dtw}} finds an optimal alignment between two time series by minimizing the cumulative distance between their matched points.

\define{\emph{\gls{dtw}} alignment}
Given two time series, a source scenario time series \(S^t = \{s^t_1, s^t_2, \dots, s^t_n\} \) and a follow-up scenario time series \(Q^t = \{q^t_1, q^t_2, \dots, q^t_m\} \), \emph{\gls{dtw}} constructs a set of matched points \(W^t\) that form the optimal alignment between two time series, constrained by the following conditions:
\begin{itemize}
    \item \emph{Boundary Constraints}: The start and end points must align.
    \item \emph{Monotonicity}: The alignment maintains the temporal order of points.
    \item \emph{Warping Window}: A predefined constraint to limit excessive stretching or compressing.
\end{itemize}
The optimal alignment under these constraints is achieved by minimizing the cumulative distance between matched points:

\begin{equation*}
    DTW(S^t, Q^t) = \min_{W^t} \sum_{(i,j) \in W^t} {(s^t_i - q^t_j)}^2
\end{equation*}
}

Our approach applies \emph{\gls{dtw}} to align corresponding points in the time series of the source scenario \(s\) and the follow-up scenario \(q(s)\).
This alignment results in a set of matched points \deletedtext{\((t_s, t_q)\)}\addedtext{\((t_s, t_q) \in W^t\)} that provides an optimal alignment between the time series data of the two scenarios.
\addedtext{To ensure that the\annotate{Comment 3.7} alignment remains realistic and does not lead to extreme distortions, we impose the Sakoe-Chiba global constraint~\cite{sakoe1978dynamic} on \emph{\gls{dtw}}.
This constraint restricts how much a point in one time series can be shifted relative to the other, allowing only a limited range of alignment deviations.
Consequently, the alignment maintains the original time order while allowing for small timing differences between the time series.}

Given the set of matched points based on \emph{\gls{dtw}}, we can calculate the extent of violation between the time series of the source and follow-up scenarios.
However, not all matched points are relevant for determining the extent of the violation.
For example, if a pedestrian enters the ego vehicle's field of view but is initially far away, the ego vehicle's speed may remain unaffected at these early points.
Consequently, evaluating matched points at this stage would not accurately reflect a violation, as a slowdown is not yet expected.
To address this, we introduce the concept of \emph{critical interval}, which defines the interval during which we expect the \gls{mr} condition to be met.
The \emph{critical interval} is therefore specific to the input relation of the \gls{mr} in use.
For instance, if the input relation requires the ego vehicle to reduce speed when a pedestrian enters its field of view, the \emph{critical interval} would be the period when the pedestrian is close enough to necessitate a slowdown.
Conversely, if the input relation specifies that the ego vehicle should slow down under certain weather conditions, the \emph{critical interval} would encompass the entire scenario, as the slowdown is expected at every timestamp.
To determine such intervals in practice, simulation-based testing approaches can leverage detailed contextual information (e.g., exact positions of actors, weather conditions) provided by the simulator.
In our case, we utilize contextual information provided by the simulator to measure the distance between the ego vehicle and the objects within its field of view at each timestamp to determine the \emph{critical interval}.
In our approach, we only consider matched pairs of points identified by \emph{\gls{dtw}}, where at least one point falls within the \emph{critical interval}, discarding pairs where neither point is within this interval.
\annotate{Comment 3.4}\addedtext{
This choice ensures that we focus on the most relevant portions of the scenario where the \gls{mr} condition clearly applies, reducing the risk of false violations caused by ambiguous or borderline cases.}
Finally, we average the extent of violation across matched pairs within the \emph{critical interval}.
The extent of violation for each matched pair is quantified by the function \(diff_{or}\), which depends on the specific output relation \(or\) of the \gls{mr} in use.
This function captures the difference between the actual outcome in the follow-up scenario and the expected outcome derived from the source scenario.
The definition of \(diff_{or}\) varies according to the type of output relation, as described in \cref{sec:mr}, and thus the extent of violation is computed accordingly.
For an \emph{invariance} output relation, the difference between the source point \(t_s\) and the follow-up point \(t_q\) should remain within a specified threshold.
This threshold can be defined either as a percentage \({\theta}\) of \(t_s\) or as an absolute value \({\phi}\), with the extent of violation calculated as \(diff_{or} \coloneq |t_q - t_s| - t_s * \theta \) or \(diff_{or} \coloneq |t_q - t_s| - \phi \), respectively.
For an \emph{increasing} output relation, the follow-up point \(t_q\) is expected to exceed the source point \(t_s\) by a specified percentage \({\theta}\) or by an absolute value \({\phi}\).
In this case, the extent of violation is calculated as \(diff_{or} \coloneq  t_s * (1 + \theta) - t_q\) or \(diff_{or} \coloneq  t_s + \phi - t_q\), respectively.
For a \emph{decreasing} output relation, the follow-up point \(t_q\) should be lower than the source point \(t_s\) by a specified percentage \({\theta}\) or by an absolute value \({\phi}\), with the extent of violation defined as \(diff_{or} \coloneq  t_q - t_s * (1 - \theta)\) or \(diff_{or} \coloneq  t_q - t_s + \phi \), respectively.

\define{Extent of \gls{mr} violation}
Given a source scenario \(s\) and a perturbation \(q\) derived from a set of \glspl{mr} with the same output relation \(or\), the extent of violation of the output relation \(or\) is defined as follows:

\begin{equation}\label{eq:violation}
    E(s, q) \coloneq \frac{1}{|T_{ci}|} \sum_{(t_s, t_q) \in T_{ci}} diff_{or}(t_s, t_q)
\end{equation}
where \(T_{ci}\) denotes the set of pairs of matched points \((t_s, t_q)\) such that at least one of \(t_s\) or \(t_q\) is within the \emph{critical interval (ci)}.
In \cref{eq:violation}, the pair \((s, q)\) forms a complete solution, and \(E\) can be regarded as the joint fitness function, which should be maximized in our problem.
\annotate{Comment 1.6}\addedtext{
\Cref{eq:violation} quantifies how much the follow-up scenario deviates from the expected behavior defined by the \gls{mr}. It does so by comparing the outputs of the source and follow-up scenarios at critical moments, i.e., time intervals where the \gls{mr} should hold.
The function \(diff_{or}(t_s, t_q)\) measures how much the actual response in the follow-up scenario differs from the expectation. The formula then averages these differences over the entire \emph{critical interval} to produce a single number that summarizes the extent of \gls{mr} violation. A higher value means a greater deviation from the expected behavior, signaling a more critical issue.
}

\annotate{Comment 1.6}\addedtext{
Following the definition of the violation extent, or the joint fitness function for complete solutions, we introduce the fitness function that evaluates each individual's contribution within the two populations.
This fitness function is particularly valuable for mechanisms that operate on each population separately, allowing optimization of each component based on its unique role in the complete solution.
}

\define{Violation detection fitness function}
Given an individual \(i\), the violation detection fitness function \(fitness\) is defined as follows:

\begin{equation}\label{eq:fitness}
    fitness(i) \coloneq \begin{cases}
        \max_j(E(i, q_j)) & \text{if \(i\) is a scenario}     \\
        \max_j(E(s_j, i)) & \text{if \(i\) is a perturbation}
    \end{cases}
\end{equation}
where \(j\) indexes individuals in the other population.
In other words, the violation detection fitness is the maximum joint fitness value across all complete solutions involving individual \(i\), and \(fitness\) can be regarded as the individual fitness function.
\annotate{Comment 3.11}\addedtext{Previous studies~\cite{panait2006archive,bull1997evolutionary,bull1998evolutionary,wiegand2001empirical} have demonstrated the effectiveness of using the maximum joint fitness value with collaborators from the other population, rather than the average or minimum.}

To make the search more exploratory and prevent the algorithm from converging too rapidly to suboptimal solutions, there are several methods to optimize the diversity of solutions.
\emph{Fitness Sharing}~\cite{cioppa2004role,goldberg1987genetic} \addedtext{and \emph{Fitness Clearing}~\cite{petrowski1996clearing} are}\deletedtext{is} amongst the best known and prevalent diversity mechanisms.
The basic idea of fitness sharing is that individuals in the same niche (within a certain niche radius) of the search space should be penalized for being too similar.
Nonetheless, fitness sharing faces challenges in problems with multiple optima because it struggles to distinguish global optima (the best possible solutions) from local optima (suboptimal solutions that are better than their immediate neighbors but not globally optimal)~\cite{sareni1998fitness}.
This difficulty arises due to the way fitness sharing distributes resources, i.e., allocates fitness values, among similar individuals, which can cause local optima to receive undue attention, making it harder for the search process to prioritize and converge on the true global optima.
Fitness clearing resembles fitness sharing; however, it is grounded in the concept of limited environmental resources~\cite{petrowski1996clearing}, aligning with our requirements mentioned in \cref{sec:problem}.
Instead of distributing resources among all individuals within the same niche, fitness clearing allocates them solely to the best members in the niche.
Practically, the capacity \(\kappa \in \mathbb{N}^\texttt{*}\) of a niche is used to indicate the maximum number of individuals it can accept.
Consequently, fitness clearing maintains the fitness of the top \({\kappa}\) individuals in the niche while removing the remaining individuals from the population.
\annotate{Comment 3.17}\addedtext{
Individuals whose fitness is cleared are largely excluded from selection in the next generation, preventing them from effectively contributing offspring.
This process eliminates competition within the niche, allowing only the dominant individuals to propagate their genetic material.
While it is theoretically possible for cleared individuals to be selected through mechanisms like tournament selection, the likelihood of this happening is very low due to their cleared fitness, which significantly reduces their chances of being favored during selection.
}
Fitness clearing significantly reduces the impact of genetic drift~\cite{sareni1998fitness}, which causes the population to trend towards individuals resembling those with high fitness~\cite{eigen1971selforganization,goldberg1987genetic,horn1998timing,mahfoud1995population}.
Furthermore, fitness clearing can be combined with elitism strategies to retain the top-performing individuals within the niches that are maintained throughout generations, and it is less complex than fitness sharing.
As demonstrated by previous studies~\cite{sareni1998fitness,covantesosuna2017analysis}, fitness clearing proves to be one of the most effective approaches.

To determine whether the distance between two individuals (either scenarios or perturbations) falls within a niche radius, we define a heterogeneous distance~\cite{wilson1997improved} function to quantify the distance between them.
In the representation we have defined, a scenario is an amalgamation of numerical and categorical values, as elaborated in \cref{sec:representation}.
A perturbation is characterized by a series of metamorphic transformations, which themselves are also composed of numerical and categorical values.

\define{Heterogeneous distance}
Given two individuals \(i\) and \(j\), the heterogeneous distance between \(i\) and \(j\) is defined as follows:

\begin{equation}\label{eq:hd}
    HD(i, j) \coloneq \sqrt{\sum_{L \in \{B, D\}}HD_{set}^2(i_L, j_L) + HD_{attr}^2(i_A, j_A)}
\end{equation}
where \(HD_{set}\) is the heterogeneous distance between two object sets, i.e., the set of static objects \(B\) or dynamic objects \(D\) of a scenario, which is defined in \cref{eq:hd_set}.
\(HD_{attr}\) is the heterogeneous distance between two attributes sets, e.g., global attributes \(A\) of a scenario, which is defined in \cref{eq:hd_attr}.

\begin{equation}\label{eq:hd_set}
    HD_{set}(I, J) \coloneq \sqrt{\sum_{i \in I}\min_{j \in J}(HD_{attr}^2(i, j))}
\end{equation}
where \(i\) and \(j\) denote the objects in set \(I\) and \(J\), respectively.
Note that the size of set \(I\) needs to be no smaller than that of set \(J\) for this equation to be applicable.
In short, this equation calculates the sum of the shortest heterogeneous distances between each object of the longer set and each object of another set, where the heterogeneous distance of objects is computed by \(HD_{attr}\).

\begin{equation}\label{eq:hd_attr}
    HD_{attr}(i, j) \coloneq \sqrt{\sum_{a=1}^m dist_a^2(i_a, j_a)}
\end{equation}
where \(m\) is the number of attributes.
The function \(dist_a(i, j)\) returns a distance between the two values \(i\) and \(j\) for attribute \(a\), which is defined as follows:

\begin{equation}\label{eq:dist}
    dist_a(i, j) \coloneq \begin{cases}
        \frac{|i - j|}{a_{\max} - a_{\min}} & \text{if \(a\) is numerical}                    \\
        \phi                                & \text{if \(a\) is categorical and \(i \neq j\)} \\
        0                                   & \text{if \(a\) is categorical and \(i = j\)}
    \end{cases}
\end{equation}
where \(a_{\max}\) and \(a_{\min}\) represent the maximum and minimum values for attribute \(a\), respectively.
When considering categorical attributes, a distance \(\phi \in \interval[open left, soft open fences]{0}{1}\) is assigned when the values are distinct, while a zero distance is assigned when the values are the same.

To penalize an individual's fitness when other individuals are in close proximity, we need to first define a clearing radius and a niche capacity, as proposed by Petrowski~\cite{petrowski1996clearing}.
Specifically, the fitness of the dominant individuals (i.e., winners) is preserved while the fitness of all the other individuals is cleared so that only the winners can survive to the next generation.
To leverage this concept, we introduce the notion of fitness clearing.

\define{Fitness clearing}
Given a population \(P\), a clearing radius \(\sigma \in \mathbb{R}^\texttt{+}\), and a niche capacity \(\kappa \in \mathbb{N}^\texttt{*}\), the fitness clearing procedure for the population \(P\) is defined in \cref{alg:clearing}.

\begin{algorithm}
    \small
    \caption{Fitness clearing procedure.}\label{alg:clearing}
    \LinesNumbered{}
    \KwIn{
        Population \(P\)\newline
        Clearing radius \({\sigma}\)\newline
        Niche capacity \({\kappa}\)
    }

    \(sortFitness(P)\)\; 
    \For{\(i \leftarrow 1\) \KwTo{} \(|P|\)}{
        \If{\(isValid(fitness(P[i]))\)}{
            Number of winners \(nbWinners \leftarrow 1\)\; 
            \For{\(j \leftarrow i + 1\) \KwTo{} \(|P|\)}{
                \If{\(isValid(fitness(P[j]))\) and \(HD(P[i], P[j]) < {\sigma}\)}{
                    \eIf{\(nbWinners < {\kappa}\)}{
                        \(nbWinners \leftarrow nbWinners + 1\)\; 
                    }{
                        Clear \(fitness(P[j])\)\; 
                    }
                }
            }
        }
    }
\end{algorithm}

The algorithm begins by sorting the population \(P\) in descending order of fitness (line 1), so the highest-fitness individuals are processed first.
For each individual \(P[i]\) in \(P\) (line 2), the algorithm checks if its fitness is valid (line 3), ensuring that previously cleared individuals are not reconsidered.
If the fitness of \(P[i]\) is valid, it is counted as the first winner in its niche by initializing a counter \(nbWinners\) to 1 (line 4).
The algorithm then iterates over subsequent individuals \(P[j]\) where \(j > i\) (line 5) to check if they have valid fitness values and their heterogeneous distance from \(P[i]\) is within the clearing radius \({\sigma}\) (line 6).
For each such individual \(P[j]\), the \(nbWinners\) counter is incremented (line 8) until it reaches the niche capacity \({\kappa}\), limiting the number of individuals that can occupy the same niche.
Once \(nbWinners\) reaches \({\kappa}\), any additional individuals within \({\sigma}\) have their fitness values cleared (line 10), meaning that their values are set to undefined, thus preventing them from being selected or archived.
This process continues until all individuals in \(P\) have been processed, ensuring that no niche has more than \({\kappa}\) high-fitness individuals, thus maintaining diversity while preserving top solutions.

However, the clearing radius \({\sigma}\) is usually unknown in many optimization problems~\cite{coello1999comprehensive}, making it difficult to determine an appropriate value without prior knowledge or extensive tuning.
To address this challenge, we borrow the concept of a dynamic fitness sharing method proposed by Tan et al.~\cite{tan2003evolutionary}, which adaptively adjusts the clearing radius based on the characteristics of the population at each generation.

\define{Dynamic clearing radius}
Given a generation \(n\), we define the dynamic clearing radius \(\sigma^{(n)}\) at generation \(n\) as follows:

\begin{equation}
    \sigma^{(n)} \coloneq \frac{HD_{\max}^{(n)}}{2 \times N^{(n)}}
\end{equation}
where \(N^{(n)}\) represents the population size at generation \(n\), while \(HD_{\max}^{(n)}\) represents the maximal \(HD\) value among individuals in the population at generation \(n\), reflecting the spread of the population.
The dynamic clearing radius adapts by scaling with the population's spread and inversely adjusting with the population size.
By scaling with \(HD_{\max}^{(n)}\), the clearing radius adapts to the overall distribution of individuals.
This ensures that larger spreads result in proportionally larger radius, promoting wider dispersion of individuals across the search space\annotate{Comment 3.18}\addedtext{, and encouraging exploration. As the population converges and spreads shrink, the radius becomes smaller, allowing finer distinctions between individuals in high-potential regions of the search space. This adaptive behavior helps the algorithm remain responsive to different search phases without requiring manual tuning.}

Dividing by \(2 \times N^{(n)}\) inversely scales the clearing radius with the population size.\deletedtext{ As the population grows, the niches become smaller, allowing individuals to occupy more distinct areas in the search space.}
\addedtext{This serves as a scaling factor that adapts competition pressure to the number of individuals, ensuring the radius is small enough to retain a reasonably large proportion as winners.}

To develop an intuition for the dynamic clearing radius \(\sigma^{(n)}\), consider a scenario where a population of \(N\) individuals is confined within a sphere of radius \(R\).
In this setup, the maximum heterogeneous distance in the population \(HD_{\max}^{(n)}\) can be approximated by the diameter of the sphere \(2R\). Substituting this approximation into the formula for \(\sigma^{(n)}\), we obtain:

\begin{equation}
    \sigma^{(n)} \approx \frac{2R}{2 N} = \frac{R}{N}.
\end{equation}
This indicates that the clearing radius for each individual is approximately \(\frac{1}{N}\) of the sphere's radius. As the population size \(N\) increases, the clearing radius decreases proportionally, dividing the sphere into finer niches and ensuring that the individuals are distributed across the sphere in a balanced way.

By eliminating the need for a fixed parameter, the dynamic clearing radius \(\sigma^{(n)}\) simplifies the optimization process and reduces dependency on prior knowledge about the often unpredictable distances between individuals in the search space~\cite{tan2003evolutionary}.

\subsection{Algorithm}\label{sec:algorithm}

\begin{figure*}[htbp]
    \centering
    \includegraphics[width=.7\linewidth]{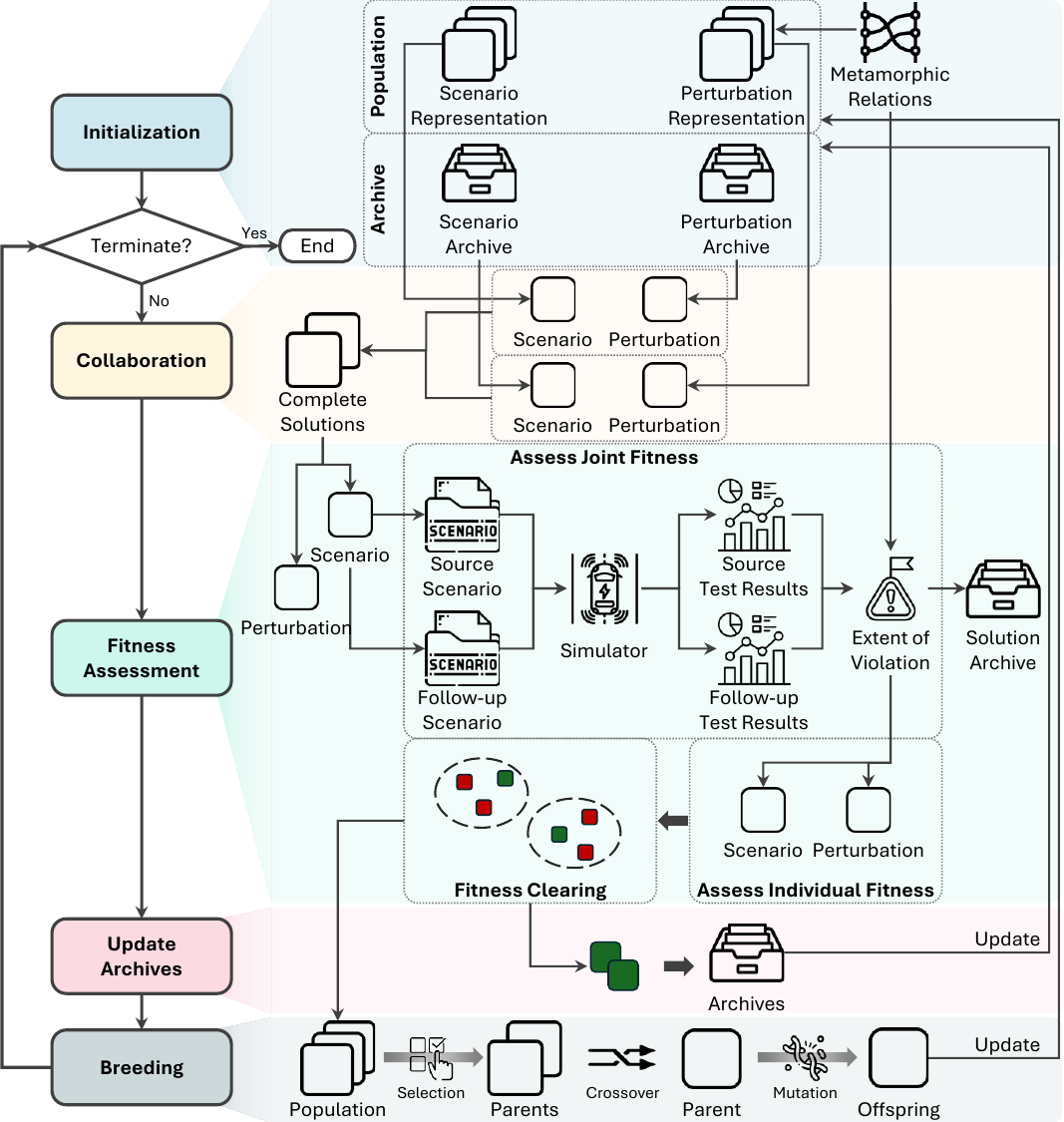}
    \caption{\protect\addedtext{Overview of \emph{\gls{ours}}.}}\label{fig:workflow}
\end{figure*}

Building upon the scenario and perturbation representations detailed in \cref{sec:representation} and the violation detection fitness function introduced in \cref{sec:fitness}, this section presents our proposed algorithm, \emph{\acrfull{ours}}.
\addedtext{The overall workflow of \emph{\gls{ours}} is shown in \cref{fig:workflow}.}
This approach combines an archive-based \gls{ccea} with \gls{mt} to effectively and efficiently explore high-dimensional and complex search spaces.
\emph{\gls{ours}} operates by decomposing the complex search space into separate populations for scenarios and perturbations, enabling these populations to evolve independently but collaborate to form complete solutions.
The evolution of each population is guided by the \glspl{mr}, which not only address the oracle problem by providing a structured way to evaluate solutions but also accelerate search convergence by focusing on meaningful variations.
Furthermore, diversity optimization strategies are applied during the archiving process to ensure that the archived individuals promote a broad exploration of the search space, effectively capturing a diverse set of complete solutions.

\Cref{alg:ccea} shows the pseudocode of \emph{\gls{ours}}.
As input, it receives a population size \(n\) and a maximum archive population size \(l\) and returns an archive \(X\) of distinct complete solutions.

\begin{algorithm}
    \small
    \caption{Cooperative co-evolutionary algorithm.}\label{alg:ccea}
    \LinesNumbered{}
    \KwIn{
        Population size \(n\)\newline
        Maximum size of archive population \(l\)
    }
    \KwOut{Archive of complete solutions \(X\)}

    Population of scenarios \(P_s \leftarrow initPopulation(n)\)\; 
    Population of perturbations \(P_q \leftarrow initPopulation(n)\)\; 
    Archive of scenarios \(X_s \leftarrow P_s\)\; 
    Archive of perturbations \(X_q \leftarrow P_q\)\; 
    Archive of complete solutions \(X \leftarrow \emptyset \)\; 

    \While{not(stopping condition)}{ 
        \(P_s, P_q, X \leftarrow assessFitness(P_s, P_q, X_s, X_q, X)\)\; 
        \(X_s \leftarrow updatePopulationArchive(P_s, l)\)\; 
        \(X_q \leftarrow updatePopulationArchive(P_q, l)\)\; 
        \(P_s \leftarrow breed(P_s) \cup X_s\)\; 
        \(P_q \leftarrow breed(P_q) \cup X_q\)\; 
    }
    \KwRet{\(X\)}\;
\end{algorithm}

The algorithm commences by randomly initializing populations of scenarios \(P_s\) and perturbations \(P_q\) at lines 1 and 2, respectively.
Corresponding population archives, \(X_s\) and \(X_q\), are also initialized at lines 3 and 4.
Additionally, an empty archive of complete solutions \(X\) is created at line 5.
The algorithm then co-evolves \(P_s\) and \(P_q\), guided by \(X_s\) and \(X_q\), until a \(stopping~condition\) is reached (line 6).
This co-evolutionary process involves three iterative steps:
\begin{enumerate*}
    \item evaluating the fitness of individuals in both \(P_s\) and \(P_q\), and updating the archive \(X\) with complete solutions and their joint fitness values as determined by the simulator (line 7);
    \item updating \(X_s\) and \(X_q\) based on individual fitness values and \(l\) (lines 8--9); and
    \item breeding \(P_s\) and \(P_q\), and merging the resulting populations with \(X_s\) and \(X_q\) to form the next generation of \(P_s\) and \(P_q\) (lines 10--11).
\end{enumerate*}
The algorithm concludes by returning the final archive \(X\) (line 13).

\subsubsection{Fitness assessment}

The function \(assessFitness\) initially computes joint fitness values of complete solutions constructed by combining individuals from \(P_s\), \(P_q\), \(X_s\), and \(X_q\).
To optimize computational efficiency, complete solutions identical to those in \(X\) (i.e., previously generated complete solutions) are excluded from re-simulation.
Subsequently, the fitness value of each individual is determined based on the joint fitness of the complete solutions in which it is included.
Finally, fitness clearing is performed for each population as detailed in \cref{sec:fitness}.

The pseudocode for \(assessFitness\) is provided in \cref{alg:fitness}, with inputs the population of scenarios \(P_s\), the population of perturbations \(P_q\), the population archive of scenarios \(X_s\), the population of perturbations \(X_q\), and the niche capacity \({\kappa}\).
The function returns updated populations \(P_s\) and \(P_q\) with individual fitness values, along with an updated archive \(X\) incorporating newly created complete solutions and their corresponding fitness values.

\begin{algorithm}
    \small
    \caption{assessFitness}\label{alg:fitness}
    \LinesNumbered{}
    \KwIn{
        Population of scenarios \(P_s\)\newline
        Population of perturbations \(P_q\)\newline
        Archive of scenarios \(X_s\)\newline
        Archive of perturbations \(X_q\)\newline
        Archive of complete solutions \(X\)\newline
        Niche capacity \({\kappa}\)
    }
    \KwOut{
        Updated population of scenarios \(P_s\)\newline
        Updated population of perturbations \(P_q\)\newline
        Updated archive of complete solutions \(X\)
    }
    Set of complete solutions \(CS \leftarrow collaborate(P_s, P_q, X_s, X_q)\)\; 
    \ForEach{Complete solution \(cs \in CS\)}{ 
        \If{\(cs \notin X\)}{
            \(simulate(cs)\)\; 
            \(cs.fitness \leftarrow assessJointFitness(cs)\)\; 
            \(X \leftarrow X \cup \{cs\}\)\; 
        }
    }
    \ForEach{Population \(P \in \{P_s, P_q\} \)}{
        \ForEach{Individual \(i \in P\)}{
            \(i.fitness \leftarrow assessIndividualFitness(i, X)\)\; 
        }
        \(fitnessClearing(P, \kappa)\)\; 
    }
    \KwRet{\(P_s, P_q, X\)}\;
\end{algorithm}

The algorithm initiates by generating an initial set of complete solutions through combinations of individuals from both populations and population archives (line 1).
To be precise, each individual in \(P_s\) collaborates with every individual in \(X_q\), and vice versa for \(P_q\) and \(X_s\).
For each generated complete solution \(cs\) (line 2), if \(cs \notin X\), i.e., \(cs\) has not been previously evaluated (line 3), its joint fitness is computed using the simulator (line 4--5) and \(cs\) with its evaluated result is added to \(X\) (line 6).
\annotate{Comment 3.22}\addedtext{
In practice, it is possible that the simulator fails to evaluate a complete solution due to conflicts among its parameters, such as overlapping object locations. In such cases, the complete solution is considered as \emph{invalid} and its fitness value is deemed undefined. Such invalid solutions will then be ignored in subsequent steps.
}
To avoid redundant simulations, the algorithm stores the results of both source and follow-up scenarios derived from a complete solution.
Upon updating \(X\) using \(CS\), for each population \(P \in \{P_s, P_q\} \) (line 9) and for each individual \(i \in P\) (line 10), the algorithm first assigns the maximum joint fitness value of the complete solutions involving \(i\) as the individual fitness of \(i\) (line 11).
This strategy of assessing individual fitness using an elitist approach (i.e., choosing the highest fitness value) aligns with findings from previous experimental research~\cite{luke2013metaheuristics,ma2019survey}.
Subsequently, the algorithm performs fitness clearing for \(P\) with a given niche capacity \({\kappa}\).
The algorithm finishes by returning the revised \(P_s\), \(P_q\), and \(X\) (line 15).

\subsubsection{Update population archive}\label{sec:update-archive}

The function \(updatePopulationArchive\) updates the population archives \(X_s\) and \(X_q\) for the next generation.
These archives are instrumental in guiding the search process, as each individual must collaborate with archive members to form complete solutions whose fitness is subsequently evaluated.
To balance exploitation and exploration, the function selects one best individual based on fitness values to ensure that high-fitness solutions are preserved and refined.
Additionally, it selects other individuals to maximize the diversity of the population archive, encouraging broader exploration of the search space and reducing the risk of premature convergence to suboptimal solutions.

The pseudocode for updating the population archive is outlined in \cref{alg:archive}.
The algorithm accepts a target population \(P\) and a maximum size of population archive \(l\) as input, returning a population archive \(X_p\) of \(P\) with a size less than or equal to \(l\).

\begin{algorithm}
    \small
    \caption{updatePopulationArchive}\label{alg:archive}
    \LinesNumbered{}
    \KwIn{
        Population \(P\)\newline
        Maximum size of population archive \(l\)
    }
    \KwOut{Population archive \(X_p\)}

    \(X_p \leftarrow \{popBestFitnessIndividual(P)\} \)\; 
    \While{\(|X_p| < l\) and \(|P| > 0\)}{
        Best individual \(best \leftarrow \square \)\; 
        \ForEach{Individual \(i \in P\)}{
            \If{\(div(X_p \cup \{i\}) > div(X_p \cup \{best\})\)}{
                \(best \leftarrow i\)\; 
            }
        }
        \(X_p \leftarrow X_p \cup \{popIndividual(best)\} \)\; 
    }
    \KwRet{\(X_p\)}\;
\end{algorithm}

The algorithm begins by creating a population archive \(X_p\) for population \(P\) utilizing the individual that possesses the highest fitness value from all the individuals in \(P\) (line 1).
As long as the size of archive \(X_p\) is below the limit \(l\) and \(P\) is not empty (line 2), the algorithm repeatedly selects the most dissimilar individual leading to maximal population diversity (\(div\)) in the archive once added (line 3--8).
Several metrics have been proposed to measure population diversity~\cite{li2019quality}, e.g., \emph{Spacing}~\cite{schott1995fault,vanveldhuizen2000measuring}, \emph{Maximum Spread (MS)}~\cite{zitzler2000comparison} and \emph{Pure Diversity (PD)}~\cite{wang2017diversity}.
These metrics are designed to require no additional problem-specific knowledge and maintain a computational complexity that is manageable, at most quadratic time complexity w.r.t.\ the population size.
Among these metrics, \emph{Spacing} evaluates diversity by calculating the minimum Euclidean distance of each individual to all others within the population.
Specifically, it measures the standard deviation of these minimum distances, thereby assessing the uniformity of the population without considering the overall spread across the search space~\cite{li2019quality,tian2019diversity}.
Although this metric provides insight into how evenly individuals are distributed~\cite{yen2014performance}, it does not address the extent of spread across the search space.
Thus, a population may exhibit high uniformity with a small spread if individuals are closely clustered, indicating limited exploration of the search space.
In contrast, \emph{MS} captures diversity by focusing solely on the extreme individuals within the population~\cite{li2019quality,wang2017diversity}, representing the widest range that the population spans in the search space.
However, this metric lacks information about the distribution of individuals between these extremes and cannot fully evaluate how well the population covers the search space. This limitation makes \emph{MS} insufficient for assessing overall population diversity~\cite{wang2017diversity}.
To address these shortcomings, we select \emph{PD} as the preferred metric for gauging population diversity.
\emph{PD} measures the total spread by summing the dissimilarities of individuals relative to the rest of the population in a greedy order, with priority given to the individual exhibiting maximal dissimilarity.
This approach provides a holistic view of diversity, effectively reflecting the primary dissimilarities within the population and the extent to which it covers the search space.

\subsubsection{Breed}

The breeding algorithm is designed to generate new individuals by combining the genetic material of selected parents from the current populations. This process is crucial for exploring the search space and finding better solutions over successive generations. The algorithm leverages crossover and mutation operators to create offspring that inherit traits from their parents while introducing variability to enhance diversity.

\begin{algorithm}
    \small
    \caption{breed}\label{alg:breed}
    \LinesNumbered{}
    \KwIn{Population \(P\)}
    \KwOut{Population of Offspring \(O\)}
    \(O \leftarrow \emptyset \)\; 
    \While{\(|O| < |P|\)}{
        Parents \(p_1, p_2 \leftarrow selection(P)\)\; 
        Offspring \(o_1, o_2 \leftarrow crossover(p_1, p_2)\)\; 
        \(o_1 \leftarrow mutate(o_1)\)\; 
        \(o_2 \leftarrow mutate(o_2)\)\; 
        \eIf{\(div(O \cup \{o_1\}) > div(O \cup \{o_2\})\)}{
            \(O \leftarrow O \cup \{o_1\} \)\; 
        }{
            \(O \leftarrow O \cup \{o_2\} \)\; 
        }
    }
    \KwRet{\(O\)}\;
\end{algorithm}

The algorithm starts by initializing an empty set \(O\) to store the offspring (line 1). The goal is to fill this set with new individuals until its size matches the original population \(P\) (line 2). First, two parent individuals, \(p_1\) and \(p_2\), are chosen from the existing population \(P\) (line 3). The selection process typically uses tournament selection, which provides individuals with higher fitness a greater likelihood of being chosen. Next, the selected parents undergo a crossover operation to produce two offspring, \(o_1\) and \(o_2\) (line 4).
The crossover operator combines parts of the parents' genetic material to create new individuals that inherit traits from both parents.
Each offspring \(o_1\) and \(o_2\) is then subjected to mutation (line 5--6).
The mutation operator applies minor, random modifications to the genetic material of the offspring. This step is essential for preserving genetic diversity in the population and enabling the algorithm to investigate new regions of the search space.
The details of the genetic operators are introduced later in \cref{sec:operator}.

After the offspring are mutated, the algorithm evaluates the diversity contribution of each offspring to the current set \(O\) (line 7), where the function \(div\) measures the diversity of a given population.
The objective is to maximize diversity within the offspring population to avoid premature convergence to suboptimal solutions. The offspring that contributes more to the diversity of the population is added to \(O\) (line 7--10), ensuring that the new population maintains a good balance between exploring new solutions and the exploitation of established effective solutions.
These steps are repeated until the set \(O\) contains as many individuals as the original population \(P\). This ensures that the population size remains constant across generations. Finally, the algorithm returns the new population \(O\) (line 13), which will replace the old population \(P\) in the next generation.

\subsection{Genetic Operators}\label{sec:operator}

Given the complexity of the representation, conventional crossover and mutation operators designed for simple vector representations are inadequate.
Therefore, we devised novel operators for scenario and perturbation representations.

\subsubsection{Crossover}

In this section, we define the crossover operator for scenarios as well as for perturbations.

\define{Scenario crossover}
Given two parent scenario representations \(s_1 = (e_1, W_1, A_1, B_1, D_1)\) and \(s_2 = (e_2, W_2, A_2, B_2, D_2)\), the offspring scenarios \(s'_1\) and \(s'_2\) are defined as follows:

\begin{align*}
    s_1 & = (e_1, W_1, A'_1, B'_1, D'_1)  \\
    s_2 & = (e_2, W_2, A'_2, B'_2, D'_2)
\end{align*}
where \(A'_1\), \(A'_2\) are two offspring resulting from a \addedtext{uniform}\deletedtext{multi-point (aka scattered)} crossover~\cite{luke2013metaheuristics,whitley2018next} between \(A_1\) and \(A_2\) with randomly selected crossover points, since there are no preferred specific crossover points.
\addedtext{
This method enables\annotate{Comment 3.20} a fine-grained exchange of genetic material, making it well-suited to our heterogeneous and complex representation.
By independently mixing genes, uniform crossover enhances diversity in offspring and enables a broader exploration of the search space.
}
Similarly, \(B'_1\) and \(B'_2\) are derived from \(B_1\) and \(B_2\), \(D'_1\) and \(D'_2\) are derived from \(D_1\) and \(D_2\).

The crossover operator for scenarios is designed to combine characteristics from two parent scenarios while maintaining the integrity of essential components.
Specifically, the parameters of the ego vehicle (\(e_1\), \(e_2\)) and the trajectory (\(W_1\), \(W_2\)) remain unchanged, while a \addedtext{uniform}\deletedtext{multi-point} crossover is performed between the global attributes (\(A_1\), \(A_2\)), the static objects (\(B_1\), \(B_2\)) and dynamic objects (\(D_1\), \(D_2\)).
\annotate{Comment 1.6}\addedtext{
Intuitively, the scenario crossover function generates new test scenarios by swapping global attributes (e.g., weather, brightness) and exchanging objects (e.g., vehicles, pedestrians, signs) between parent scenarios.
}
This approach ensures that the offspring inherit traits from both parents while preserving the overall structure of the scenarios.

\define{Perturbation crossover}
Given two parent perturbation representations \(q_1 = \langle c^1_1, c^1_2, \ldots, c^1_k \rangle \) and \(q_2 = \langle c^2_1, c^2_2, \ldots, c^2_k \rangle \), the offspring perturbations \(q'_1\) and \(q'_2\) are generated through pairwise crossover between \(q_1\) and \(q_2\), which is defined as follows:

\begin{align*}
    q'_1 & = \langle c'^1_1, c'^1_2, \ldots, c'^1_k \rangle  \\
    q'_2 & = \langle c'^2_1, c'^2_2, \ldots, c'^2_k \rangle
\end{align*}
where each pair of metamorphic transformations \(c^1_i\) and \(c^2_i\) undergoes \addedtext{uniform}\deletedtext{multi-point} crossover with randomly selected crossover points.
\annotate{Comment 1.6}\addedtext{
For example, if \(c^1_i\) represents adding a vehicle \(d^1_i\) to the scenario and \(c^2_i\) represents adding a vehicle \(d^2_i\), the uniform crossover randomly swaps some of the parameters of \(d^1_i\) and \(d^2_i\) to generate two offspring, \(c'^1_i\) and \(c'^2_i\). Similarly, if both \(c^1_i\) and \(c^2_i\) modify a global attribute, the crossover operation may simply swap them. However, if \(c^1_i\) and \(c^2_i\) modify different types of parameters (e.g., static objects, dynamic objects, or global attributes), the crossover operator preserves them as they are.
}

\subsubsection{Mutation}

Similar to the crossover, the mutation operator is defined for scenario and perturbation representations separately.

\define{Scenario mutation}
Given a scenario representation \(s = (e, W, A, B, D)\) with both numerical and categorical attributes, the polynomial mutation~\cite{deb1995simulated} and integer randomization mutation~\cite{luke2013metaheuristics} are applied to scenario attributes, respectively.
Similar to the scenario crossover, the parameters of the ego vehicle \(e\) and the trajectory \(W\) remain unchanged.
Consequently, the scenario after mutation is represented as \(s' = (e, W, A', B', D')\), where \(A'\), \(B'\), and \(D'\) denote the mutated global attributes, static objects, and dynamic objects, respectively.
Furthermore, we also designed two additional operators that in turns \emph{add} or \emph{remove} dynamic or static objects in a scenario.
The probability of applying an operator and triggering a mutation are both configurable.

The \emph{add} operator incrementally incorporates multiple objects based on a hyperbolic distribution.
Specifically, it adds a randomly generated object into the scenario \(s\) with a probability of \({\eta}\).
Upon its addition, a second randomly generated object is added with a probability of \(\eta^2\).
This process continues until no further objects are appended.
Generally, during each mutation iteration \(n\), a new object is added with a probability of \(\eta^n\).
This non-linear operator draws inspiration from the add operator used in EvoSuite~\cite{fraser2011evosuite,panichella2015reformulating} for unit-test suite or case generation.

The \emph{remove} operator randomly removes multiple dynamic or static objects from the scenario \(s\).
Similar to the \emph{add} operator, the \emph{remove} operator also removes objects following a hyperbolic distribution.
This operator is designed to counteract the potential bloating effect~\cite{fraser2013evosuite,panichella2018automated}, i.e., the length of scenarios incrementally expanding over successive generations.

\addedtext{
Intuitively,\annotate{Comment 1.6} scenario mutation may introduce random changes to a scenario by modifying global attributes (e.g., adjusting weather conditions or brightness) and altering actors (e.g., shifting vehicle positions, changing pedestrian speed, or adding/removing objects).
These subtle variations help explore a wider range of driving conditions while preserving the core structure of the scenario.
}

\define{Perturbation mutation}
Given a perturbation representation \(q = \langle c_1, c_2, \ldots, c_k \rangle \), the polynomial mutation and integer randomization mutation are applied to each metamorphic transformation \(c_i\) with a certain mutation probability, as the algorithm does to mutate scenarios.
Since there are \deletedtext{four}\addedtext{three} distinct operations, namely \emph{addition}, \emph{deletion}, \addedtext{and} \emph{replacement}\deletedtext{, and \emph{preservation}}, as mentioned in \cref{sec:perturbation}, the mutation process varies.
\annotate{Comment 1.6}\addedtext{
If the transformation \(c_i\) represents a change to a global attribute, the algorithm applies a polynomial or an integer randomization mutation to generate \(c'_i\), which assigns a new random value to the same global attribute.
In case \(c_i\) represents an operation on static or dynamic objects, perturbation mutation introduces random changes to \(c_i\). This can involve altering the parameters of a perturbation (e.g., changing a vehicle's initial speed) or disabling/enabling a perturbation entirely. This process ensures a diverse set of scenario variations while maintaining meaningful modifications.}%
\deletedtext{The algorithm mutates the object when the operation is \emph{addition} or \emph{replacement}, while it remains unchanged without mutation when the operation is \emph{deletion} or \emph{preservation}, since no objects are involved.}

\subsubsection{Selection}

The selection operator for both populations is standard tournament selection, the most prevalent selection technique in evolutionary algorithms~\cite{whitley2018next}.
It selects candidate individuals for breeding based on a simple ranking of their fitness values.
Note that hyperparameter values for crossover, mutation, and selection (e.g., tournament size, crossover rate, and mutation rate) can significantly impact breeding performance.
A detailed discussion of hyperparameter tuning in our evaluation is provided in \cref{sec:setting}.

%% file: Evaluation.tex
\section{Empirical Evaluation}\label{sec:evaluation}
In this section, we empirically assess the effectiveness and efficiency of \emph{\gls{ours}} against two baseline methods in generating test cases for \gls{mr} violation detection.
Specifically, we answer the following \glspl{rq}.

\begin{enumerate}[label={RQ\arabic*}, font=\bfseries, labelsep=0pt, wide=0pt, listparindent=\parindent]
    \item: How \emph{effectively} can \emph{\gls{ours}} find test cases violating \glspl{mr} compared to baseline methods?\label{rq1}

          To answer \labelcref{rq1}, we evaluate the performance of different search methods in terms of the number and diversity of complete solutions identified that violate the predefined \glspl{mr} within a given search budget.
    \item: How \emph{efficiently} can \emph{\gls{ours}} find test cases violating \glspl{mr} compared to baseline methods?\label{rq2}

          To answer \labelcref{rq2}, we examine the speed at which different search methods identify complete solutions that violate the predefined \glspl{mr}.
    \item: \annotate{Comment 2.3}\addedtext{How \emph{effectively} do the diversity mechanisms of \emph{\gls{ours}} enhance the identification of diverse test cases violating \glspl{mr}?}\label{rq3}

          \addedtext{To answer \labelcref{rq3}, we assess the performance of \emph{\gls{ours}} with and without diversity mechanisms, in terms of the number and diversity of complete solutions which violate the predefined \glspl{mr} within a given search budget.}
\end{enumerate}

\subsection{Baseline Methods}
To the best of our knowledge, this is the first study to generate test cases using multiple \glspl{mr} to identify \gls{mr} violations in \glspl{ads}, and there are no direct baselines.
Consequently, we compare \emph{\gls{ours}} against two baseline methods, i.e., \emph{\gls{rs}} and \emph{\gls{sga}}~\cite{holland1992adaptation}.

\begin{itemize}
    \item \textbf{\emph{\acrfull{rs}:}} It randomly generates complete solutions.
          \emph{\gls{rs}} is commonly used as a baseline for evaluating search-based methods, given its simplicity and effectiveness for simple problems.
          The results of \emph{\gls{rs}} will thus reveal the difficulty of the search problem.

    \item \textbf{\emph{\acrfull{sga}:}} It generates complete solutions without utilizing distinct populations for scenarios and perturbations.
          The results of \emph{\gls{sga}} will show how effective \emph{\gls{ours}} is compared to a standard and simpler search method.
\end{itemize}

\annotate{Comment 3.29}\addedtext{
We include \emph{\gls{rs}} as a simple, non-optimized baseline to provide a lower-bound comparison. While it is not an advanced optimization approach, \emph{\gls{rs}} serves as a ``sanity check'' that highlights whether \emph{\gls{ours}} yields significant improvements beyond mere chance. This ensures that any gains in performance can be attributed to our method's evolutionary design rather than random variation alone.
We also compare \emph{\gls{ours}} with \emph{\gls{sga}}, which does not separate scenarios and perturbations into distinct populations. This helps isolate the effect of cooperative co-evolution in \emph{\gls{ours}}, demonstrating whether the additional complexity of evolving two populations independently truly enhances exploration and solution quality.
}

\subsection{Experimental Setup}
We now describe the experimental setup used to address the defined \glspl{rq}, including the simulation platform utilized, the involved \glspl{mr}, and the settings of \emph{\gls{ours}} and baseline methods.

\subsubsection{Simulation platform}
For the experiments, we utilized \textsc{Carla}~\cite{dosovitskiy2017carla}, a widely adopted open-source simulator for research in autonomous driving.
\textsc{Carla} offers realistic urban environments, configurable weather conditions, and dynamic objects such as vehicles and pedestrians, allowing for the simulation of diverse driving scenarios.
In addition, we integrated \textsc{InterFuser}~\cite{shao2022safety}, a safety-enhanced autonomous driving framework designed to improve scene understanding through multi-modal sensor fusion, which is one of the top-performing \glspl{ads} on the \textsc{Carla} leaderboard~\cite{carlaleaderboard} during our evaluation period.
\textsc{InterFuser} combines data from LiDAR and multi-view cameras to generate a comprehensive understanding of the driving environment.

Our resource-intensive experiments were carried out on a remote server equipped with 8 CPU cores and 16 threads, alongside 2 high-performance GPUs, each with 24GB of memory.
To maximize computational efficiency, we parallelized scenario execution by running 6 Docker containers, each hosting a separate \textsc{Carla} instance (version 0.9.10.1) and leveraging CUDA 11.8 for GPU-accelerated computations, allowing multiple simulations to run concurrently.

\subsubsection{MRs involved in the experiments}
We employed the \glspl{mr} from \cref{tab:mrs}, which were extracted from the relevant literature~\cite{deng2023declarative,pan2021metamorphic,zhang2018deeproad,deng2021bmt,iqbal2024metamorphic} and are both applicable to \gls{ads} testing and compatible with the simulation platform we use.
For instance, an \gls{mr} that specifies replacing buildings with trees is incompatible with our platform due to its limitations.

\annotate{Comment 2.2}In our evaluation, we categorized \(MR_1\) to \(MR_5\)\deletedtext{ and}\addedtext{,} \(MR_6\) to \(MR_7\)\addedtext{, and \(MR_8\) to \(MR_{13}\)} into \deletedtext{two}\addedtext{three} distinct groups\deletedtext{, i.e., the \emph{decreasing} \gls{mr} group and the \emph{invariance} \gls{mr} group}, respectively.
Each group is characterized by a common output relation.
This categorization allows our approach to effectively identify \gls{mr} violations that impact multiple \glspl{mr} within these groups.
For \deletedtext{the \emph{decreasing} \gls{mr} group}\addedtext{\(GP_1\)}, the output relation was defined as a decrease in the speed of the ego vehicle by at least 20\%.
For \deletedtext{the \emph{invariance} \gls{mr} group}\addedtext{\(GP_2\) and \(GP_3\)}, the output relation was defined as maintaining a steering angle within one degree.

\addedtext{
The\annotate{Comment 3.5} thresholds of output relations used in our study are derived from the established literature and refined through a pilot experiment conducted prior to the main study. This preliminary step helped identify and resolve potential discrepancies and ensure proper calibration of the \glspl{mr} for our experiments. While our focus is not on optimizing these thresholds, they were chosen to support experimentation based on practical considerations and empirical observations from the pilot experiment. In practice, domain experts determine these thresholds based on industry standards and domain knowledge to ensure that the \glspl{mr} accurately reflect expected system behavior.
}

\begin{table}[htbp]
    \scriptsize
    \caption{\glspl{mr} involved in the experiments.}\label{tab:mrs}
    \begin{threeparttable}
        \begin{tabularx}{\linewidth}{p{2.5em}lX}
            \toprule
            \textbf{Category} & \textbf{\gls{mr}}                       & \textbf{Description}                                                                                                                                                                                                                                         \\
            \midrule
            \multirow{10}{*}{\rotatebox[origin=c]{90}{\addedtext{\makecell{\(GP_1\)                                                                                                                                                                                                                                                    \\ Speed Reduction}}\deletedtext{Decreasing}}} & \(MR_1\)\tnote{*}                       & If a pedestrian appears on the roadside, then the ego vehicle should slow down.                                                                                                                                                                                       \\
                              & \(MR_2\)\tnote{*,\({\S}\)}              & If the driving time changes into night, then the ego vehicle should slow down.                                                                                                                                                                               \\
                              & \(MR_3\)\tnote{\({\S}\)}                & Adding a vehicle in the front of the ego vehicle, the speed of the ego vehicle should decrease in t1\% to t2\%.                                                                                                                                              \\
                              & \(MR_4\)\tnote{\({\S}\)}                & Adding a pedestrian in the front of the ego vehicle, the speed of the ego vehicle should decrease in t1\% to t2\%.                                                                                                                                           \\
                              & \(MR_5\)\tnote{\({\S}\)}                & Changing from sunny to rainy, the speed of the ego vehicle should decrease in t1\% to t2\%.                                                                                                                                                                  \\
            \midrule
            \multirow{6}{*}{\rotatebox[origin=c]{90}{\makecell{\addedtext{\(GP_2\)}                                                                                                                                                                                                                                                    \\\addedtext{Environmental}                                                                                                                                                                                                                                                                                                                                      \\Invariance}}}                 & \(MR_6\)\tnote{\({\dagger}\)}           & The density of fog (light fog, heavy fog, dense fog, strong fog, and extra strong fog) in a foggy driving scene will not affect the steering angle of the autonomous driving systems.                                                                                 \\
                              & \(MR_7\)\tnote{\({\ddagger}\)}          & No matter how the driving scenes are synthesized to cope with different weather conditions (sunny and rainy), the driving steering angle is expected to be consistent.                                                                                       \\
            \midrule
            \multirow{25}{*}{\rotatebox[origin=c]{90}{\makecell{\addedtext{\(GP_3\)}                                                                                                                                                                                                                                                   \\\addedtext{Actor Invariance}}}}                        & \addedtext{\(MR_8\)\tnote{\({\|}\)}}    & \addedtext{In a given driving scenario where the ego vehicle detects a target obstacle and attempts to avoid a collision, the ego vehicle should keep the steering unchanged when the speed of the ego vehicle changes (increased or decreased by a certain factor).} \\
                              & \addedtext{\(MR_9\)\tnote{\({\|}\)}}    & \addedtext{In a given driving scenario where the ego vehicle detects a target obstacle and attempts to avoid a collision, the ego vehicle should keep the steering unchanged the same when adjusting (i.e., scaled down or up) the size of the ego vehicle.} \\
                              & \addedtext{\(MR_{10}\)\tnote{\({\|}\)}} & \addedtext{In a given driving scenario where the ego vehicle detects a target obstacle and attempts to avoid a collision, the ego vehicle should keep the steering unchanged when adjusting (i.e., scaled down or up) the size of the target obstacle.}      \\
                              & \addedtext{\(MR_{11}\)\tnote{\({\|}\)}} & \addedtext{In a given driving scenario where the ego vehicle detects a target obstacle and attempts to avoid a collision, the ego vehicle should keep the steering unchanged when changing the position of the ego vehicle.}                                 \\
                              & \addedtext{\(MR_{12}\)\tnote{\({\|}\)}} & \addedtext{In a given driving scenario where the ego vehicle detects a target obstacle and attempts to avoid a collision, the ego vehicle should keep the steering unchanged when changing the speed of the target obstacle.}                                \\
                              & \addedtext{\(MR_{13}\)\tnote{\({\|}\)}} & \addedtext{In a given driving scenario where the ego vehicle detects a target obstacle and attempts to avoid a collision, the ego vehicle should keep the steering unchanged when adding additional actors.}                                                 \\
            \bottomrule
        \end{tabularx}
        \begin{tablenotes}
            \footnotesize
            \item[*] These \glspl{mr} are derived from study~\cite{deng2023declarative}.
            \item[\({\dagger}\)] This \gls{mr} is derived from study~\cite{pan2021metamorphic}.
            \item[\({\ddagger}\)] This \gls{mr} is derived from study~\cite{zhang2018deeproad}.
            \item[\({\S}\)] These \glspl{mr} are derived from study~\cite{deng2021bmt}.
            \item[\({\|}\)] \addedtext{These \glspl{mr} are derived from study~\cite{iqbal2024metamorphic}.}
        \end{tablenotes}
    \end{threeparttable}
\end{table}

\subsubsection{Settings of CoCoMEGA and baseline methods}\label{sec:setting}
For all methods, we established a small population size of seven individuals.
This decision takes into account the high cost of fitness evaluation, which averages two minutes per individual (scenario).
The literature commonly recommends a small population size for costly fitness functions~\cite{benabdessalem2020automated,chugh2017survey}.
Regarding the selection operator utilized in \emph{\gls{ours}} and \emph{\gls{sga}}, we employed tournament selection\annotate{Comment 3.25}\addedtext{, which is the most commonly used selection method in modern genetic algorithms~\cite{whitley2019handbook}. This method is particularly useful as it allows direct control over selection pressure by adjusting the tournament size. In our case, we selected} \deletedtext{with }a tournament size of 3.%
\annotate{Comment 3.34}\deletedtext{facilitating improved exploitation and a rapid convergence rate~\mbox{\cite{sette2001genetic}}.}
\addedtext{This choice is supported by prior studies, as small tournament sizes (2 or 3) are widely used in the literature~\cite{lavinas2018experimental} to moderate selection pressure, balance exploration and exploitation, and thereby prevent early convergence to suboptimal solutions. Furthermore, when using genetic algorithms in practice, a tournament size of 3 has been observed to potentially enhance efficiency compared to a size of 2 by reducing function evaluations and improving CPU usage~\cite{kaelo2007integrated}.}
The mutation rate used in \emph{\gls{sga}} is set to 0.2, which is considered relatively high; however, it helps to prevent genetic drift in the context of a small population size~\cite{sette2001genetic}.
Lastly, we established the crossover rate to 0.8, a value commonly recommended~\cite{mirjalili2018genetic}.
For the hyperparameters used in \emph{\gls{ours}}, since there are no suggested values for \glspl{ccea}, we decided to tune them by performing a pilot experiment with smaller budgets.
As a result, we used the following hyperparameters for \emph{\gls{ours}}: maximum population archive size \(=3\), crossover rate for individuals \(=0.8\), mutation rate for individuals \(=0.2\).

For all methods, we established the total number of simulations to 200 as the search budget, which is sufficient to observe the convergence of effectiveness metrics based on our pilot experiment.
\annotate{Comment 3.12}\addedtext{
Specifically, we monitored convergence by tracking the average fitness value of the population and archives across generations, identifying stagnation when improvements became negligible over successive generations, thus signaling convergence.}
As the majority of the execution cost is associated with running simulations, the computational budget for the experiments primarily relies on the number of simulations conducted.
Therefore, we consider the total number of simulations to be the search budget.
It is worth noting that for \emph{\gls{ours}} and \emph{\gls{sga}}, the actual number of simulations may slightly exceed the predefined limit, due to population-based methods checking whether the search budget has been exhausted only after completing one generation.
Furthermore, since the number of simulations per generation can vary across methods, the results cannot be perfectly aligned with the search budget.
To address this misalignment, we use linear interpolation based on the search budget to align the results across methods.
Specifically, for a given metric, we interpolate the values at points where the actual number of simulations exceeds or falls short of the predefined search budget.
Linear interpolation estimates the metric value at the exact search budget by assuming a linear relationship between the metric and the number of simulations conducted.
This approach provides a fair basis for evaluating all methods under a consistent computational budget.

\subsection{RQ1: Effectiveness of CoCoMEGA}

\begin{figure*}[htbp]
    \centering
    \includegraphics[width=.8\linewidth]{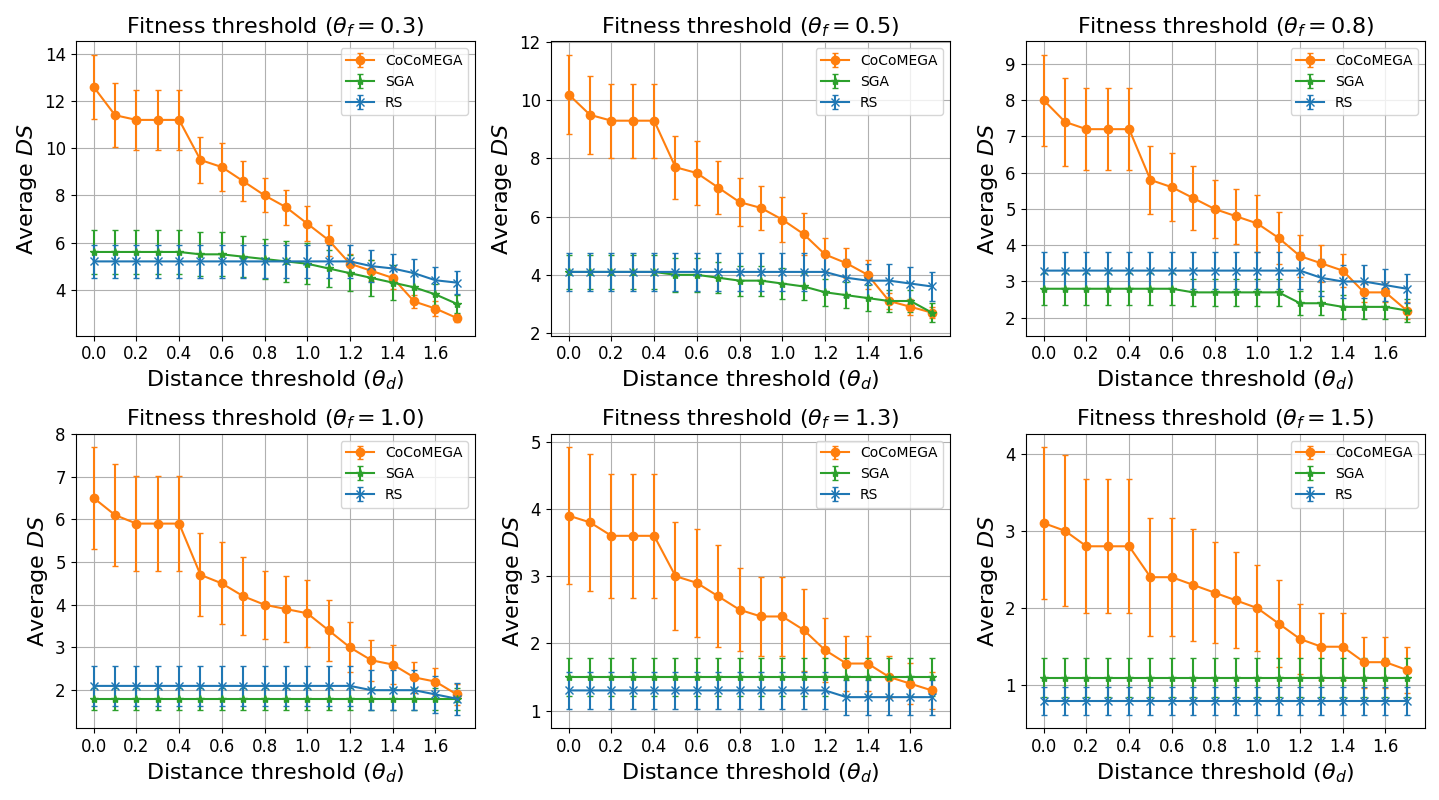}
    \caption{
        \protect\annotate{Comment 1.9}\protect\addedtext{
        \emph{\acrfull{ds}} vs.\ distance threshold (\(\theta_d\)) across different fitness thresholds (\(\theta_f\)).
        A higher \emph{\gls{ds}} indicates more distinct violating solutions discovered by the method.
        Each curve plots the average \emph{\gls{ds}} at different \(\theta_d\) settings, under a specific fitness threshold \(\theta_f\).
        This figure reveals how each method balances the quantity and diversity of solutions as \(\theta_f\) varies.
        }%
        \protect\deletedtext{The relationship between \emph{\gls{ds}}, fitness threshold \(\theta_f\), and distance threshold \(\theta_d\) for \emph{\gls{ours}}, \emph{\gls{sga}}, and \emph{\gls{rs}}.}
    }\label{fig:ds}
\end{figure*}

\subsubsection{Methodology}
To answer \labelcref{rq1}, we generate sets of complete solutions violating the predefined \glspl{mr} using \emph{\gls{ours}} and other baseline methods within the same search budget.
We compare the methods in terms of \emph{\gls{ds}} and \emph{\gls{sd}}.

\begin{itemize}
    \item \textbf{\emph{\acrfull{ds}}} denotes the number of distinct complete solutions the method finds that violate the given \glspl{mr} within a certain search budget.
          Specifically, given a fitness threshold \(\theta_f\) and a distance threshold \(\theta_d\), let \(CS_V\) denote the set of complete solutions produced by a search method \(V\), satisfying the following conditions:
          \begin{enumerate*}
              \item the fitness value of each complete solution in \(CS_V\) is greater than \(\theta_f\) and
              \item the pairwise distance between any two complete solutions in \(CS_V\) exceeds  \(\theta_d\).
          \end{enumerate*}
          Then, \emph{\gls{ds}} of \(V\) is defined as \(DS(V) \coloneq |CS_V|\).
    \item \textbf{\emph{\acrfull{sd}}} measures the diversity of the distinct complete solutions identified by the method, based on applied fitness thresholds \(\theta_f\) and distance thresholds \(\theta_d\).
          Specifically, we quantify \emph{\gls{sd}} using the following three metrics:
          \begin{itemize}
              \item \textbf{\emph{\gls{apd}}} represents the average \annotate{Comment 3.36}\addedtext{heterogeneous distance (cf.\ \cref{eq:hd})} between all pairs of distinct complete solutions identified under a given search budget.
                    \addedtext{The heterogeneous distance between two complete solutions is the heterogeneous distance between their follow-up scenarios, generated by applying each solution's perturbation to its source scenario.}
                    This metric provides insight into the degree of variation among solutions in terms of scenario representation.
              \item \textbf{\emph{\gls{mrc}}} measures the percentage of predefined \glspl{mr} covered by the complete solutions found within a specified search budget.
                    This metric reflects how thoroughly the solutions explore potential violations relative to the predefined \glspl{mr}.
              \item \textbf{\emph{\gls{cmr}}}: indicates the number of unique \gls{mr} combinations that the complete solutions can violate within a given search budget.
                    This metric offers insight into the capability of the complete solutions to expose a variety of undesirable behaviors characterized by combinations of MR violations, highlighting the breadth of testing.
          \end{itemize}
          \addedtext{Each\annotate{Comment 2.4} of these metrics captures a distinct aspect of solution diversity.
          \emph{\gls{apd}} measures spatial diversity by quantifying how spread out solutions are in the search space but does not indicate whether different types of \gls{mr} violations are covered.
          \emph{\gls{mrc}} addresses this by measuring the number of violated \glspl{mr}, ensuring broad behavioral exploration. However, \emph{\gls{mrc}} does not account for cases where multiple \glspl{mr} are violated simultaneously, which is captured by \emph{\gls{cmr}}. By combining these three metrics, we provide a more comprehensive assessment of solution diversity.}

\end{itemize}

To gain a better understanding of how \emph{\gls{ds}} and \emph{\gls{sd}} varies based on different \(\theta_f\) and \(\theta_d\) thresholds, we vary \(\theta_f\) and \(\theta_d\) to analyze their impact.
We set \(\theta_f\) to 0.3, 0.5, 0.8, 1.0, 1.3, and 1.5.
This range covers the median fitness value (approximately 0.3) to the 90th percentile (approximately 1.5) of the identified complete solutions.
We excluded solutions with fitness values below 0.3, as they correspond to trivial violations, and those above 1.5, as they are few in number and exhibit a wide range of fitness values.
For \(\theta_d\), we selected values from 0.0 to 1.7, with increments of 0.1.
This range encompasses the median pairwise distance (approximately 1.7) of the identified complete solutions.
Additionally, to address the randomness in search-based methods, we conduct each experiment 10 times per method.

\subsubsection{Results}

Based on \deletedtext{the \emph{decreasing} \gls{mr} group}\addedtext{\(GP_1\)}, \cref{fig:ds} visually compares the \emph{\gls{ds}} achieved by \emph{\gls{ours}}, \emph{\gls{sga}}, and \emph{\gls{rs}} across 10 experimental runs, varying by fitness threshold \(\theta_f\) and distance threshold \(\theta_d\).
Each subplot displays \(\theta_d\) on the x-axis and the average \emph{\gls{ds}} on the y-axis, with 95\% confidence intervals represented as error bars.

Overall, for all three methods, \emph{\gls{ds}} values decrease as \(\theta_f\) increases, reflecting that higher \(\theta_f\) values led to only retaining more stringent solution quality, thus limiting the number of complete solutions meeting the criteria.
When \(\theta_f\) is set high (e.g., \(\theta_f = 1.5\)), \emph{\gls{ds}} values are low across all methods, indicating a lack of high-fitness solutions with severe violations.
This observation implies that within the allocated search budget, the methods were unable to explore solutions with fitness values beyond 1.5.

For\annotate{Comment 3.39} \emph{\gls{ours}}, \emph{\gls{ds}} values \deletedtext{tend to decrease with increasing \(\theta_d\), as identifying sufficiently distinct complete solutions becomes more challenging at higher \(\theta_d\) values.}
\addedtext{
naturally decrease as \(\theta_d\) increases, since stricter distinctiveness constraints inherently filter out more solutions.}
In contrast, \emph{\gls{sga}} and \emph{\gls{rs}} show more consistent trends, especially at \(\theta_f \geqslant 1.0\), where the algorithms can occasionally identify sufficiently different complete solutions, though the total number of the solutions remains limited.
This implies that identifying \gls{mr} violations is a challenging problem, which \emph{\gls{rs}} and \emph{\gls{sga}} struggle to address effectively.
\emph{\gls{ours}}, by comparison, identifies complete solutions across a broader range of distances.

Additionally, for a given \(\theta_d\), the gap between \emph{\gls{ours}} and the baseline methods (i.e., \emph{\gls{sga}} and \emph{\gls{rs}}) generally widens as \(\theta_f\) decreases.
However, an exception occurs when \(\theta_d\) is relatively high (\(\theta_d \geqslant 1.2\)), especially for low fitness thresholds (\(\theta_f \leqslant 0.8\)), where the average \emph{\gls{ds}} of \emph{\gls{ours}} is lower than that of the baseline methods.
This outcome suggests that \emph{\gls{sga}} and \emph{\gls{rs}} are more likely to find more trivial violations at greater distances from each other in these cases.
In contrast, \emph{\gls{ours}} focuses on severe violations, resulting into more guided testing rather than maximizing diversity.
This targeted approach is advantageous when maximizing diversity must be balanced with thoroughly exploring severe violations.

Our analysis shows that \emph{\gls{ours}} consistently outperforms \emph{\gls{sga}} and \emph{\gls{rs}} in identifying severe violations when \(\theta_f \geqslant 1.0\).
\annotate{Comment 1.8}\addedtext{For instance, under moderate thresholds (\(\theta_f = 1.0, \theta_d = 1.0\)), \emph{\gls{ours}} discovers on average 3.8 distinct \gls{mr}-violating solutions within the specified budget, whereas \emph{\gls{sga}} and \emph{\gls{rs}} detect about 1.8 and 2.1, respectively, corresponding to over a twofold improvement over \emph{\gls{sga}} and more than 80\% over \emph{\gls{rs}}.
Across all fitness and distance thresholds, the advantage remains consistent, with \emph{\gls{ours}} discovering, on average, 87\% (\(p\)-value \(< 10^{-31}\)) more distinct solutions than \emph{\gls{rs}} and 83\% (\(p\)-value \(< 10^{-35}\)) more than \emph{\gls{sga}}.}
\annotate{Comment 3.41}\addedtext{
To assess the statistical significance of the observed differences, we conducted non-parametric Mann-Whitney U tests with Fisher's method for \(p\)-value correction. The results confirm \emph{\gls{ours}}'s clear advantage in discovering more distinct \gls{mr}-violating solutions across most threshold configurations.}
Although \addedtext{in certain conditions with low fitness thresholds (\(\theta_f \leqslant 0.8\)) and high distance thresholds (\(\theta_d \geqslant 1.2\)),} \emph{\gls{rs}} and \emph{\gls{sga}} may excel at exploring the search space and yield higher \emph{\gls{ds}} values, \deletedtext{at lower fitness thresholds (\(\theta_f \leqslant 0.8\)) and higher distance thresholds (\(\theta_d \geqslant 1.2\)) }these solutions tend to be of lower quality, as reflected by their low fitness, and thus of less practical interest.

While \emph{\gls{ds}} provides a measure of the quantity and quality of identified complete solutions based on fitness and distance thresholds, \emph{\gls{apd}} offers insight into the spatial diversity of these solutions by assessing how dispersed they are within the search space.
To analyze the \emph{\gls{apd}} results, we examine two aspects:
\begin{enumerate*}
    \item the relationship between \emph{\gls{apd}}, fitness threshold \(\theta_f\), and distance threshold \(\theta_d\) for \emph{\gls{ours}}, \emph{\gls{sga}}, and \emph{\gls{rs}} across generations, and
    \item the evolution of \emph{\gls{apd}} across generations under different \(\theta_f\) and \(\theta_d\) values.
\end{enumerate*}
The first analysis helps us evaluate each method's ability to generate a diverse set of final solutions, while the second allows us to understand how effectively each method explores diverse regions of the search space over time.

\begin{figure*}[htbp]
    \centering
    \includegraphics[width=.8\linewidth]{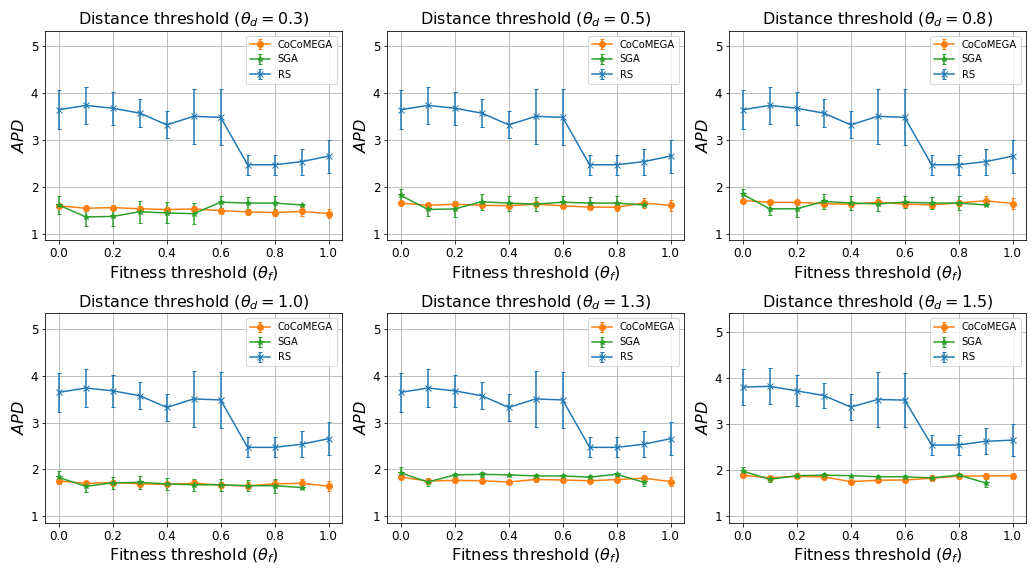}
    \caption{
        \protect\annotate{Comment 1.9}\protect\addedtext{
        \emph{\acrfull{apd}} vs.\ fitness threshold (\(\theta_f\)) across different distance thresholds (\(\theta_d\)).
        \emph{\gls{apd}} quantifies how spread out the solutions are.
        A higher \emph{\gls{apd}} indicates greater diversity among the found solutions.
        Each subplot corresponds to a certain distance threshold \(\theta_d\), and the y-axis shows the \emph{\gls{apd}} of discovered solutions under various \(\theta_f\) values.
        This figure demonstrates how each method maintains diversity while striving for higher fitness (i.e., more severe violations).
        }%
        \protect\deletedtext{The relationship between \emph{\gls{apd}}, fitness threshold \(\theta_f\), and distance threshold \(\theta_d\) for \emph{\gls{ours}}, \emph{\gls{sga}}, and \emph{\gls{rs}}.}
    }\label{fig:apd-by-gen}
\end{figure*}

\Cref{fig:apd-by-gen} shows the relationship between \emph{\gls{apd}}, \(\theta_f\), and \(\theta_d\) for \emph{\gls{ours}}, \emph{\gls{sga}}, and \emph{\gls{rs}} across 10 experimental runs. Each subplot corresponds to a specific \(\theta_d\) value, with \(\theta_f\) values plotted along the x-axis. The y-axis represents the \emph{\gls{apd}}, with 95\% confidence intervals represented by error bars.
It is worth mentioning that the maximum fitness threshold has been adjusted to 1.0, as the number of solutions significantly diminishes beyond this fitness threshold, making it infeasible to measure \emph{\gls{apd}} accurately at higher \(\theta_f\) values.
Across all thresholds, \emph{\gls{rs}} consistently achieves higher \emph{\gls{apd}} values compared to both \emph{\gls{ours}} and \emph{\gls{sga}}, suggesting that \emph{\gls{rs}} produces more widely spaced solutions. However, \emph{\gls{rs}}'s higher \emph{\gls{apd}} may not translate to solution quality or relevance, as \emph{\gls{rs}} lacks a guided search mechanism and generates solutions randomly.

When comparing \emph{\gls{ours}} and \emph{\gls{sga}}, we observe that \emph{\gls{ours}} achieves \emph{\gls{apd}} values that are generally on par with \emph{\gls{sga}} across all tested distance thresholds \(\theta_d\) and fitness thresholds \(\theta_f\). This suggests that both methods produce complete solutions with similar levels of diversity within the search space. The comparable diversity between \emph{\gls{ours}} and \emph{\gls{sga}} should be interpreted by considering the different mechanisms through which they aim to preserve diversity. Since \emph{\gls{ours}} uses separate populations for scenarios and perturbations, its candidate solutions at each generation are formed by combining individuals from these two populations. This setup results in significant commonalities among candidate solutions, as each candidate solution consists of a specific pairing of a scenario and a perturbation from their corresponding populations. In contrast, \emph{\gls{sga}} operates within a single population, allowing each candidate solution to evolve independently. Consequently, \emph{\gls{ours}} is inherently more inclined to produce a set of complete solutions with lower diversity compared to \emph{\gls{sga}}.
However, \cref{fig:apd-by-gen} suggests that \emph{\gls{ours}} effectively compensates for this potential disadvantage, likely through its diversity optimization mechanisms, such as the population archive update strategy (cf.\ \cref{sec:update-archive}) and fitness clearing (cf.\ \cref{sec:fitness}). This indicates that \emph{\gls{ours}}'s design is robust enough to maintain high diversity even within the interdependent structure of its co-evolutionary framework.

\begin{figure*}[htbp]
    \centering
    \includegraphics[width=.75\linewidth]{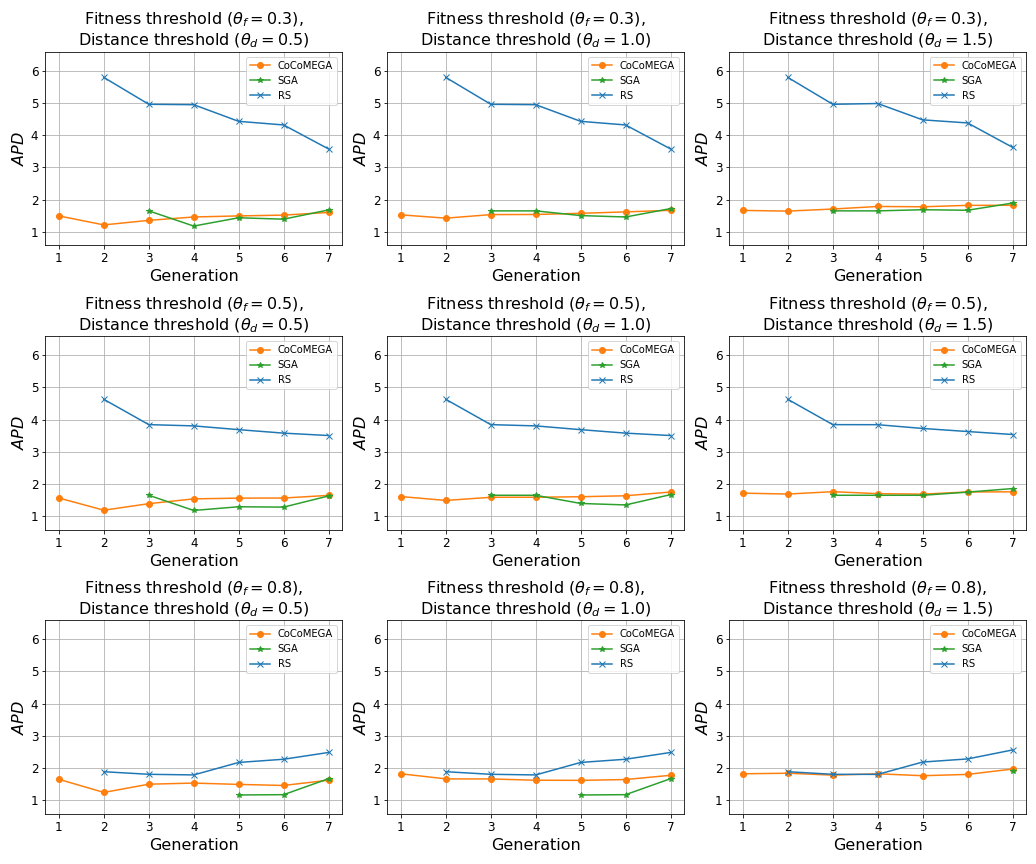}
    \caption{
        \protect\annotate{Comment 1.9}\protect\addedtext{
        Evolution of \emph{\acrfull{apd}} across generations.
        Higher \emph{\gls{apd}} values imply that solutions are more spread out in the search space, indicating broad exploration.
        Subplots differ by the chosen fitness threshold (\(\theta_f\)) and distance threshold (\(\theta_d\)).
        Each curve tracks the \emph{\gls{apd}} for a specific method over several generations of search.
        This figure illustrates how each method maintains or improves solution diversity over generations.
        }%
        \protect\deletedtext{The \emph{\gls{apd}} across generations for \emph{\gls{ours}}, \emph{\gls{sga}}, and \emph{\gls{rs}} under varying fitness thresholds \(\theta_f\) and distance thresholds \(\theta_d\).}
    }\label{fig:apd-over-gen}
\end{figure*}

\Cref{fig:apd-over-gen} displays the \emph{\gls{apd}} across generations for \emph{\gls{ours}}, \emph{\gls{sga}}, and \emph{\gls{rs}} under varying fitness thresholds \(\theta_f\) and distance thresholds \(\theta_d\). Each subplot corresponds to a combination of \(\theta_f\) and \(\theta_d\), showing the progression of \emph{\gls{apd}} values over seven generations. It is notable that for some methods, certain \emph{\gls{apd}} values are absent in the initial generations, particularly at higher \(\theta_d\) and \(\theta_f\) levels, due to an insufficient number of complete solutions meeting the specified thresholds in the early generations. To address this limitation, the maximum fitness threshold has been capped at 0.8, as the number of solutions diminishes significantly beyond this point in the early generations, making further measurement impractical.

As expected, given that it does not include any form of guidance, across all settings, \emph{\gls{rs}} maintains consistently higher \emph{\gls{apd}} values compared to \emph{\gls{ours}} and \emph{\gls{sga}}, aligning with the trends observed in \cref{fig:apd-by-gen}. However, as \(\theta_f\) increases, the gap between \emph{\gls{rs}} and the other methods gradually closes, suggesting that \emph{\gls{rs}}'s broad exploratory advantage diminishes under stricter fitness constraints, suggesting that many of the solutions it produces are less interesting from a practical standpoint.
For \emph{\gls{ours}} and \emph{\gls{sga}}, the \emph{\gls{apd}} remains relatively stable across generations, suggesting that both methods effectively maintain diversity levels while progressively enhancing solution quality. Under all tested configurations, \emph{\gls{ours}} and \emph{\gls{sga}} demonstrate comparable \emph{\gls{apd}} values, reflecting similar levels of diversity within the structured, guided search frameworks of each method. This consistency in diversity indicates that both methods effectively balance exploration and refinement in their search processes, allowing them to preserve solution diversity over generations.

\begin{table*}[htbp]
    \scriptsize
    \setlength\tabcolsep{3.5pt}
    \caption{\gls{mrc} and \gls{cmr} values for different methods at different values of \(\theta_f\) and \(\theta_d\).}\label{tab:mrc-cmr}
    \begin{tabularx}{\linewidth}{llYYYYYY}
        \toprule
        \multirow{2}{2em}{\({\boldsymbol\theta_d}\)} & \multirow{2}{*}{\textbf{Method}} & \multicolumn{6}{c}{\textbf{Average \emph{\gls{mrc}}\(\boldsymbol{\pm CI_{0.95}}\) (Average \emph{\gls{cmr}}\(\boldsymbol{\pm CI_{0.95}}\))}}                                                                                                                                                                                                                                          \\
        \cmidrule{3-8}
                                                     &                                  & \(\theta_f=0.3\)                                                                                                                             & \(\theta_f=0.5\)                            & \(\theta_f=0.8\)                            & \(\theta_f=1.0\)                             & \(\theta_f=1.3\)                             & \(\theta_f=1.5\)                             \\
        \midrule
        \multirow{4}{2em}{\(\theta_d=0.0\)}          & \emph{\gls{ours}}                & \(\boldsymbol{100.0\pm0.0}\) (\(3.0\pm0.3\))                                                                                                 & \(\boldsymbol{98.0\pm1.8}\) (\(2.6\pm0.3\)) & \(\boldsymbol{90.0\pm7.2}\) (\(2.1\pm0.3\)) & \(\boldsymbol{78.0\pm10.9}\) (\(1.8\pm0.4\)) & \(\boldsymbol{66.0\pm12.3}\) (\(1.3\pm0.3\)) & \(\boldsymbol{56.0\pm13.1}\) (\(1.0\pm0.3\)) \\
        \cmidrule{2-8}
                                                     & \emph{\gls{sga}}                 & \(88.0\pm7.2\) (\(4.2\pm0.5\))                                                                                                               & \(84.0\pm8.8\) (\(3.4\pm0.5\))              & \(82.0\pm8.7\) (\(2.6\pm0.4\))              & \(66.0\pm10.8\) (\(1.8\pm0.3\))              & \(56.0\pm11.7\) (\(1.5\pm0.3\))              & \(40.0\pm11.4\) (\(1.1\pm0.2\))              \\
        \cmidrule{2-8}
                                                     & \emph{\gls{rs}}                  & \(92.0\pm7.2\) (\(1.0\pm0.0\))                                                                                                               & \(90.0\pm9.0\) (\(0.9\pm0.1\))              & \(90.0\pm9.0\) (\(0.9\pm0.1\))              & \(64.0\pm13.4\) (\(0.8\pm0.1\))              & \(48.0\pm12.9\) (\(0.8\pm0.1\))              & \(22.0\pm8.3\) (\(0.7\pm0.1\))               \\
        \midrule
        \multirow{4}{2em}{\(\theta_d=1.0\)}          & \emph{\gls{ours}}                & \(\boldsymbol{100.0\pm0.0}\) (\(2.9\pm0.3\))                                                                                                 & \(\boldsymbol{98.0\pm1.8}\) (\(2.6\pm0.3\)) & \(\boldsymbol{90.0\pm7.2}\) (\(2.1\pm0.3\)) & \(\boldsymbol{78.0\pm10.9}\) (\(1.8\pm0.4\)) & \(\boldsymbol{66.0\pm12.3}\) (\(1.3\pm0.3\)) & \(\boldsymbol{56.0\pm13.1}\) (\(1.0\pm0.3\)) \\
        \cmidrule{2-8}
                                                     & \emph{\gls{sga}}                 & \(88.0\pm7.2\) (\(4.1\pm0.5\))                                                                                                               & \(84.0\pm8.8\) (\(3.3\pm0.4\))              & \(82.0\pm8.7\) (\(2.5\pm0.4\))              & \(66.0\pm10.8\) (\(1.8\pm0.3\))              & \(56.0\pm11.7\) (\(1.5\pm0.3\))              & \(40.0\pm11.4\) (\(1.1\pm0.2\))              \\
        \cmidrule{2-8}
                                                     & \emph{\gls{rs}}                  & \(92.0\pm7.2\) (\(1.0\pm0.0\))                                                                                                               & \(90.0\pm9.0\) (\(0.9\pm0.1\))              & \(90.0\pm9.0\) (\(0.9\pm0.1\))              & \(64.0\pm13.4\) (\(0.8\pm0.1\))              & \(48.0\pm12.9\) (\(0.8\pm0.1\))              & \(22.0\pm8.3\) (\(0.7\pm0.1\))               \\
        \midrule
        \multirow{4}{2em}{\(\theta_d=1.7\)}          & \emph{\gls{ours}}                & \(\boldsymbol{98.0\pm1.8}\) (\(2.0\pm0.3\))                                                                                                  & \(\boldsymbol{94.0\pm3.8}\) (\(2.1\pm0.2\)) & \(86.0\pm7.6\) (\(1.6\pm0.1\))              & \(\boldsymbol{78.0\pm10.9}\) (\(1.3\pm0.2\)) & \(\boldsymbol{64.0\pm12.0}\) (\(1.1\pm0.2\)) & \(\boldsymbol{56.0\pm13.1}\) (\(1.0\pm0.3\)) \\
        \cmidrule{2-8}
                                                     & \emph{\gls{sga}}                 & \(88.0\pm7.2\) (\(3.1\pm0.3\))                                                                                                               & \(84.0\pm8.8\) (\(2.5\pm0.4\))              & \(80.0\pm8.5\) (\(2.1\pm0.3\))              & \(66.0\pm10.8\) (\(1.8\pm0.3\))              & \(56.0\pm11.7\) (\(1.5\pm0.3\))              & \(40.0\pm11.4\) (\(1.1\pm0.2\))              \\
        \cmidrule{2-8}
                                                     & \emph{\gls{rs}}                  & \(92.0\pm7.2\) (\(1.0\pm0.0\))                                                                                                               & \(90.0\pm9.0\) (\(0.9\pm0.1\))              & \(\boldsymbol{90.0\pm9.0}\) (\(0.9\pm0.1\)) & \(64.0\pm13.4\) (\(0.8\pm0.1\))              & \(48.0\pm12.9\) (\(0.8\pm0.1\))              & \(22.0\pm8.3\) (\(0.7\pm0.1\))               \\
        \bottomrule
    \end{tabularx}
\end{table*}

While \emph{\gls{apd}} captures the spatial diversity of solutions within the search space, \emph{\gls{mrc}} and \emph{\gls{cmr}} focus on behavioral diversity by measuring how thoroughly the complete solutions test predefined \glspl{mr} and explore unique combinations of \gls{mr} violations.
\Cref{tab:mrc-cmr} presents the relationship between \emph{\gls{mrc}} and \emph{\gls{cmr}} values across varying \(\theta_f\) and \(\theta_d\) values for each method under the same search budget.
To capture the range of distance constraints, we selected three representative values for \(\theta_d \in \{0.0, 1.0, 1.7\} \).

Across all methods and settings, except when \(\theta_f=0.8\), \emph{\gls{ours}} consistently achieves the highest \emph{\gls{mrc}} values, indicating it covers a greater percentage of predefined \glspl{mr} than \emph{\gls{sga}} and \emph{\gls{rs}}.
This performance remains strong even as \(\theta_f\) increases, though \emph{\gls{mrc}} declines, reflecting the increasing difficulty in covering \glspl{mr} under more stringent fitness requirements.

While \emph{\gls{sga}} has a slight edge in \emph{\gls{cmr}} values across certain settings (notably, with low fitness thresholds like \(\theta_f = 0.3\)), \emph{\gls{ours}} generally achieves competitive \emph{\gls{cmr}} values, particularly as \(\theta_f\) increases.
The \emph{\gls{cmr}} values for \emph{\gls{rs}} remain low across all settings, indicating its relative inefficiency in discovering diverse \gls{mr} combinations.

Changing \(\theta_d\) has a relatively small effect on the \emph{\gls{mrc}} values for \emph{\gls{ours}}, as it maintains high coverage across nearly all distance thresholds, although there is a slight decline at \(\theta_d = 1.7\) and higher \(\theta_f\) values.
In contrast, \emph{\gls{rs}} shows significant decreases in both \emph{\gls{mrc}} and \emph{\gls{cmr}} as \(\theta_f\) and \(\theta_d\) increase, underscoring its limitations in maintaining \emph{\gls{mrc}} under more restrictive conditions.

\emph{\gls{ours}} stands out for its robustness across a range of \(\theta_f\) and \(\theta_d\) values, consistently achieving high \emph{\gls{mrc}}.
This demonstrates its effectiveness in identifying complete solutions that thoroughly explore and violate the predefined \glspl{mr}, especially under varying fitness and distance thresholds.

For \deletedtext{the \emph{invariance} \gls{mr} group}\addedtext{\(GP_2\)}, none of the three methods were able to identify complete solutions.
A possible reason is that \textsc{Interfuser} may be particularly adept at handling weather variations and such changes have thus minimal impact on the steering angle of the ego vehicle.

\annotate{Comment 2.2}\addedtext{
For \(GP_3\), the results align closely with those of \(GP_1\), reinforcing our conclusions.
Notably, some performance metrics exhibit further improvements, underscoring the robustness of \emph{\gls{ours}}.
Given the similar trends and to avoid redundancy, a detailed breakdown, including all relevant figures and tables, is provided in the appendix. 
}

\Finding{
    \annotate{Comment 1.8}\addedtext{\emph{\gls{ours}} outperforms \emph{\gls{sga}} by 83\% and \emph{\gls{rs}} by 87\% in identifying severe \gls{mr} violations, as measured by the \emph{\gls{ds}} metric.}%
    \deletedtext{\emph{\gls{ours}} is more effective than \emph{\gls{sga}} and \emph{\gls{rs}} in identifying severe \gls{mr} violations.}
    Furthermore, it also achieves both good spatial and behavioral diversity, effectively covering diverse regions within the search space.
}

\subsection{RQ2: Efficiency of CoCoMEGA}

\subsubsection{Methodology}
To address \labelcref{rq2}, we compare \emph{\gls{ours}} with the baseline methods in terms of \emph{\gls{ds}} and \emph{\gls{mrc}} as a function of the search budget.
In particular, we evaluate both \emph{\gls{ds}} and \emph{\gls{mrc}} across various search methods, gradually increasing the search budget from 10\% to 100\% in 10\% increments.
The methodology mirrors that of \labelcref{rq1}, using identical hyperparameters and performing 10 repetitions for each method, with the only difference being the varying search budget.
By analyzing how \emph{\gls{ds}} and \emph{\gls{mrc}} values evolve across different methods and budget levels, we gain insights into the relationship between the search budget and the effectiveness of each method.

\subsubsection{Results}

\begin{figure*}[htbp]
    \centering
    \includegraphics[width=.737\linewidth]{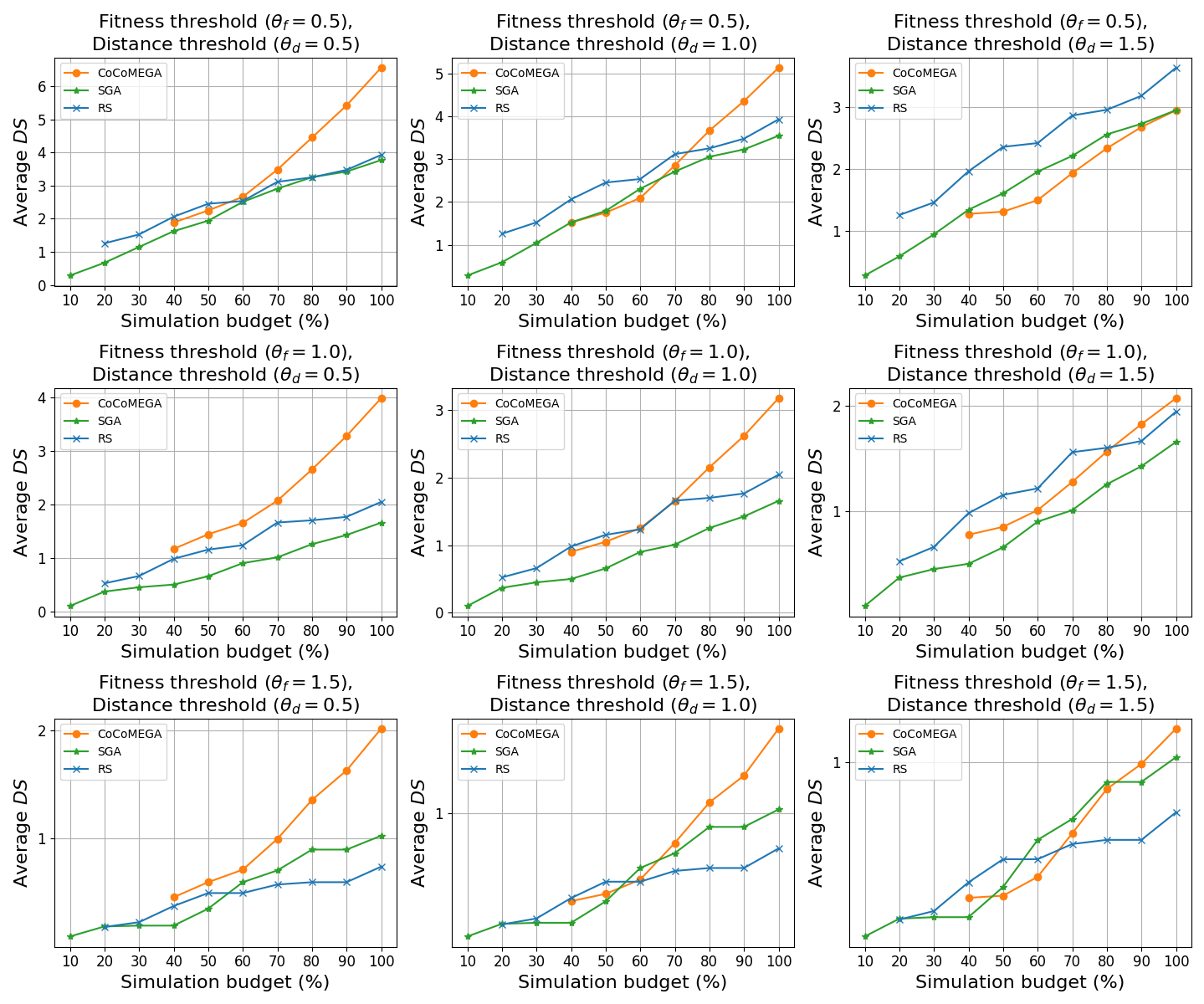}
    \caption{
        \protect\annotate{Comment 1.9}\protect\addedtext{
        \emph{\acrfull{ds}} vs.\ search budget.
        A higher \emph{\gls{ds}} indicates more distinct violating solutions discovered by the method.
        Each subplot shows \emph{\gls{ds}} at incremental percentages of the total simulation budget.
        Different curves represent different methods, compared in distinct fitness threshold (\(\theta_f\)) and distance threshold (\(\theta_d\)) settings.
        This figure highlights each method's efficiency: steeper or higher curves mean the algorithm finds more unique solutions earlier in the budget.
        }%
        \protect\deletedtext{The relationship between average \emph{\gls{ds}}, fitness threshold \(\theta_f\), and distance threshold \(\theta_d\) for \emph{\gls{ours}}, \emph{\gls{sga}}, and \emph{\gls{rs}} at various search budgets.}
    }\label{fig:dss}
\end{figure*}

We analyzed how the relationship between average \emph{\gls{ds}} and the percentage of the search budget used is affected by varying threshold values for \(\theta_f\) and \(\theta_d\).
Based on the experimental results in \labelcref{rq1}, we selected three representative \(\theta_f \in \{0.5, 1.0, 1.5\} \) and \(\theta_d \in \{0.5, 1.0, 1.5\} \) values that span the range of fitness and distance constraints.
Each subplot of \cref{fig:dss} tracks the effectiveness of these methods as the search budget incrementally increases from 10\% to 100\% of the total.

Across all subplots, a consistent trend is observed: \emph{\gls{ds}} values increase as the search budget grows.
This is expected, as a higher budget provides more opportunities for each method to explore the search space, thus finding a greater number of \emph{\gls{ds}}.
Across nearly all threshold settings, \emph{\gls{ours}} outperforms \emph{\gls{sga}} and \emph{\gls{rs}}, achieving higher \emph{\gls{ds}} values even with smaller budgets.
This indicates that \emph{\gls{ours}} is more efficient in exploring the search space and finding diverse complete solutions compared to the other two methods.
Exceptions occur primarily for \(\theta_d = 1.5\), where \emph{\gls{ours}} performs comparably to other methods.
A possible explanation for this behavior is that a higher \(\theta_d\) reduces the likelihood of forming complete solutions, thus limiting the advantage of \emph{\gls{ours}} in this specific setting.

The fitness threshold \(\theta_f\) has a notable impact on the \emph{\gls{ds}} values across methods.
However, the pattern of trends remains similar across all \(\theta_f\) values.
In general, with the increasing of \(\theta_f\), \emph{\gls{ds}} values drop for all methods, reflecting the difficulty in finding severe violations regardless of distance constraints.

The distance threshold \(\theta_d\) also influences the efficiency of each method.
At a low \(\theta_d = 0.5\), \emph{\gls{ours}} exhibits a clear advantage over \emph{\gls{sga}} and \emph{\gls{rs}}, almost consistently achieving higher \emph{\gls{ds}} values.
This suggests that \emph{\gls{ours}} is highly effective in identifying complete solutions that are distinct.
As \(\theta_d\) increases to 1.0, \emph{\gls{ours}} maintains its lead above a 70\% search budget, though the gap with \emph{\gls{sga}} and \emph{\gls{rs}} starts to narrow slightly.
At the highest \(\theta_d = 1.5\), the gap between \emph{\gls{ours}} and the baseline methods diminishes further.
\emph{\gls{ours}} ultimately identifies more \emph{\gls{ds}} than the baseline methods, except when \(\theta_f = 0.5\), where \emph{\gls{rs}} outperforms the other two methods. This suggests that \emph{\gls{rs}} is particularly effective at finding a greater number of diverse but trivial violations.
For \(\theta_f \geqslant 1.0\) and \(\theta_d = 1.5\), all three methods display similarly low \emph{\gls{ds}} values, suggesting that a combination of high fitness and large distance requirements severely limits the number of viable complete solutions.
In these cases, \emph{\gls{rs}} occasionally achieves \emph{\gls{ds}} values comparable to \emph{\gls{ours}}, likely due to its simpler approach to exploring the search space.

At higher search budgets (70--100\%), \emph{\gls{ours}} achieves the highest \emph{\gls{ds}} values across most configurations.
However, its \emph{\gls{ds}} values are sometimes similar to or slightly lower than those of the baseline methods.
This may be due to the initial overhead of \emph{\gls{ours}}, as it exhaustively simulates potential complete solutions by combining scenarios and perturbations in the first generation, completing only one search generation while \emph{\gls{sga}} completes seven or more.
Nonetheless, \emph{\gls{ours}} continues to find new, distinct complete solutions as the budget increases.
Although the maximum search budget was capped at 200 due to computational limits, \emph{\gls{ours}} shows a faster growth in \emph{\gls{ds}} values than the baseline methods, suggesting it would find even more complete solutions with a larger budget.

\begin{figure*}[htbp]
    \centering
    \includegraphics[width=.75\linewidth]{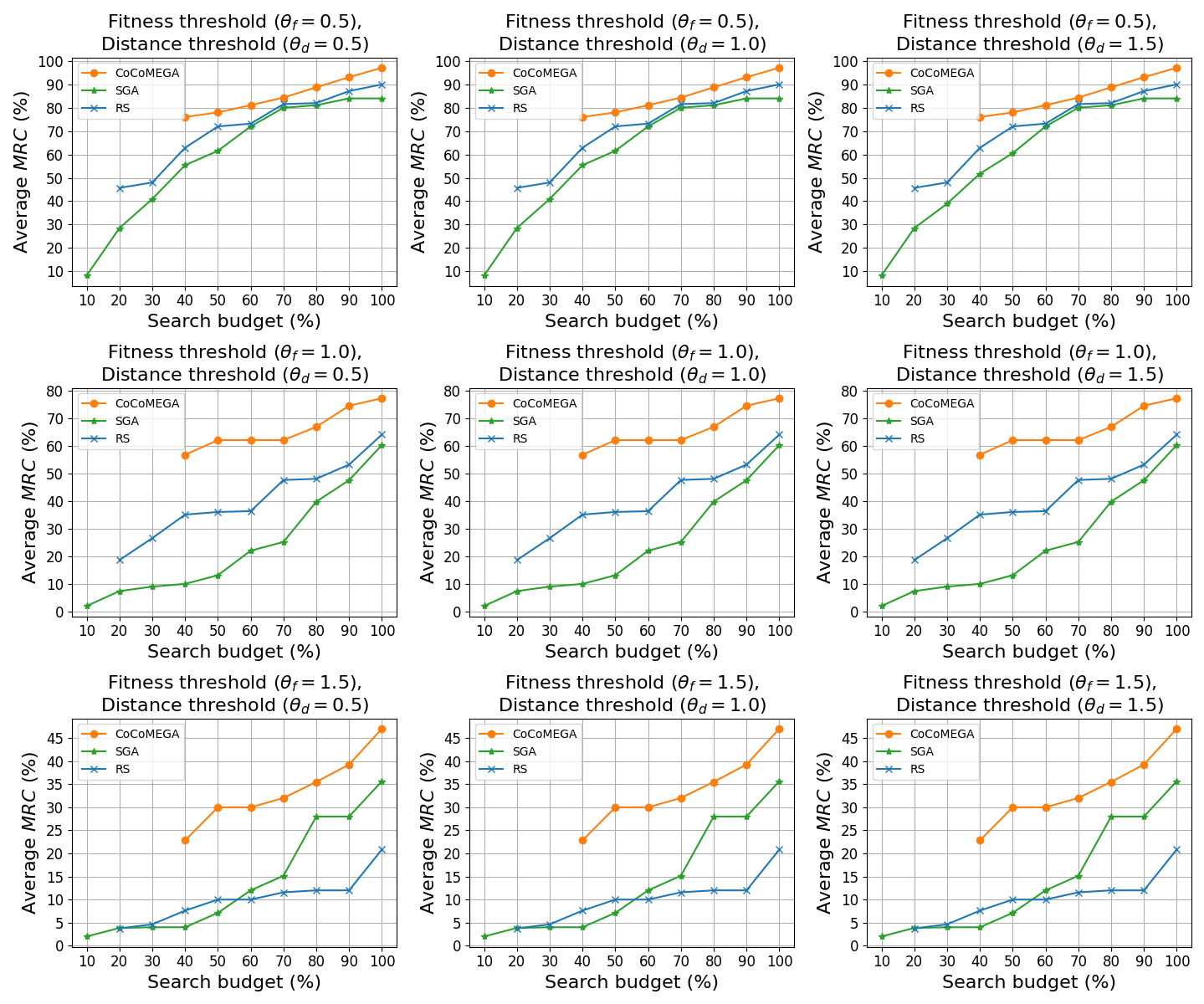}
    \caption{
        \protect\annotate{Comment 1.9}\protect\addedtext{
        \emph{\acrfull{mrc}} vs.\ search budget.
        \emph{\gls{mrc}} indicates the percentage of defined \glspl{mr} that are violated by at least one discovered solution.
        Higher \emph{\gls{mrc}} means the method finds solutions that violate more of the predefined \glspl{mr}, indicating thorough behavioral exploration.
        The x-axis represents how much of the simulation budget has been consumed.
        This figure shows how quickly each method covers all relevant \gls{mr} violations as the search progresses.
        }%
        \protect\deletedtext{The relationship between average \emph{\gls{mrc}}, fitness threshold \(\theta_f \in \{0.5, 1.0, 1.5\} \), and distance threshold \(\theta_d \in \{0.5, 1.0, 1.5\} \) for \emph{\gls{ours}}, \emph{\gls{sga}}, and \emph{\gls{rs}} at various simulation budgets.}
    }\label{fig:mrcs}
\end{figure*}

\Cref{fig:mrcs} illustrates how \emph{\gls{mrc}} evolves as the percentage of search budget consumed increases.
Across all threshold settings, \emph{\gls{ours}} significantly and consistently outperforms \emph{\gls{sga}} and \emph{\gls{rs}} in terms of \emph{\gls{mrc}}, indicating that \emph{\gls{ours}} can identify complete solutions covering more \glspl{mr} than the other two baseline methods.
Furthermore, when \(\theta_f = 0.5\), the \emph{\gls{mrc}} of \emph{\gls{ours}} approaches nearly 100\%.
While \emph{\gls{mrc}} declines as \(\theta_f\) increases, \(\theta_d\) has minimal impact on \emph{\gls{mrc}}.
At the highest \(\theta_f = 1.5\), \emph{\gls{ours}} further widens the gap over \emph{\gls{sga}} and \emph{\gls{rs}}, demonstrating its capability to find severe violations that cover a broader range of \glspl{mr}.

\annotate{Comment 1.8}\addedtext{To quantitatively compare the efficiency of \emph{\gls{ours}}, \emph{\gls{sga}} and \emph{\gls{rs}} algorithms, we employed the \emph{\gls{auc}} measure, a widely accepted and arguably the most useful performance aggregation method in optimization research~\cite{hansen2010comparing}. Specifically, we computed the area under each algorithm's performance curve over the entire budget range (0\%--100\%) separately for each of the two metrics, i.e., \emph{\gls{ds}} over budget (\cref{fig:dss}) and \emph{\gls{mrc}} over budget (\cref{fig:mrcs}). This area can be interpreted as an average performance over the budget range and thus provides a comprehensive efficiency measure across all budget levels into a single value.
\annotate{Comment 3.41}To compare \emph{\gls{ours}} with the baselines, we calculated the average differences in their \emph{\gls{auc}} values across different threshold configurations, and assessed the statistical significance of these improvements using Wilcoxon signed-rank tests.

For the \emph{\gls{ds}} over budget curves, \emph{\gls{ours}} achieved an average \emph{\gls{auc}} improvement of 46.13\% (\(p\)-value \(\approx 0.023\)) compared to \emph{\gls{sga}} and 31.95\% (\(p\)-value \(\approx 0.037\)) compared to \emph{\gls{rs}}, across moderate and high fitness thresholds (\(\theta_f \geq 1\)). Further, for the \emph{\gls{mrc}} over budget curves, the efficiency gains of \emph{\gls{ours}} were even more substantial, with average \emph{\gls{auc}} improvements of 90.07\% (\(p\)-value \(\approx 0.036\)) over \emph{\gls{sga}} and 85.86\% (\(p\)-value \(\approx 0.035\)) over \emph{\gls{rs}}. These results confirm that \emph{\gls{ours}} consistently outperforms \emph{\gls{sga}} and \emph{\gls{rs}} in terms of overall efficiency.}

\begin{figure}[htbp]
    \centering
    \includegraphics[width=.9\linewidth]{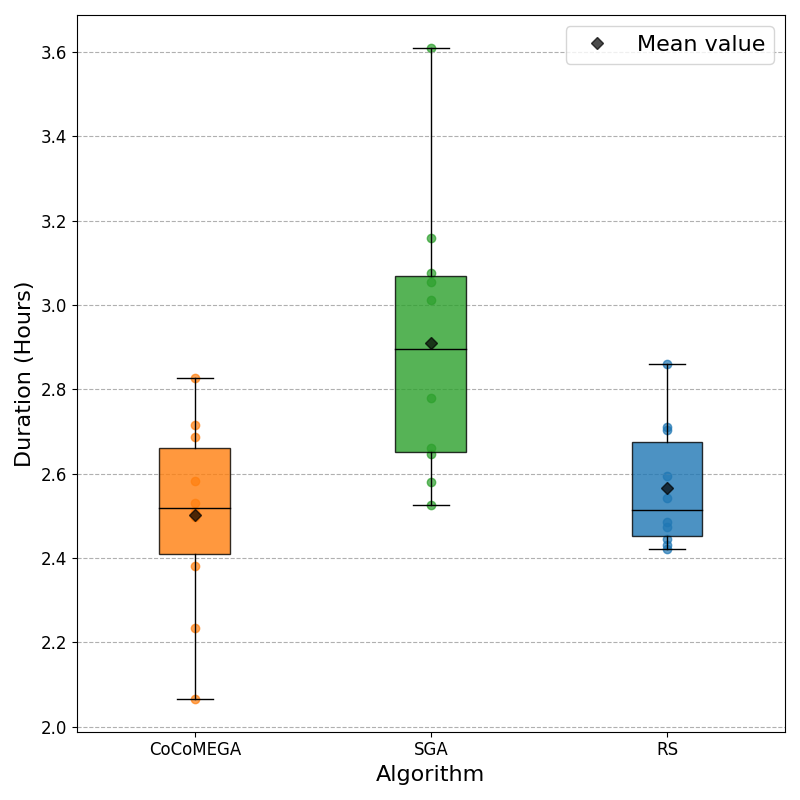}
    \caption{
        \protect\addedtext{Comparison of computational efficiency (execution time) for \emph{\gls{ours}}, \emph{\gls{sga}}, and \emph{\gls{rs}}. Diamonds indicate mean execution times. Lower execution times indicate higher computational efficiency.}
    }\label{fig:exec_time}
\end{figure}

\annotate{Comment 3.35}\addedtext{
Following the analysis of the \emph{\gls{ds}} and \emph{\gls{mrc}} metrics, we also compared the computational efficiency of \emph{\gls{ours}}, \emph{\gls{sga}}, and \emph{\gls{rs}} directly in terms of execution time, within a fixed search budget. \Cref{fig:exec_time} illustrates the distribution of execution times (in hours) for each algorithm across 10 experimental runs.
As depicted in \cref{fig:exec_time}, \emph{\gls{ours}} demonstrates superior computational efficiency with an average execution time of approximately 2.50 hours, compared to \emph{\gls{rs}} with 2.57 hours and \emph{\gls{sga}} with 2.91 hours. In particular, \emph{\gls{ours}} consistently achieves faster execution times than \emph{\gls{sga}}, indicating greater overall efficiency.
The longer execution time observed in \emph{\gls{sga}} is likely due to the generation of more invalid complete solutions throughout its search process. Invalid solutions are solutions whose parameters violate simulation constraints, such as overlapping object locations, causing the simulator to fail at the beginning of the simulation. While these invalid solutions do not count against the simulation budget and will then be ignored in the evolutionary process, the attempt to simulate them incurs additional computational overhead, thus contributing to increased overall execution time.
}

In summary, the results indicate that \emph{\gls{ours}} is particularly well-suited for tasks requiring diverse, severe violations across a range of search budgets and thresholds.
\addedtext{The \emph{\gls{auc}} analysis further reinforces this observation, showing a clear efficiency advantage for \emph{\gls{ours}} in terms of both \emph{\gls{ds}} and \emph{\gls{mrc}} metrics.}
Both \emph{\gls{sga}} and \emph{\gls{rs}} struggle to maintain competitive \emph{\gls{ds}} values, especially at higher \(\theta_f\), where the search space becomes more constrained.
\emph{\gls{rs}} occasionally performs comparably to \emph{\gls{ours}} at high \(\theta_d\) and low \(\theta_f\), which is likely due to its pure exploratory nature.

\Finding{
    \emph{\gls{ours}} is more efficient than \emph{\gls{sga}} and \emph{\gls{rs}} in identifying diverse, severe violations across a range of search budgets and thresholds, \annotate{Comment 1.8}\addedtext{achieving an average \emph{\gls{auc}} improvement of 46\% over \emph{\gls{sga}} and 32\% over \emph{\gls{rs}} for the \emph{\gls{ds}} metric, and 90\% over \emph{\gls{sga}} and 86\% over \emph{\gls{rs}} for the \emph{\gls{mrc}} metric.}\deletedtext{demonstrating superior efficiency in exploring the search space.}
    The performance advantage of \emph{\gls{ours}} persists across various threshold settings, with its lead slightly diminishing only at the highest distance thresholds (enforcing very high diversity) or under low-fitness thresholds.
}

\addedtext{
\subsection{RQ3: Effectiveness of Diversity Mechanisms in CoCoMEGA}\annotate{Comment 2.3}

\subsubsection{Methodology}
To answer \labelcref{rq3}, we evaluate the impact of the diversity mechanisms in \emph{\gls{ours}} by comparing the performance of the full method with a variant that excludes the diversity mechanisms, i.e., fitness clearing and diversity optimization for population archives.
The variant, denoted as \emph{\gls{ours}\textbackslash{}d}, is identical to \emph{\gls{ours}} in all aspects except that it does not include the diversity mechanisms.
We compare the two methods in terms of \emph{\gls{ds}} and \emph{\gls{sd}} across different fitness and distance thresholds.
The methodology mirrors that of \labelcref{rq1}, using identical hyperparameters and performing 10 repetitions for each method, with the only difference being the presence of diversity mechanisms.
}

\begin{figure*}[htbp]
    \centering
    \includegraphics[width=.75\linewidth]{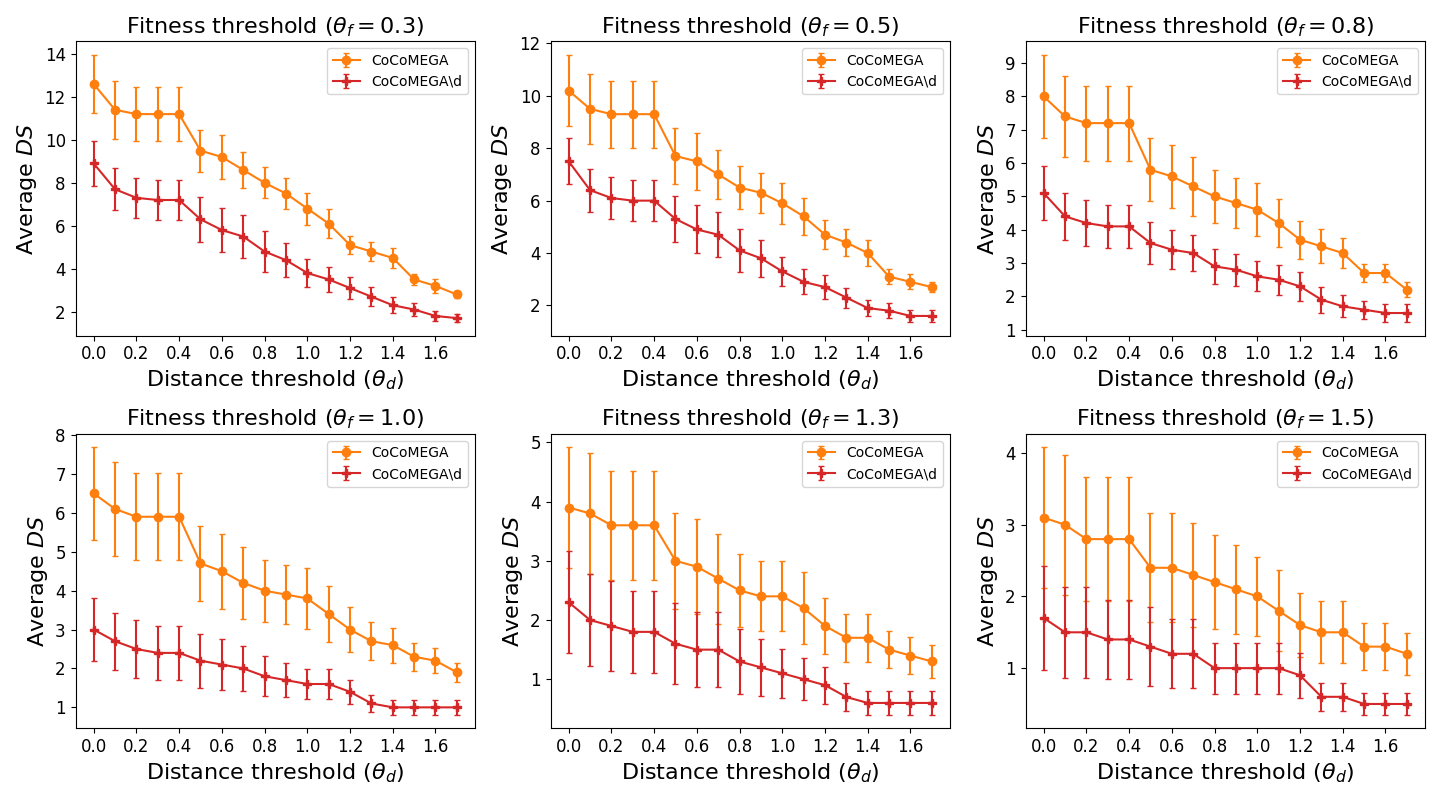}
    \caption{
        \protect\addedtext{Comparison of \emph{\gls{ours}} with and without diversity mechanisms:
        \emph{\acrfull{ds}} vs.\ distance threshold (\(\theta_d\)) across different fitness thresholds (\(\theta_f\)).
        A higher \emph{\gls{ds}} indicates more distinct violating solutions discovered by the method.
        Each curve plots the average \emph{\gls{ds}} at different \(\theta_d\) settings, under a specific fitness threshold \(\theta_f\).
        This figure reveals how diversity mechanisms impact the quantity and diversity of solutions as \(\theta_f\) varies.}
    }\label{fig:ds-diversity}
\end{figure*}

\begin{figure*}[htbp]
    \centering
    \includegraphics[width=.75\linewidth]{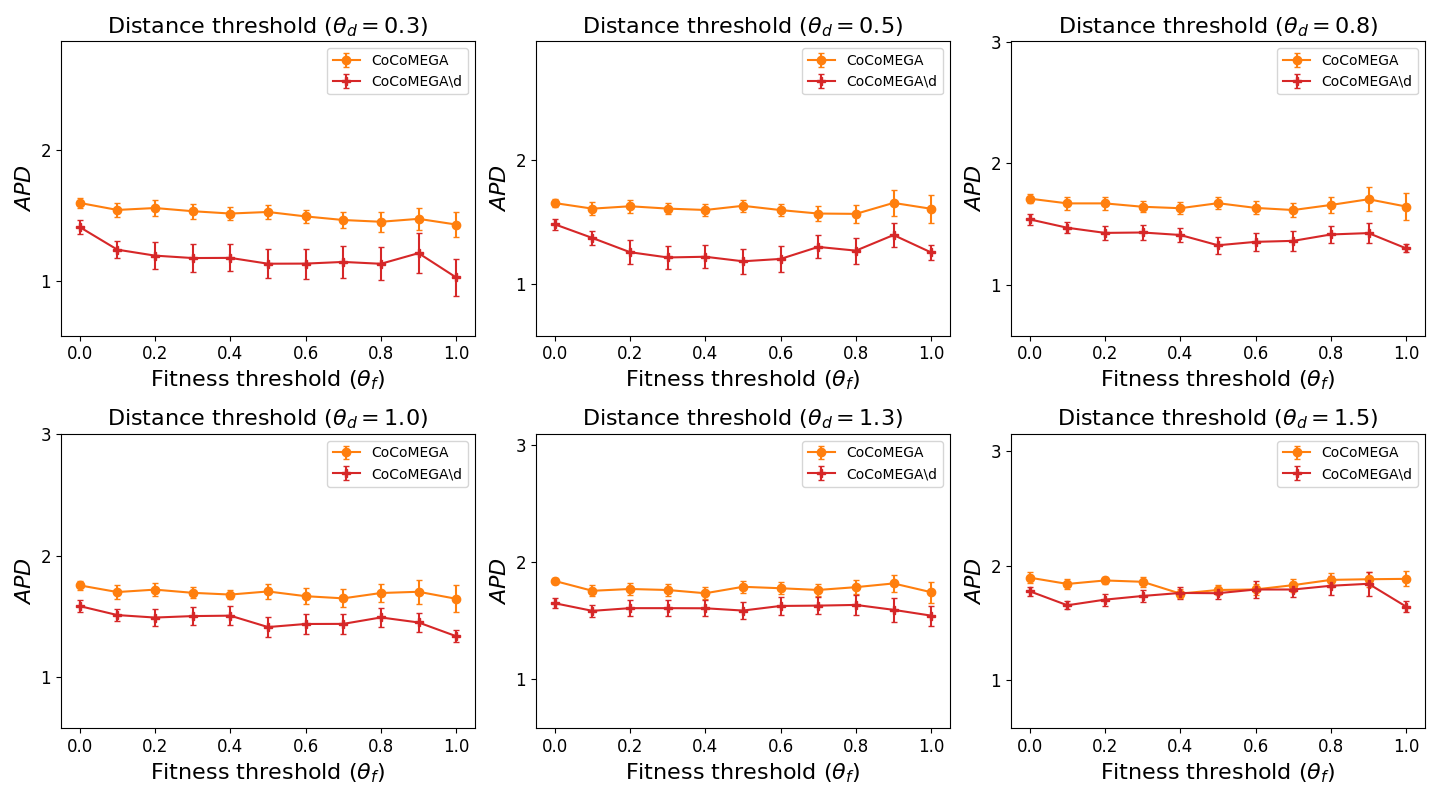}
    \caption{
        \protect\addedtext{Comparison of \emph{\gls{ours}} with and without diversity mechanisms:
        \emph{\acrfull{apd}} vs.\ fitness threshold (\(\theta_f\)) across different distance thresholds (\(\theta_d\)).
        \emph{\gls{apd}} quantifies how spread out the solutions are.
        A higher \emph{\gls{apd}} indicates greater diversity among the found solutions.
        Each subplot corresponds to a certain distance threshold \(\theta_d\), and the y-axis shows the \emph{\gls{apd}} of discovered solutions under various \(\theta_f\) values.
        This figure demonstrates how diversity mechanisms maintain diversity while striving for higher fitness (i.e., more severe violations).}
    }\label{fig:apd-diversity}
\end{figure*}

\begin{table*}[htbp]
    \scriptsize
    \setlength\tabcolsep{3.5pt}
    \caption{\protect\addedtext{\gls{mrc} and \gls{cmr} values for \gls{ours} with and without diversity mechanisms at different values of \(\theta_f\) and \(\theta_d\).}}\label{tab:mrc-cmr-diversity}
    \begin{tabularx}{\linewidth}{p{1.2em}p{5.3em}YYYYYY}
        \toprule
        \multirow{2}{1.2em}{\(\boldsymbol{\theta_d}\)} & \multirow{2}{*}{\textbf{Method}}   & \multicolumn{6}{c}{\textbf{Average \emph{\gls{mrc}}\(\boldsymbol{\pm CI_{0.95}}\) (Average \emph{\gls{cmr}}\(\boldsymbol{\pm CI_{0.95}}\))}}                                                                                                                                                                                                                                          \\
        \cmidrule{3-8}
                                                       &                                    & \(\theta_f=0.3\)                                                                                                                             & \(\theta_f=0.5\)                            & \(\theta_f=0.8\)                            & \(\theta_f=1.0\)                             & \(\theta_f=1.3\)                             & \(\theta_f=1.5\)                             \\
        \midrule
        \multirow{2.5}{1.2em}{\(\theta_d=0.0\)}        & \emph{\gls{ours}}                  & \(\boldsymbol{100.0\pm0.0}\) (\(3.0\pm0.3\))                                                                                                 & \(\boldsymbol{98.0\pm1.8}\) (\(2.6\pm0.3\)) & \(\boldsymbol{90.0\pm7.2}\) (\(2.1\pm0.3\)) & \(\boldsymbol{78.0\pm10.9}\) (\(1.8\pm0.4\)) & \(\boldsymbol{66.0\pm12.3}\) (\(1.3\pm0.3\)) & \(\boldsymbol{56.0\pm13.1}\) (\(1.0\pm0.3\)) \\
        \cmidrule{2-8}
                                                       & \emph{\gls{ours}\textbackslash{}d} & \(86.0\pm7.6\) (\(2.9\pm0.4\))                                                                                                               & \(86.0\pm7.6\) (\(2.5\pm0.4\))              & \(82.0\pm9.1\) (\(2.2\pm0.3\))              & \(56.0\pm12.0\) (\(1.4\pm0.4\))              & \(34.0\pm11.7\) (\(1.1\pm0.4\))              & \(34.0\pm11.7\) (\(0.9\pm0.3\))              \\
        \midrule
        \multirow{2.5}{1.2em}{\(\theta_d=1.0\)}        & \emph{\gls{ours}}                  & \(\boldsymbol{100.0\pm0.0}\) (\(2.9\pm0.3\))                                                                                                 & \(\boldsymbol{98.0\pm1.8}\) (\(2.6\pm0.3\)) & \(\boldsymbol{90.0\pm7.2}\) (\(2.1\pm0.3\)) & \(\boldsymbol{78.0\pm10.9}\) (\(1.8\pm0.4\)) & \(\boldsymbol{66.0\pm12.3}\) (\(1.3\pm0.3\)) & \(\boldsymbol{56.0\pm13.1}\) (\(1.0\pm0.3\)) \\
        \cmidrule{2-8}
                                                       & \emph{\gls{ours}\textbackslash{}d} & \(84.0\pm8.4\) (\(2.3\pm0.3\))                                                                                                               & \(84.0\pm8.4\) (\(2.1\pm0.3\))              & \(80.0\pm9.7\) (\(1.9\pm0.3\))              & \(54.0\pm12.0\) (\(1.2\pm0.3\))              & \(32.0\pm11.5\) (\(0.9\pm0.3\))              & \(32.0\pm11.5\) (\(0.8\pm0.3\))              \\
        \midrule
        \multirow{2.5}{1.2em}{\(\theta_d=1.7\)}        & \emph{\gls{ours}}                  & \(\boldsymbol{98.0\pm1.8}\) (\(2.0\pm0.3\))                                                                                                  & \(\boldsymbol{94.0\pm3.8}\) (\(2.1\pm0.2\)) & \(\boldsymbol{86.0\pm7.6}\) (\(1.6\pm0.1\)) & \(\boldsymbol{78.0\pm10.9}\) (\(1.3\pm0.2\)) & \(\boldsymbol{64.0\pm12.0}\) (\(1.1\pm0.2\)) & \(\boldsymbol{56.0\pm13.1}\) (\(1.0\pm0.3\)) \\
        \cmidrule{2-8}
                                                       & \emph{\gls{ours}\textbackslash{}d} & \(66.0\pm9.3\) (\(1.5\pm0.2\))                                                                                                               & \(70.0\pm9.0\) (\(1.4\pm0.1\))              & \(68.0\pm10.1\) (\(1.2\pm0.2\))             & \(50.0\pm11.8\) (\(0.9\pm0.2\))              & \(28.0\pm10.5\) (\(0.6\pm0.2\))              & \(28.0\pm10.5\) (\(0.5\pm0.2\))              \\
        \bottomrule
    \end{tabularx}
\end{table*}

\addedtext{
\subsubsection{Results}
\Cref{fig:ds-diversity} compares the \emph{\gls{ds}} values of \emph{\gls{ours}} with and without diversity mechanisms across various fitness and distance thresholds.
The results indicate that \emph{\gls{ours}} consistently achieves higher \emph{\gls{ds}} values than \emph{\gls{ours}\textbackslash{}d}, highlighting the significant contribution of diversity mechanisms to identifying distinct solutions.
This performance gap remains consistent across all fitness and distance thresholds.
While both methods experience a decrease in \emph{\gls{ds}} values as \(\theta_d\) increases, \emph{\gls{ours}} maintains its superiority, underscoring the effectiveness of its diversity mechanisms in preserving solution diversity.
These findings suggest that diversity mechanisms are crucial for promoting a broadly distributed solution set and maintaining an expansive search space.

\Cref{fig:apd-diversity} depicts the \emph{\gls{apd}} between solutions across different fitness and distance thresholds.
\emph{\gls{ours}} consistently produces solutions with higher \emph{\gls{apd}} compared to its counterpart without diversity, indicating a broader exploration of the solution space and a more distributed solution set.
The difference is especially significant at lower \(\theta_d\) values, where \emph{\gls{ours}\textbackslash{}d} produces solutions with lower \emph{\gls{apd}}, suggesting solutions that are more clustered together in the search space.
This confirms that the diversity mechanisms contribute to exploring more varied solutions.

\Cref{tab:mrc-cmr-diversity} presents the \emph{\gls{mrc}} and \emph{\gls{cmr}} values for both \emph{\gls{ours}} and \emph{\gls{ours}\textbackslash{}d}.
\emph{\gls{ours}} consistently achieves higher \emph{\gls{mrc}} across various \(\theta_f\) values than \emph{\gls{ours}\textbackslash{}d}.
The absence of diversity mechanisms results in a even larger decrease in \emph{\gls{mrc}} at higher \(\theta_f\) values, highlighting their importance in preserving solution diversity.
Similarly, \emph{\gls{cmr}} values are notably higher with diversity enforcement.
These findings confirm the effectiveness of diversity mechanisms in improving both solution coverage and variability.

\Finding{
    \addedtext{
    Diversity mechanisms play a crucial role in enhancing solution diversity and exploration.
    \emph{\gls{ours}} with diversity mechanisms consistently outperforms its counterpart without such mechanisms, identifying 91\% more \gls{mr} violations in terms of \emph{\gls{ds}} across various fitness and distance thresholds.
    Additionally, these mechanisms improve \emph{\gls{mrc}} and \emph{\gls{cmr}} values by 50\% and 33\%, respectively, ensuring broader solution coverage and variability.
    }
}
}

%% file: Discussion.tex
\section{Discussion}\label{sec:discussion}

\annotate{Comment 1.10}\addedtext{
\subsection{Developer Feedback}\label{sec:dev_feedback}

To complement our automated evaluation, we sought feedback from three domain experts with extensive safety engineering experience in \glspl{ads}. Our objective was to determine whether the detected \gls{mr} violations represented safety risks from the experts' perspective, and if so, to what extent. Informed consent was obtained from the domain experts for the use of their feedback in this study.

We supplied the experts with 50 test cases representing the most severe violations found by \gls{ours} across 10 executions.
For each test case, we provided source and follow-up videos along with velocity-change diagrams, highlighting where the violations occurred.
We then asked two questions to be answered on a Likert scale~\cite{likert1932technique}:
\begin{enumerate*}[label = (\roman*)]
    \item ``This test case entail safety risks'' (Rate from 1\,=\,Strongly Disagree to 5\,=\,Strongly Agree), and
    \item ``How severe is the issue in terms of its potential impact on driving safety?'' (Rate from 1\,=\,Negligible to 5\,=\,Critical). 
\end{enumerate*}

\begin{figure}[htbp]
    \centering
    \includegraphics[width=\linewidth]{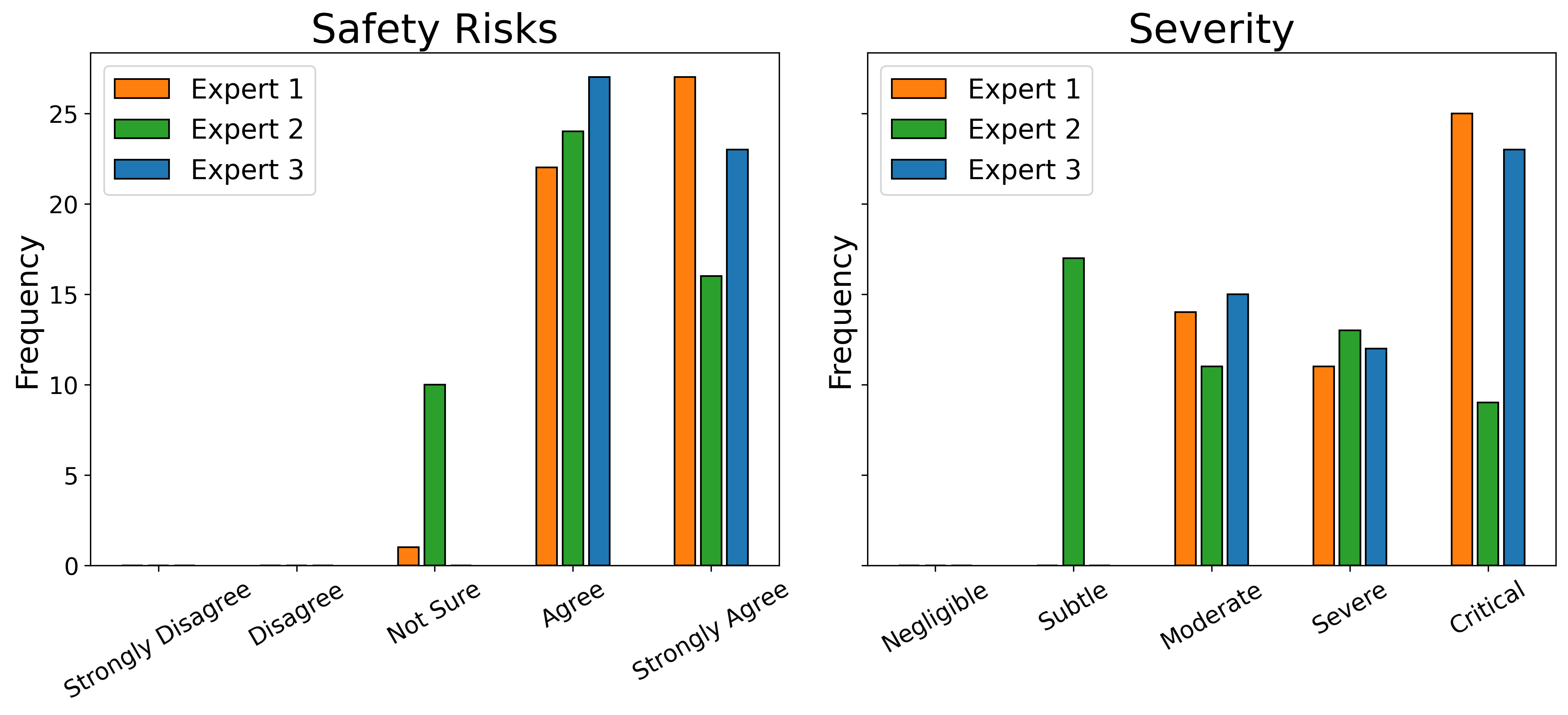}
    \caption{
        Combined histograms of expert ratings for the 50 test cases:
        Left subplot shows responses to ``This test case entail safety risks''
        and the right subplot shows responses to ``How severe is the issue in terms of its potential impact on driving safety?'' 
        Each bar color represents a different expert (orange for Expert 1, green for Expert 2, and blue for Expert 3).
    }\label{fig:dev-feedback}
\end{figure}

As seen in \cref{fig:dev-feedback}, experts tended to rank the test cases on the higher end of both scales.
For the question on safety risks (left subplot), a large proportion of ratings fell into ``Agree'' or ``Strongly Agree'' (i.e., 4 or 5), with average scores of 4.52, 4.12, and 4.46 for expert 1, 2, and 3, respectively. This suggests that most of the \gls{mr}-violating scenarios were perceived as posing genuine safety concerns.
Similarly, for severity (right subplot), many violations were deemed ``Severe'' or ``Critical'' (i.e., 4 or 5), with average scores of 4.22, 3.28, and 4.16, respectively, underlining that these issues could have a substantial impact on driving safety. These expert assessments confirm that the test cases generated by our approach are indeed viewed by practitioners as indicative of noteworthy safety risks, thereby underscoring the practical importance of the discovered violations.
}

\subsection{Interpretation of Solution Diversity}
The metrics of \emph{\gls{apd}}, \emph{\gls{mrc}}, and \emph{\gls{cmr}} provide complementary perspectives on the diversity of identified complete solutions, together forming a robust framework for assessing diversity in the test cases generated for \glspl{ads}.

The \emph{\gls{apd}} metric, representing the average distance between pairs of complete solutions, provides a straightforward quantification of diversity in the scenario space.
Higher \emph{\gls{apd}} values indicate that complete solutions are spread out and capture a broader array of possible scenarios.
This metric is particularly useful for understanding the variation in complete solutions regarding scenario representation, as it directly reflects how far solutions diverge from one another.
However, \emph{\gls{apd}} has limitations.
For instance, methods like \emph{\gls{rs}} may yield a few widely spaced solutions, inflating \emph{\gls{apd}} despite limited diversity.
Thus, \emph{\gls{apd}} alone may not fully capture diversity when solutions are few and highly dispersed.

The \emph{\gls{mrc}} and \emph{\gls{cmr}} metrics, on the other hand, measure the percentage of predefined \glspl{mr} covered by the complete solutions and the unique combinations of \glspl{mr} that the solutions violate, providing different aspects of diversity by assessing how comprehensively the solutions explore the possible ways in which system behaviors can lead to \gls{mr} violations and examining how solutions interact with multiple \glspl{mr} simultaneously.
Unlike \emph{\gls{mrc}}, which is concerned with the total coverage of individual \glspl{mr}, \emph{\gls{cmr}} captures further the complexity and richness of the testing scenarios, indicating that complete solutions cover not only isolated violations but also complex, interrelated ones.
This capability is crucial for understanding potential compounded risks in \glspl{ads} that may emerge when multiple \glspl{mr} are violated simultaneously.
High values in \emph{\gls{mrc}} or \emph{\gls{cmr}} reflect a robust exploration of critical behaviors, essential for evaluating system resilience under more nuanced and complex conditions.

In summary, together, \emph{\gls{apd}}, \emph{\gls{mrc}}, and \emph{\gls{cmr}} offer a multifaceted approach to evaluating solution diversity.
\emph{\gls{apd}} provides a measure of structural diversity, while \emph{\gls{mrc}} and \emph{\gls{cmr}} focus on behavioral diversity, reflecting how comprehensively the complete solutions challenge the system's response to diverse conditions and potential violations.
This combined approach ensures that the test cases not only cover a wide range of possible scenarios but also uncover complex, high-risk scenarios, which are critical for a comprehensive assessment of \glspl{ads}.

\subsection{Threats to Validity}
In this section, we analyze possible threats to validity of our results.
These threats are grouped into four categories: Internal, External, Construct, and Conclusion~\cite{wohlin2012experimentation}.

\subsubsection{Internal validity}
Internal validity refers to the degree to which a study reliably demonstrates a cause-and-effect relationship between the experiment and its outcome.
In our research, one of the primary threats to internal validity is the sensitivity of population-based methods (i.e., \emph{\gls{ours}} and \emph{\gls{sga}}) to hyperparameter values, as the current settings may not reflect optimal configurations.
To address this, we adopted widely recommended hyperparameter values~\cite{mirjalili2018genetic}, which are frequently employed in the literature.
For parameters without recommended values, we determined them through a pilot experiment.

Another internal validity threat concerns the correctness of the \glspl{mr} used in the experiment, as they play a central role in identifying \gls{mr} violations.
Any incorrect or overly restrictive \glspl{mr} could lead to false positives or negatives, potentially affecting the reliability of test results.
For instance, an inaccurate threshold in an output relation could either fail to detect a true violation or detect one where there is not.
To mitigate this, we selected \glspl{mr} that are well-established in the literature and validated them through a pilot experiment before full implementation in experiments.
This phase allowed us to identify and analyze discrepancies or unexpected results, refining the \glspl{mr} as needed, e.g., thresholds in output relations were adjusted.
This process ensured the \glspl{mr} were well-calibrated and reliable for the extensive experiments.

A further threat to internal validity arises from minor discrepancies between the actual number of executed simulations and the allocated simulation budget for the population-based methods.
Ensuring consistent budgets across different methods is essential for a fair comparison and we addressed this issue by using linear interpolation to align simulation budgets.
However, linear interpolation introduces its own risks, as it assumes that the relationship between the search budget and metrics is linear and continuous.
This assumption may not hold in cases where metrics exhibit non-linear or abrupt changes across budgets, potentially introducing inaccuracies in the aligned results.
To mitigate this threat, we validated the interpolation approach by examining trends in metrics and confirming that the linear assumption holds within the tested budget range.

\subsubsection{External validity}
External validity addresses the extent to which the results can be generalized to various contexts and examines the gap between the simulation environment and real-world conditions.
In our research, the primary concern regarding external validity was the potential limitations arising from the use of a specific \gls{ads} (\textsc{Interfuser}) and simulator (\textsc{Carla}).
However, \textsc{Carla} is highly regarded and widely used open-source simulator known for its high fidelity, and \textsc{Interfuser} was a top-performing \gls{ads} on the \textsc{Carla} leaderboard~\cite{carlaleaderboard} during our evaluation period.
Moreover, the considerable implementation overhead required to enable scenario execution with \textsc{Carla} and \textsc{Interfuser} significantly constrained our ability to evaluate additional technologies.
As part of our future research, we intend to extend our approach to other \glspl{ads} and explore its applicability to domains such as autonomous mobile robotic systems and unmanned aerial systems.
This will contribute to a more comprehensive understanding of the effectiveness and limitations of our methodology in various real-world contexts.

Another potential threat to the external validity of our study lies in the limited variety of \glspl{mr} we considered, as we focused only on two types of common output relations, i.e.,\annotate{Comment 3.6} \emph{invariance} and \emph{decreasing} output relations.
\addedtext{
These \glspl{mr} are not representative of more complex (e.g., nonlinear or non-monotonic) relationships, thus potentially restricting} the generalizability of our findings, as other types of input and output relations exist that could uncover different types of \gls{mr} violations.
However, we selected the most widely used \glspl{mr} in system-level testing of \glspl{ads}, from the literature, thus capturing common types of system behaviors relevant to \glspl{ads}.
Expanding the study to include a broader variety of \glspl{mr}, which remain to be defined, could provide a more comprehensive assessment and further validate the robustness of our approach across diverse testing scenarios.

\subsubsection{Construct validity}
Construct validity addresses the extent to which the object of study truly reflects the underlying theory behind it.
In our research, a key concern related to construct validity is the suitability of the \emph{\gls{ds}} and \emph{\gls{sd}} metrics for representing the effectiveness of each method in identifying severe \gls{mr} violations and exploring diverse regions of the search space.
While these metrics are chosen to capture the concept of effectiveness, they may not fully account for all dimensions of solution quality and diversity.

\subsubsection{Conclusion validity}
Conclusion validity refers to the extent to which a research conclusion could be trusted.
In our experiments, we encountered limitations due to the relatively small number of repetitions due to enormous computation costs, i.e., 10 runs per method were conducted.
To address the potential impact of statistical error arising from this limited sample size, we have systematically reported descriptive statistics in conjunction with their corresponding confidence intervals, providing insights into the central tendency and variations of results.

%% file: Conclusion.tex
\section{Conclusion}\label{sec:conclusion}
In this paper, we introduce \emph{\gls{ours}}, a novel automated testing method that effectively combines, for the first time, \acrfull{mt} and advanced search-based testing to support the system-level assessments of \glspl{ads}.
By integrating \gls{mt}, \emph{\gls{ours}} effectively detects test cases that violate \acrfullpl{mr}, capturing desirable properties expected to hold in \glspl{ads}, even when safety requirements are not violated. This helps uncover subtle safety risks.
Our evaluation on the \textsc{Carla} simulator using the \textsc{Interfuser} \gls{ads} demonstrated that \emph{\gls{ours}} outperforms baseline methods across various budget and threshold configurations, efficiently generating severe and diverse \gls{mr} violations that explore wider regions of the search space.
\addedtext{Expert assessments of these violations confirmed that most pose real safety risks, with many deemed severe or critical, underscoring their practical relevance to improving \gls{ads} safety.}
These results suggest \emph{\gls{ours}} is an effective and efficient solution to the challenges of \gls{ads} testing by decomposing the high-dimensional search space using a \acrfull{ccea}, thus expanding the test coverage of potential safety risks.
Future work could extend this approach to additional simulators, broaden its scope to other complex autonomous systems, and further explore strategies for improved testing efficiency such as surrogate modeling.

%% file: Appendices.tex
{\appendix[Experimental Results on \texorpdfstring{\(GP_3\)}{GP3}]\label{sec:appx}

Based on \(GP_3\), \cref{fig:ds-set2} visually compares the \emph{\gls{ds}} achieved by \emph{\gls{ours}}, \emph{\gls{sga}}, and \emph{\gls{rs}} across 10 experimental runs, varying by fitness threshold \(\theta_f\) and distance threshold \(\theta_d\).
Each subplot displays \(\theta_d\) on the x-axis and the average \emph{\gls{ds}} on the y-axis, with 95\% confidence intervals represented as error bars.

\Cref{fig:apd-by-gen-set2} illustrates the relationship between \emph{\gls{apd}}, \(\theta_f\), and \(\theta_d\) for the methods across 10 experimental runs.
Each subplot corresponds to a specific \(\theta_d\) value, with \(\theta_f\) along the x-axis and \emph{\gls{apd}} on the y-axis, also displaying 95\% confidence intervals.

\Cref{fig:apd-over-gen-set2} tracks the \emph{\gls{apd}} over seven generations for the methods under varying \(\theta_f\) and \(\theta_d\).
Each subplot represents a unique \(\theta_f\)-\(\theta_d\) combination, showing \emph{\gls{apd}} progression over generations.
Notably, for \emph{\gls{sga}} and \emph{\gls{rs}}, some \emph{\gls{apd}} values are missing in certain or all generations due to an insufficient number of complete solutions meeting the specified thresholds.

\Cref{tab:mrc-cmr-set2} summarizes the relationship between \emph{\gls{mrc}} and \emph{\gls{cmr}} values across different \(\theta_f\) and \(\theta_d\) values for each method under the same search budget.

\Cref{fig:dss-set2} tracks the effectiveness of these methods as the search budget incrementally increases from 10\% to 100\% of the total, while \cref{fig:mrcs-set2} illustrates the evolution of \emph{\gls{mrc}} relative to the search budget consumed.

\begin{figure*}[htbp]
    \centering
    \includegraphics[width=.8\linewidth]{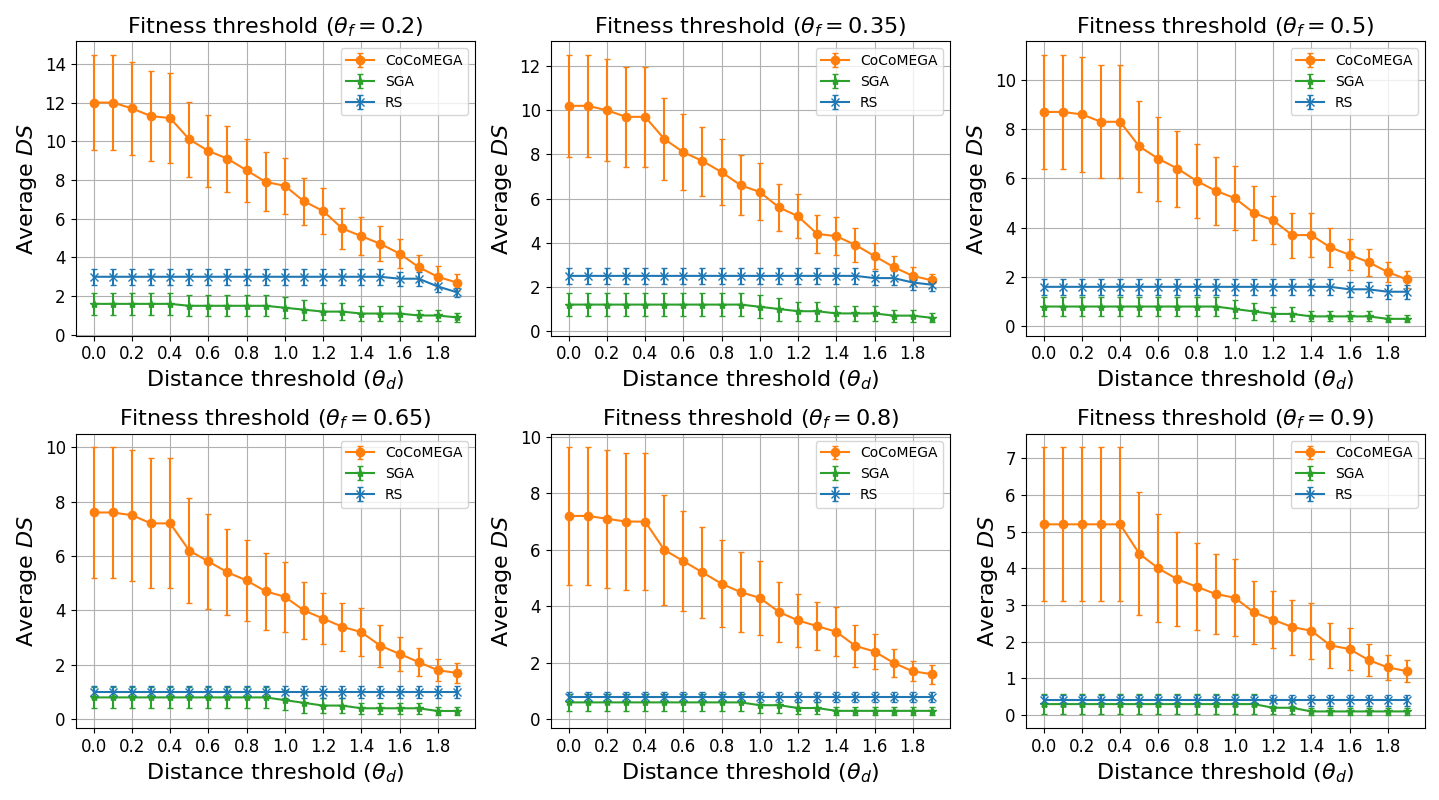}
    \caption{
        \emph{\acrfull{ds}} vs.\ distance threshold (\(\theta_d\)) across different fitness thresholds (\(\theta_f\)).
        A higher \emph{\gls{ds}} indicates more distinct violating solutions discovered by the method.
        Each curve plots the average \emph{\gls{ds}} at different \(\theta_d\) settings, under a specific fitness threshold \(\theta_f\).
        This figure reveals how each method balances the quantity and diversity of solutions as \(\theta_f\) varies.
    }\label{fig:ds-set2}
\end{figure*}

\begin{figure*}[htbp]
    \centering
    \includegraphics[width=.8\linewidth]{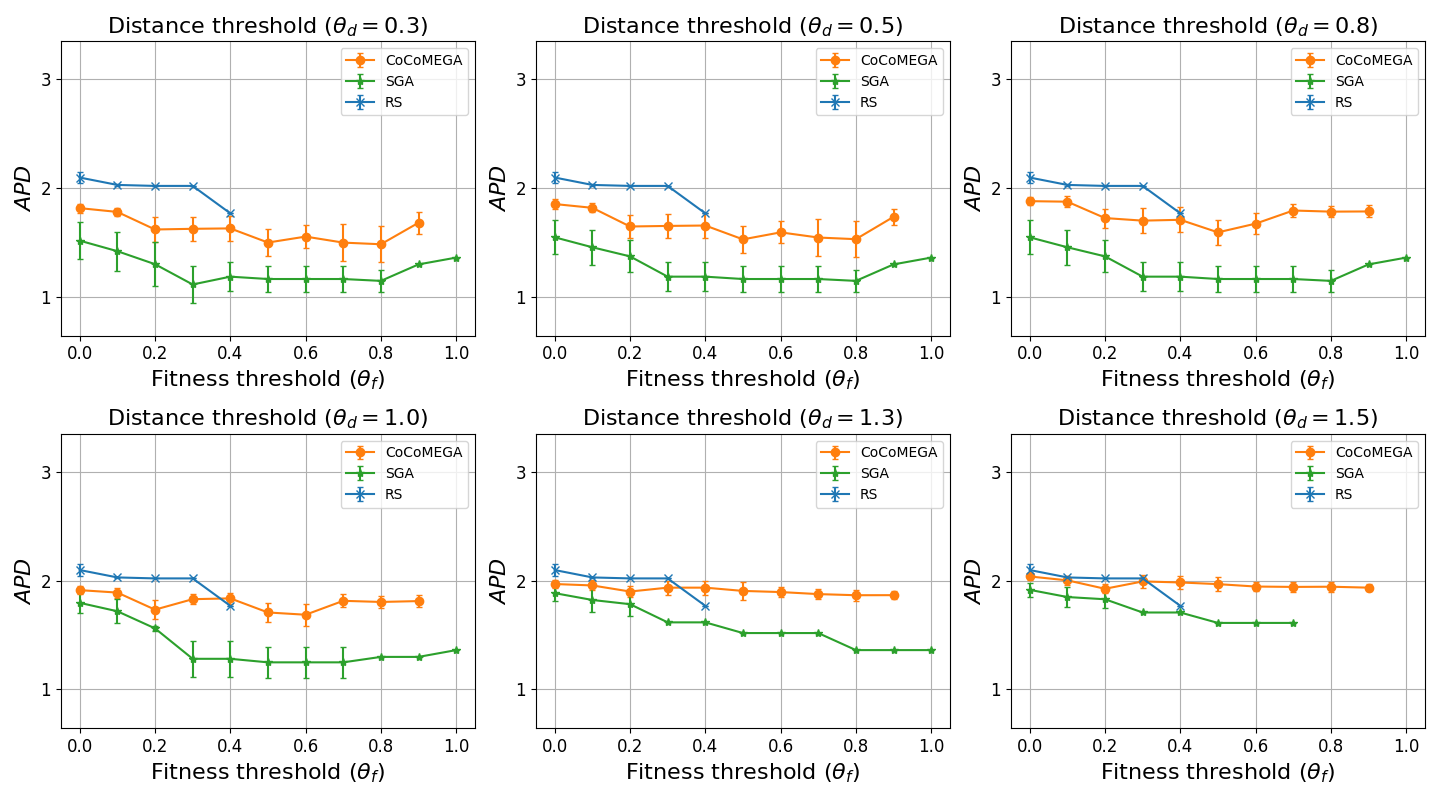}
    \caption{
        \emph{\acrfull{apd}} vs.\ fitness threshold (\(\theta_f\)) across different distance thresholds (\(\theta_d\)).
        \emph{\gls{apd}} quantifies how spread out the solutions are.
        A higher \emph{\gls{apd}} indicates greater diversity among the found solutions.
        Each subplot corresponds to a certain distance threshold \(\theta_d\), and the y-axis shows the \emph{\gls{apd}} of discovered solutions under various \(\theta_f\) values.
        This figure demonstrates how each method maintains diversity while striving for higher fitness (i.e., more severe violations).
    }\label{fig:apd-by-gen-set2}
\end{figure*}

\begin{figure*}[htbp]
    \centering
    \includegraphics[width=.65\linewidth]{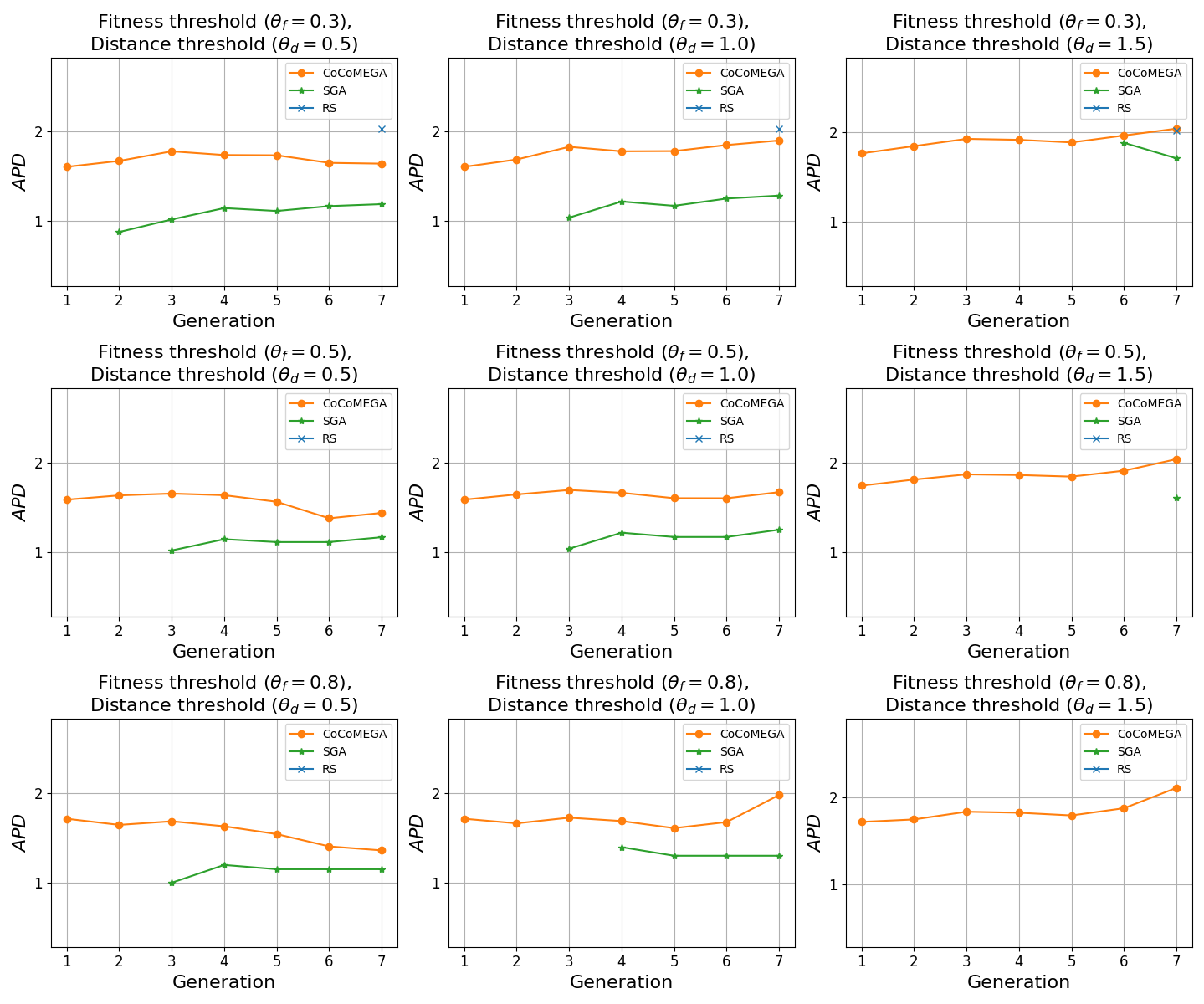}
    \caption{
        Evolution of \emph{\acrfull{apd}} across generations.
        Higher \emph{\gls{apd}} values imply that solutions are more spread out in the search space, indicating broad exploration.
        Subplots differ by the chosen fitness threshold (\(\theta_f\)) and distance threshold (\(\theta_d\)).
        Each line tracks the \emph{\gls{apd}} for a specific method over several generations of search.
        This figure illustrates how each method maintains or improves solution diversity over generations.
    }\label{fig:apd-over-gen-set2}
\end{figure*}

\begin{table*}[htbp]
    \scriptsize
    \setlength\tabcolsep{3.5pt}
    \caption{\gls{mrc} and \gls{cmr} values for different methods at different values of \(\theta_f\) and \(\theta_d\).}\label{tab:mrc-cmr-set2}
    \begin{tabularx}{\linewidth}{llYYYYYY}
        \toprule
        \multirow{2}{2em}{\(\boldsymbol{\theta_d}\)} & \multirow{2}{*}{\textbf{Method}} & \multicolumn{6}{c}{\textbf{Average \emph{\gls{mrc}}\(\boldsymbol{\pm CI_{0.95}}\) (Average \emph{\gls{cmr}}\(\boldsymbol{\pm CI_{0.95}}\))}}                                                                                                                                                                                                                                            \\
        \cmidrule{3-8}
                                                     &                                  & \(\theta_f=0.2\)                                                                                                                             & \(\theta_f=0.35\)                            & \(\theta_f=0.5\)                             & \(\theta_f=0.65\)                            & \(\theta_f=0.8\)                             & \(\theta_f=0.9\)                             \\
        \midrule
        \multirow{4}{2em}{\(\theta_d=0.0\)}          & \emph{\gls{ours}}                & \(\boldsymbol{91.7\pm7.5}\) (\(3.0\pm0.4\))                                                                                                  & \(\boldsymbol{81.7\pm9.9}\) (\(2.6\pm0.4\))  & \(\boldsymbol{80.0\pm10.9}\) (\(2.1\pm0.4\)) & \(\boldsymbol{73.3\pm12.3}\) (\(2.0\pm0.4\)) & \(\boldsymbol{65.0\pm13.0}\) (\(2.0\pm0.4\)) & \(\boldsymbol{61.7\pm14.2}\) (\(1.7\pm0.4\)) \\
        \cmidrule{2-8}
                                                     & \emph{\gls{sga}}                 & \(38.3\pm12.9\) (\(1.2\pm0.5\))                                                                                                              & \(31.7\pm12.6\) (\(1.0\pm0.4\))              & \(28.3\pm13.1\) (\(0.6\pm0.3\))              & \(28.3\pm13.1\) (\(0.6\pm0.3\))              & \(21.7\pm11.9\) (\(0.4\pm0.2\))              & \(10.0\pm9.0\) (\(0.2\pm0.2\))               \\
        \cmidrule{2-8}
                                                     & \emph{\gls{rs}}                  & \(83.3\pm10.0\) (\(1.0\pm0.0\))                                                                                                              & \(\boldsymbol{81.7\pm11.1}\) (\(0.9\pm0.1\)) & \(55.0\pm13.6\) (\(0.8\pm0.1\))              & \(36.7\pm12.6\) (\(0.7\pm0.1\))              & \(20.0\pm8.3\) (\(0.7\pm0.1\))               & \(6.7\pm2.5\) (\(0.4\pm0.1\))                \\
        \midrule
        \multirow{4}{2em}{\(\theta_d=1.0\)}          & \emph{\gls{ours}}                & \(\boldsymbol{91.7\pm7.5}\) (\(3.0\pm0.4\))                                                                                                  & \(\boldsymbol{81.7\pm9.9}\) (\(2.5\pm0.4\))  & \(\boldsymbol{80.0\pm10.9}\) (\(2.0\pm0.4\)) & \(\boldsymbol{73.3\pm12.3}\) (\(1.8\pm0.4\)) & \(\boldsymbol{65.0\pm13.0}\) (\(1.8\pm0.4\)) & \(\boldsymbol{61.7\pm14.2}\) (\(1.6\pm0.4\)) \\
        \cmidrule{2-8}
                                                     & \emph{\gls{sga}}                 & \(38.3\pm12.9\) (\(1.2\pm0.5\))                                                                                                              & \(31.7\pm12.6\) (\(1.0\pm0.4\))              & \(28.3\pm13.1\) (\(0.6\pm0.3\))              & \(28.3\pm13.1\) (\(0.6\pm0.3\))              & \(21.7\pm11.9\) (\(0.4\pm0.2\))              & \(10.0\pm9.0\) (\(0.2\pm0.2\))               \\
        \cmidrule{2-8}
                                                     & \emph{\gls{rs}}                  & \(83.3\pm10.0\) (\(1.0\pm0.0\))                                                                                                              & \(\boldsymbol{81.7\pm11.1}\) (\(0.9\pm0.1\)) & \(55.0\pm13.6\) (\(0.8\pm0.1\))              & \(36.7\pm12.6\) (\(0.7\pm0.1\))              & \(20.0\pm8.3\) (\(0.7\pm0.1\))               & \(6.7\pm2.5\) (\(0.4\pm0.1\))                \\
        \midrule
        \multirow{4}{2em}{\(\theta_d=1.7\)}          & \emph{\gls{ours}}                & \(\boldsymbol{86.7\pm7.7}\) (\(1.8\pm0.2\))                                                                                                  & \(\boldsymbol{81.7\pm9.9}\) (\(1.6\pm0.2\))  & \(\boldsymbol{80.0\pm10.9}\) (\(1.5\pm0.3\)) & \(\boldsymbol{70.0\pm12.0}\) (\(1.3\pm0.3\)) & \(\boldsymbol{61.7\pm12.5}\) (\(1.3\pm0.3\)) & \(\boldsymbol{56.7\pm13.3}\) (\(1.0\pm0.3\)) \\
        \cmidrule{2-8}
                                                     & \emph{\gls{sga}}                 & \(38.3\pm12.9\) (\(0.9\pm0.3\))                                                                                                              & \(30.0\pm12.0\) (\(0.7\pm0.3\))              & \(23.3\pm11.0\) (\(0.4\pm0.2\))              & \(23.3\pm11.0\) (\(0.4\pm0.2\))              & \(18.3\pm10.1\) (\(0.3\pm0.1\))              & \(6.7\pm6.0\) (\(0.1\pm0.1\))                \\
        \cmidrule{2-8}
                                                     & \emph{\gls{rs}}                  & \(83.3\pm10.0\) (\(1.0\pm0.0\))                                                                                                              & \(\boldsymbol{81.7\pm11.1}\) (\(0.9\pm0.1\)) & \(55.0\pm13.6\) (\(0.8\pm0.1\))              & \(36.7\pm12.6\) (\(0.7\pm0.1\))              & \(20.0\pm8.3\) (\(0.7\pm0.1\))               & \(6.7\pm2.5\) (\(0.4\pm0.1\))                \\
        \bottomrule
    \end{tabularx}
\end{table*}

\begin{figure*}[htbp]
    \centering
    \includegraphics[width=.65\linewidth]{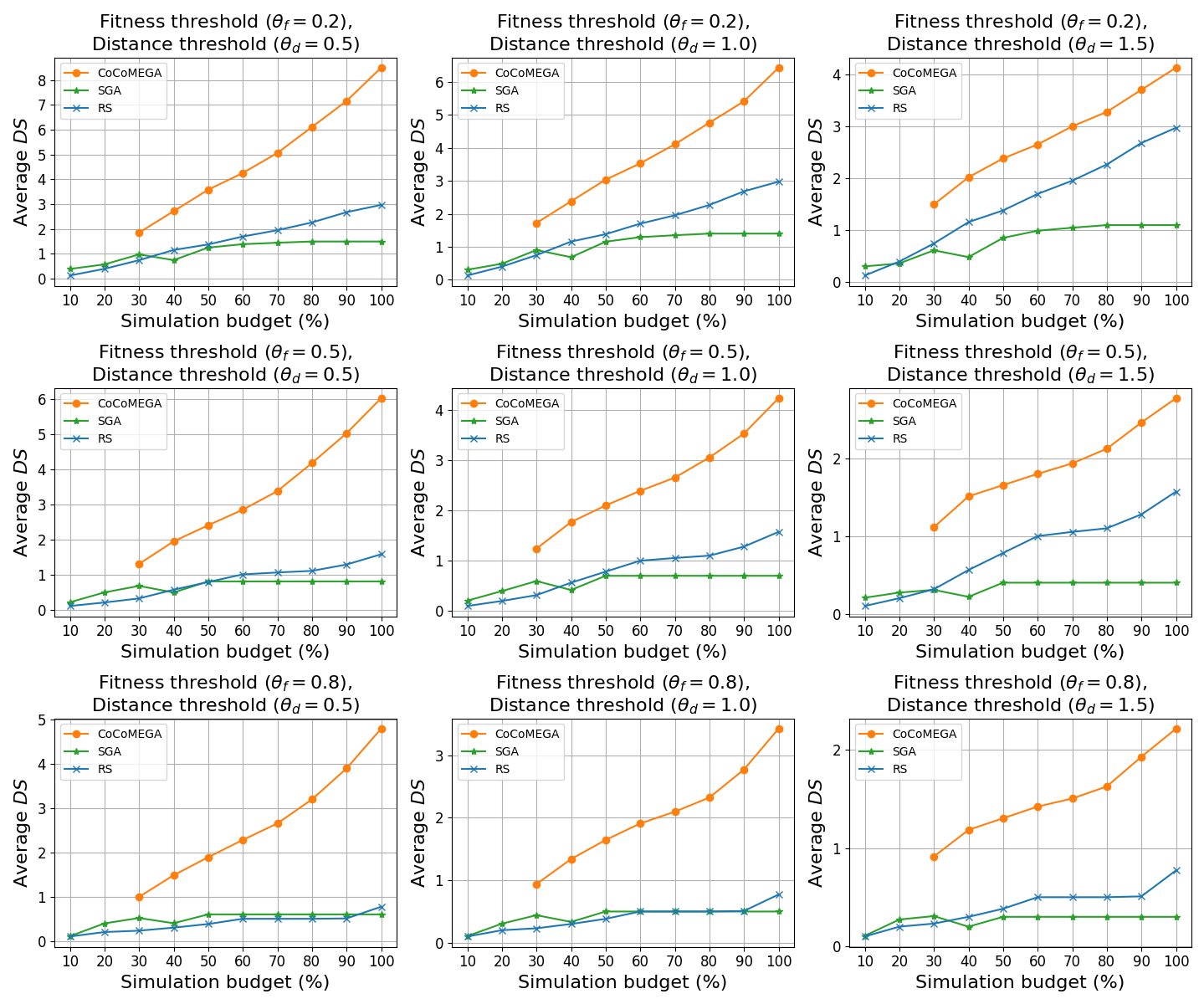}
    \caption{
        \emph{\acrfull{ds}} vs.\ search budget.
        A higher \emph{\gls{ds}} indicates more distinct violating solutions discovered by the method.
        Each subplot shows \emph{\gls{ds}} at incremental percentages of the total simulation budget.
        Different curves represent different methods, compared in distinct fitness threshold (\(\theta_f\)) and distance threshold (\(\theta_d\)) settings.
        This figure highlights each method's efficiency: steeper or higher lines mean the algorithm finds more unique solutions earlier in the budget.
    }\label{fig:dss-set2}
\end{figure*}

\begin{figure*}[htbp]
    \centering
    \includegraphics[width=.65\linewidth]{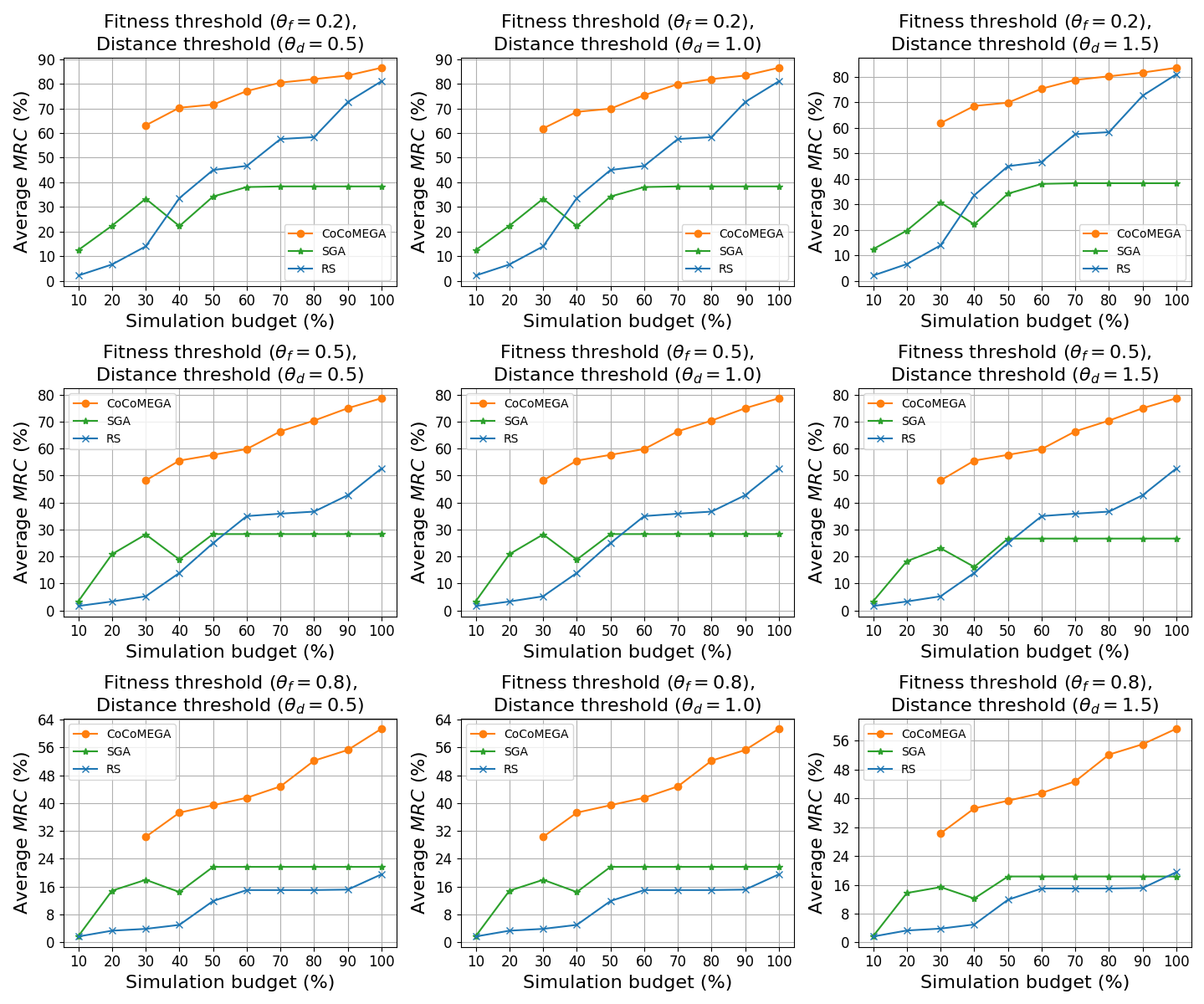}
    \caption{
        \emph{\acrfull{mrc}} vs.\ search budget.
        \emph{\gls{mrc}} indicates the percentage of defined \glspl{mr} that are violated by at least one discovered solution.
        Higher \emph{\gls{mrc}} means the method finds solutions that violate more of the predefined \glspl{mr}, indicating thorough behavioral exploration.
        The x-axis represents how much of the simulation budget has been consumed.
        This figure shows how quickly each method covers all relevant \gls{mr} violations as the search progresses.
    }\label{fig:mrcs-set2}
\end{figure*}
}
